\documentclass[preprint,preprintnumbers,aps,11pt,superscriptaddress,nofootinbib,tightenlines,floatfix]{revtex4-1}

\usepackage{amsmath,amssymb}
\usepackage{graphicx}
\usepackage{bm}
\usepackage{comment}
\usepackage{color}
\usepackage{subfigure}
\usepackage{array}
\usepackage{multirow}
\usepackage{natbib}
\usepackage{url}

\newcommand{\gsim}{\raisebox{-0.7ex}{$\stackrel{\textstyle >}{\sim}$ }}
\newcommand{\lsim}{\raisebox{-0.7ex}{$\stackrel{\textstyle <}{\sim}$ }}
\usepackage[T1]{fontenc}
\usepackage[latin9]{inputenc}
\usepackage{graphicx}
\usepackage{esint}
\usepackage{hyperref}
\usepackage{atbegshi}
\usepackage{lipsum}

\def\siii{^3 \hskip -0.025in S _1}
\def\diii{^3 \hskip -0.025in D _1}

\begin{document}

\dimen\footins=5\baselineskip\relax

\preprint{\vbox{
\hbox{INT-PUB-13-037}
\hbox{NT@UW-13-025} 
}}

\title{Two-Nucleon Systems in a Finite Volume: \\
(II) $\siii$-$\diii$ Coupled
  Channels and the Deuteron
}
\author{Ra\'ul A. Brice\~no{\footnote{\tt rbriceno@jlab.org}}}
\affiliation{Jefferson Laboratory, 12000 Jefferson Avenue, Newport
  News, VA 23606, USA}

\author{Zohreh Davoudi{\footnote{\tt davoudi@uw.edu}}
}
\affiliation{Department of Physics, University of Washington,
 Box 351560, Seattle, WA 98195, USA}
\affiliation{Institute for Nuclear Theory, Box 351550, Seattle, WA 98195-1550, USA}

\
\author{Thomas Luu{\footnote{\tt t.luu@fz-juelich.de}}
}
\affiliation{Institute for Advanced Simulation,
Forschungszentrum J\"{u}lich, D--52425 J\"{u}lich, Germany}
\affiliation{Institut f\"ur Kernphysik and J\"ulich Center for Hadron Physics,
Forschungszentrum J\"{u}lich, D--52425 J\"{u}lich, Germany}

\author{Martin J. Savage{\footnote{\tt mjs5@uw.edu}}}
\affiliation{Department of Physics, University of Washington,
 Box 351560, Seattle, WA 98195, USA}
\affiliation{Institute for Nuclear Theory, Box 351550, Seattle, WA 98195-1550, USA}

\date{\today}

\begin{abstract} 
The energy spectra of two nucleons in a cubic volume 
provide access to the two phase shifts and one mixing angle that define
the S-matrix in the $\siii$-$\diii$ coupled channels containing the deuteron.
With the aid of recently derived energy quantization conditions for such systems,
and the known  scattering parameters,
these spectra are predicted for a range of volumes.
It is found that 
extractions of the infinite-volume deuteron binding energy
and leading scattering parameters, including  
the S-D mixing angle at the deuteron pole,
are possible 
from  Lattice QCD calculations of two-nucleon 
systems with boosts  of $|\mathbf{P}| \leq \frac{2\pi}{\rm L}\sqrt{3}$
in volumes with  $10~\rm{fm} \lesssim {\rm L} \lesssim 14~\rm{fm}$.
The viability of extracting the asymptotic D/S ratio of the
deuteron wavefunction from Lattice QCD calculations is discussed.
\end{abstract}
\maketitle
\section{Introduction  \label{sec:Intro} }
\noindent
An overarching goal of modern nuclear physics is to observe and quantify the
emergence of low-energy nuclear phenomena from quantum chromodynamics (QCD).
A critical step toward this goal is a refinement of the chiral nuclear
forces beyond what has been made possible by decades of experimental
exploration, using Lattice QCD (LQCD).
This numerical technique
is making rapid progress toward predicting low-energy nuclear processes 
with fully-quantified uncertainties~\cite{Beane:2006mx,Ishii:2006ec,Aoki:2008hh,Beane:2009py,Aoki:2009ji,Beane:2010hg,Beane:2011iw,Beane:2012ey,Beane:2012vq,Beane:2013br,Yamazaki:2012hi,Yamazaki:2012fn,Murano:2013xxa}.
The lightest nucleus, the deuteron, played an important historical role in 
understanding the form of the nuclear forces and 
the developments that led to the modern phenomenological nuclear potentials,
e.g. Refs.~\cite{Wiringa:1994wb,Machleidt:2000ge}.
While challenging for LQCD calculations, postdicting the properties of the
deuteron, and other light nuclei, is a critical part of the verification of
LQCD technology
that is required in order to trust predictions of quantities
for which there is little or no experimental guidance.
In nature, the deuteron,
with  total angular momentum and parity of $J^\pi=1^+$,
is the only bound state of a neutron and proton,
bound by $B_d^{\infty}= 2.224644(34)~{\rm MeV}$.
While predominantly S-wave, the non-central components  of the nuclear forces
(the tensor force) induce a D-wave component,
and the $J^\pi=1^+$ two-nucleon (NN) sector 
that contains the deuteron
is a 
$\siii$-$\diii$ coupled-channels system.
An important consequence of the nonconservation
of orbital angular momentum
is that the deuteron is not spherical,
and possesses a non-zero quadrupole moment
(the experimentally measured value of the electric quadrupole moment of the  
deuteron is $Q_d=0.2859(3)~{\rm fm}^2$~\cite{PhysRevA.20.381}).
The S-matrix for this coupled-channels system 
can be parameterized by two phase shifts 
and one mixing angle,
with the mixing angle manifesting itself in the 
asymptotic $D/S$ ratio of the deuteron wavefunction, 
$\eta = 0.02713(6)$~\cite{PhysRev.93.1387, PhysRevC.47.473, deSwart:1995ui}.
A direct calculation of the three scattering 
parameters from QCD, at both physical and unphysical light-quark masses,
would provide important insights into the tensor components of the nuclear forces.

As LQCD calculations are performed in a finite volume (FV) with 
certain boundary conditions (BCs)
imposed upon the fields, 
precise determinations of the deuteron properties from LQCD
requires understanding FV effects. 
Corrections to the binding energy of a bound state, such as
the deuteron, depend exponentially upon the volume,
and are dictated by its size,
and also by the range of the nuclear forces.
With the assumption of a purely S-wave deuteron, 
the leading order (LO) volume corrections 
have been determined
for a deuteron at rest in a cubic volume of spatial extent ${\rm L}$ and with the
fields subject to periodic BCs in the spatial directions~\cite{Luscher:1986pf, Luscher:1990ux, Beane:2003da}.
They are found to 
scale as ${1\over {\rm L}}e^{-\kappa_d^{\infty} {\rm L}}$, where
$\kappa_d^{\infty}$ is the 
infinite-volume
deuteron binding momentum
(in the non-relativistic limit, $\kappa_d^{\infty}=\sqrt{M B_d^{\infty}}$, with $M$ being the nucleon mass).
Volume corrections beyond LO have been determined, and extended to
systems that are moving in the volume~\cite{Konig:2011nz, Bour:2011ef,Davoudi:2011md}.
As $\eta$, $Q_d$, and other observables
dictated  by the tensor interactions, are small at the physical light-quark masses, 
FV analyses of existing LQCD calculations~\cite{Beane:2006mx, Beane:2012vq,  Beane:2013br,
  Yamazaki:2012hi} using 
L\"uscher's method~\cite{Luscher:1986pf, Luscher:1990ux} 
have taken the deuteron to be purely S-wave, neglecting the D-wave admixture,
even at unphysical pion masses, introducing a
systematic uncertainty into these 
analyses.~\footnote{Recent lattice effective field theory (EFT) calculations include
the effects of higher partial waves and mixing~\cite{Lee:2008fa, Bour:2012hn}, 
and thus are able to calculate matrix
elements of non-spherical quantities like $Q_d$ up to a given order in the
low-energy EFT, but their FV
analyses treat the deuteron as a S-wave~\cite{Lee:2008fa, Bour:2012hn}.
}
Although the mixing between the S-wave and D-wave is known to be small at the
physical light-quark masses, its contribution to the 
calculated FV binding energies must be determined in order to
address this systematic uncertainty.
Further, it is not known if the mixing
between these channels remains small at unphysical quark masses.
As the central and tensor components of the nuclear forces have
different forms, their contribution to the FV effects will, in
general, differ.
The contributions from the tensor interactions 
are found to be relatively enhanced for certain  center of mass (CM) boosts in modest
volumes due to the reduced spatial symmetry of the system.
Most importantly, 
extracting the  S-D mixing angle at the deuteron binding energy,
in addition to  the S-wave scattering parameters,
requires a complete coupled-channels analysis of the FV 
spectrum.

Extending the formalism developed for coupled-channel 
systems~\cite{Detmold:2004qn,He:2005ey,Liu:2005kr,Bernard:2008ax,Lage:2009zv,Bernard:2010fp,Ishizuka:2009bx,Briceno:2012yi,Hansen:2012tf,Guo:2012hv,Li:2012bi}, 
the 
FV  formalism describing NN systems with arbitrary CM momenta, spin, angular momentum and isospin
has been developed recently,
providing expressions for the energy eigenvalues 
in irreducible representations (irreps) of the FV symmetry groups~\cite{Briceno:2013lba}.
In this work, we utilize this FV formalism to explore how the S-D mixing angle 
at the deuteron binding energy, along with the binding energy itself,
can be optimally extracted from LQCD calculations performed in cubic
volumes with fields subject to 
periodic BCs (PBCs) in the spatial directions.
Using the phase shifts and mixing angles generated by phenomenological 
NN potentials that are fit to NN scattering data~\cite{NIJMEGEN},
the expected FV energy spectra in the 
positive-parity isoscalar channels
are determined  at the physical pion mass (we assume exact
isospin symmetry throughout). 
It is found that correlation functions of 
boosted NN systems will play a key role in extracting the S-D mixing angle in future LQCD calculations. 
The FV energy shifts of the ground state of different irreps of the symmetry groups 
associated with momenta $\mathbf{P}=\frac{2\pi}{\rm L}(0,0,1)$ 
and $\frac{2\pi}{\rm L}(1,1,0)$,
are found to have enhanced sensitivity to the mixing angle in modest volumes
and to
depend both on its magnitude and sign.
A feature of the FV spectra, 
with practical implications for future LQCD calculations,
is that the contribution 
to the energy splittings from  channels with $J>1$, 
made possible by the  reduced symmetry of the volume,
are negligible for ${\rm L}\gtrsim 10~\text{fm}$ as the phase shifts in
those channels are small at low energies.
As the generation of multiple ensembles of 
gauge-field configurations at the physical light-quark masses
will require significant computational resources on capability-computing platforms,
we have investigated the viability of precision determinations of the deuteron
binding energy and scattering parameters 
from one lattice volume using 
the 
six bound-state energies
associated with 
CM momenta $|\mathbf{P}|\leq\frac{2\pi}{\rm L}\sqrt{3}$. 
We have also considered extracting the asymptotic D/S ratio from the
behavior of the deuteron FV wavefunction and its
relation to the S-D mixing angle.

\section{Deuteron and the Finite Volume Spectrum
\label{sec:DeutFV}
}
\noindent
The spectra of energy eigenvalues of two nucleons in the isoscalar
channel with positive parity in a cubic volume subject to  PBCs are dictated by
the S-matrix elements in this sector, including those defining the 
$\siii$-$\diii$ coupled channels that contain the deuteron.
The following determinant condition, 
\begin{align}
\text{det}[\mathcal{M}^{-1}+\delta \mathcal{G}^V]=0
\ \ \ ,
\label{QC}
\end{align}
provides the relation between the infinite volume
on-shell scattering amplitude 
$\mathcal{M}$ 
and the FV  CM energy of the NN system
below the inelastic threshold~\cite{Briceno:2013lba}.
In this work, we restrict ourselves to nonrelativistic (NR) quantum mechanics,
and as such the energy-momentum relation is 
${\rm E}_{NR}={\rm E}-2M={\rm E}_{NR}^* + {\rm E}_{CM} = \frac{k^{*2}}{M}+\frac{\mathbf{P}^2}{4M}$,
where $\mathbf{P}$ is the total momentum of the
system, and $\mathbf{k}^*$ is the momentum of each nucleon in the CM frame. 
The subscript will be dropped for the remainder of the paper, simply denoting
${\rm E}_{NR}^{(*)}$ by ${\rm E}^{(*)}$.
Due to the PBCs, the total momentum is  discretized, 
$\mathbf{P}=\frac{2\pi}{\rm L}\mathbf{d}$,
with $\mathbf{d}$ being an integer triplet that will be referred to as the boost
vector. 
$\delta \mathcal{G}^V$ is a matrix in the basis of $\left|JM_J(LS)\right
\rangle$ 
where $J$ is the total angular momentum, $M_J$ is the eigenvalue of the $\hat
J_z$ operator, 
and $L$ and $S$ are the orbital angular momentum and the total spin of the
channel, respectively. 
The matrix elements of $\delta \mathcal{G}^V$  in the positive-parity
isoscalar channel in this basis are,
\begin{eqnarray}
&& \left[\delta\mathcal{G}^V\right]_{JM_J,LS;J'M_J',L'S'}
=
\frac{iMk^*}{4\pi}\delta_{S1}\delta_{S'1}\left[\delta_{JJ'}\delta_{M_JM_J'}\delta_{LL'}
  +i\sum_{l,m}\frac{(4\pi)^{3/2}}{k^{*l+1}}
c_{lm}^{\mathbf{d}}(k^{*2}; {\rm L}) \right.
\nonumber\\
&& \qquad \qquad ~ \left .  \times \sum_{M_L,M_L',M_S}\langle
  JM_J|LM_L 1M_S\rangle \langle L'M_L' 1M_S|J'M_J'\rangle 
\int d\Omega~Y^*_{L M_L}Y^*_{l m}Y_{L' M_L'}\right]
\ \ ,
\label{deltaG}
\end{eqnarray}
and are evaluated at the on-shell momentum of each nucleon in the CM frame, 
$k^*=\sqrt{M {\rm E}^*-{|\mathbf{P}|^2}/{4}}$. 
$\langle JM_J|LM_L 1M_S\rangle$ and $\langle L'M_L' 1M_S|J'M_J'\rangle$ are
Clebsch-Gordan coefficients, 
and $c_{lm}^{\mathbf{d}}(k^{*2}; {\rm L})$ is a kinematic function related to
the 
three-dimensional zeta function, 
$\mathcal{Z}^\mathbf{d}_{lm}$,~\cite{Luscher:1986pf, Luscher:1990ux, Rummukainen:1995vs, Christ:2005gi,Kim:2005gf},
\begin{eqnarray}
\hspace{1cm} 
c^\mathbf{d}_{lm}(k^{*2}; {\rm L})
& = & \frac{\sqrt{4\pi}}{\rm L^3}\left(\frac{2\pi}{\rm
    L}\right)^{l-2}\mathcal{Z}^\mathbf{d}_{lm}[1;(k^* {\rm L}/2\pi)^2]
\ \ \ ,
\nonumber\\
\mathcal{Z}^\mathbf{d}_{lm}[s;x^2]
& = & \sum_{\mathbf{n}}\frac{|\mathbf{r}|^lY_{l,m}(\mathbf{r})}{(r^2-x^2)^s}
\ \ \ ,
\label{clm}
\end{eqnarray}
where $\mathbf{\mathbf{r}}=\mathbf{n}-\mathbf{d}/2$ with $\mathbf{n}$ an integer triplet.

The finite-volume matrix
$\delta\mathcal{G}^V$ is neither diagonal in the $J$
basis nor in the $LS$ basis, as is clear from the form of  Eq.~(\ref{deltaG}). 
As a result 
of the scattering amplitudes in higher partial waves 
being  suppressed at low-energies,
the infinite-dimensional matrices present in the 
determinant condition 
can be truncated 
to a finite number of partial waves. 
For the following analysis of positive-parity isoscalar channel, the
scattering in all but the S- and D-waves are neglected. 
With this truncation, the scattering amplitude matrix $\mathcal{M}$ can be written as 
\begin{align}
{\mathcal{M}}=
 \left(
\begin{array}{cccc}
  \mathcal{M}_{1,S} & \mathcal{M}_{1,SD} & 0 & 0 \\
\mathcal{M}_{1,SD} & \mathcal{M}_{1,D} & 0 & 0 \\
 0 & 0 & \mathcal{M}_{2,D} & 0 \\
 0 & 0 & 0 & \mathcal{M}_{3,D} \\
\end{array}
\right)
\ \ \ ,
\end{align}
where the first subscript of the diagonal elements, $\mathcal{M}_{J,L}$,
denotes the total angular momentum of the channel 
and the second subscript denotes 
the orbital angular momentum. The off-diagonal elements in $J=1$ sub-block are
due to the S-D mixing. 
In the $J=3$ channel, there is a mixing between $L=2$ and $L=4$
partial waves, but as 
scattering in the $L=4$ partial wave is being neglected, the scattering amplitude in this channel
remains diagonal. 
Each element of this matrix is a diagonal matrix of dimension
$(2J+1)\times(2J+1)$ dictated by the $M_J$ quantum number.

\begin{center}
\begin{table} [ht!]
\label{tab:param1}
\begin{tabular}{|ccccc|}
\hline
$\hspace{.3cm}\mathbf{d}\hspace{.3cm}$&$\hspace{.3cm}$point
group$\hspace{.3cm}$&$\hspace{.3cm}$
classification$\hspace{.3cm}$&$\hspace{.3cm}N_{\text{elements}}\hspace{.3cm}$&irreps (dimension) \\\hline \hline
$~(0,0,0)~$&${O}$&cubic&$24$&$~\mathbb{A}_1(1),\mathbb{A}_2(1),\mathbb{E}(2),\mathbb{T}_1(3),\mathbb{T}_2(3)~$\\
$~(0,0,1)~$&$D_{4}$&tetragonal&$8$&$~\mathbb{A}_1(1),\mathbb{A}_2(1),\mathbb{E}(2),\mathbb{B}_1(1),\mathbb{B}_2(1)~$\\ 
$~(1,1,0)~$&$D_{2}$&orthorhombic&$4$&$~\mathbb{A}(1),\mathbb{B}_1(1),\mathbb{B}_2(1),\mathbb{B}_3(1)~$\\
$~(1,1,1)~$&$D_{3}$&trigonal&$6$&$~\mathbb{A}_1(1),\mathbb{A}_2(1),\mathbb{E}(2)~$
\\\hline\hline
\end{tabular}
\caption{
Classification of the point groups corresponding to the symmetry
of the FV calculations with boost vectors, $\mathbf{d}$. 
The forth column shows the number of elements of each group, 
and the last column gives the irreducible representations of each point group
along with their dimensions.
}
\label{tab:groups}
\end{table}
\end{center}
With the aid of the symmetry properties of the FV calculation with different
boosts, the determinant condition providing the  
energy eigenvalues given in Eq.~(\ref{QC}) can be decomposed into separate eigenvalue equations
corresponding to 
different irreps of the point group,
\begin{eqnarray}
\det \left[\mathcal{M}^{-1}+\delta \mathcal{G}^{V}
\right]\ =\ 
\prod_{\Gamma^i}\det\left[(\mathcal{M}^{-1})_{\Gamma^i}+\delta 
\mathcal{G}^{V}_{\Gamma^i} \right]^{N(\Gamma^i)}\ =\ 0 
\ \ ,
\label{NNQC-irrep}
\end{eqnarray}
where $\Gamma^i$ labels different irreps of the corresponding point group, and
$N({\Gamma^i})$ denotes the dimensionality of each irrep. 
Table~\ref{tab:groups} summarizes some characteristics of the cubic
($O$), tetragonal ($D_{4}$), 
orthorhombic ($D_{2}$) and trigonal ($D_{3}$) point groups that correspond to
systems with boosts 
$\mathbf{d}=0$, $(0,0,1)$, $(1,1,0)$ and  $(1,1,1)$, respectively. 
Such a reduction has been carried out in Ref.~\cite{Briceno:2013lba} for all
possible NN channels with boosts
$|\mathbf{d}|\leq \sqrt{2}$. 
For the boosts considered in Ref.~\cite{Briceno:2013lba}, as well as for
$\mathbf{d} = (1,1,1)$,
the  necessary QCs for the NN system in the positive-parity isoscalar channel are given in
Appendix~\ref{app: QC}.
It is worth noting that for 
systems composed of equal-mass NR particles,
these QCs can be also utilized for boosts of the form 
$(2n_1,2n_2,2n_3)$, $(2n_1,2n_2,2n_3+1)$, 
$(2n_1+1,2n_2+1,2n_3)$ and $(2n_1+1,2n_2+1,2n_3+1)$
where $n_1,n_2,n_3$ are integers, and all cubic rotations of these vectors~\cite{Briceno:2013lba}.

\begin{figure}[ht!]
\begin{center}  
\subfigure[]{
\includegraphics[scale=0.145]{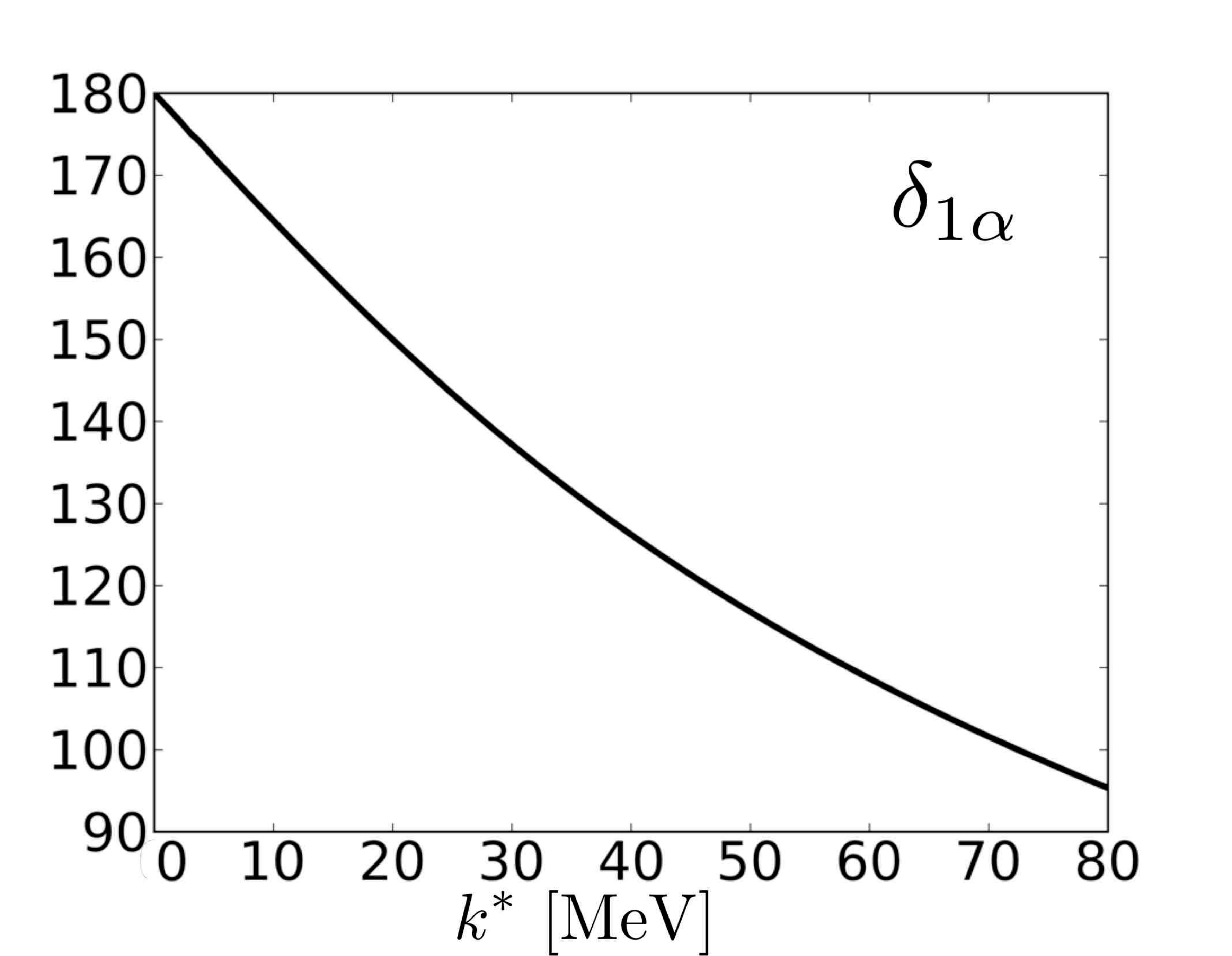}}
\subfigure[]{
\includegraphics[scale=0.145]{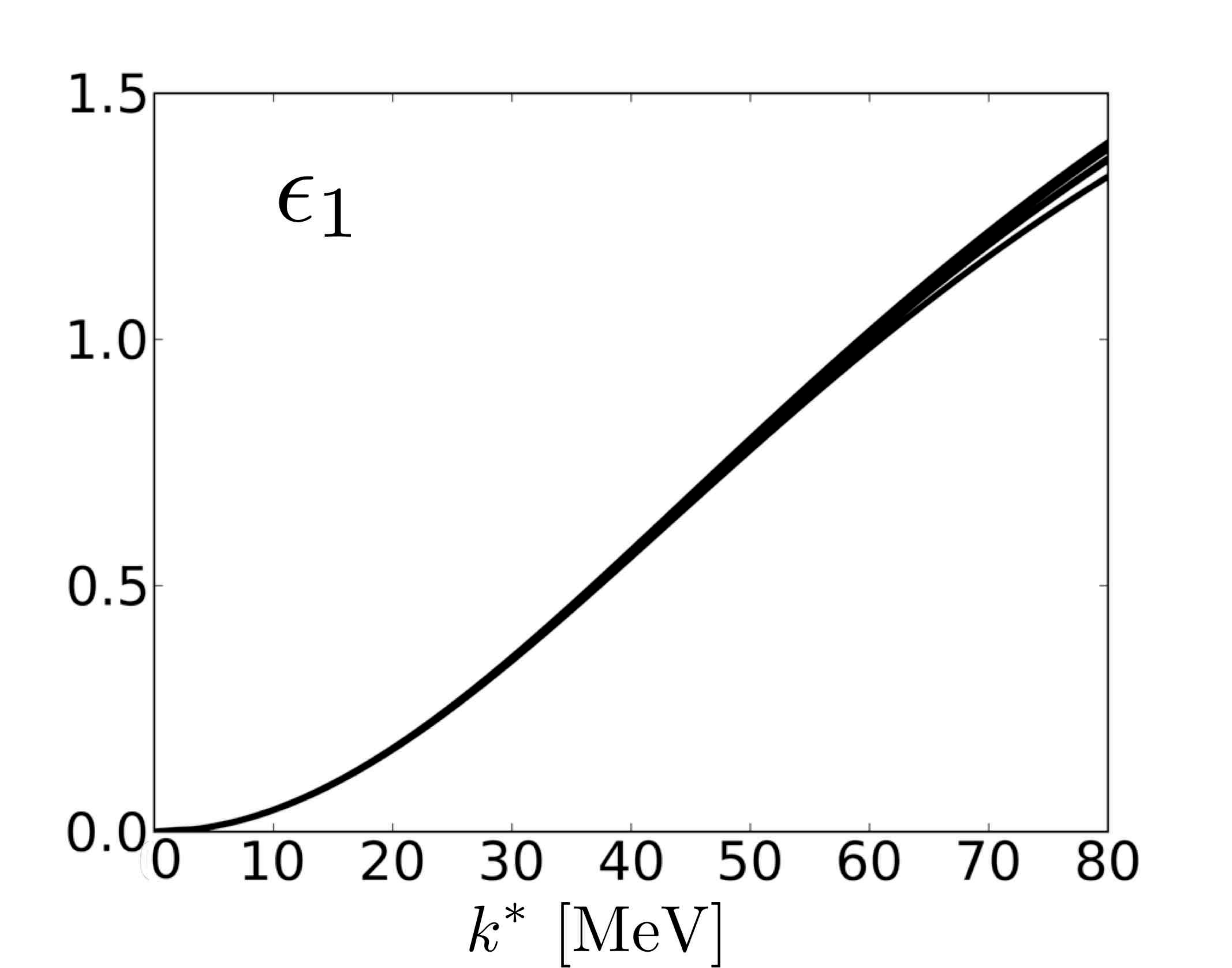}}
\subfigure[]{
\includegraphics[scale=0.145]{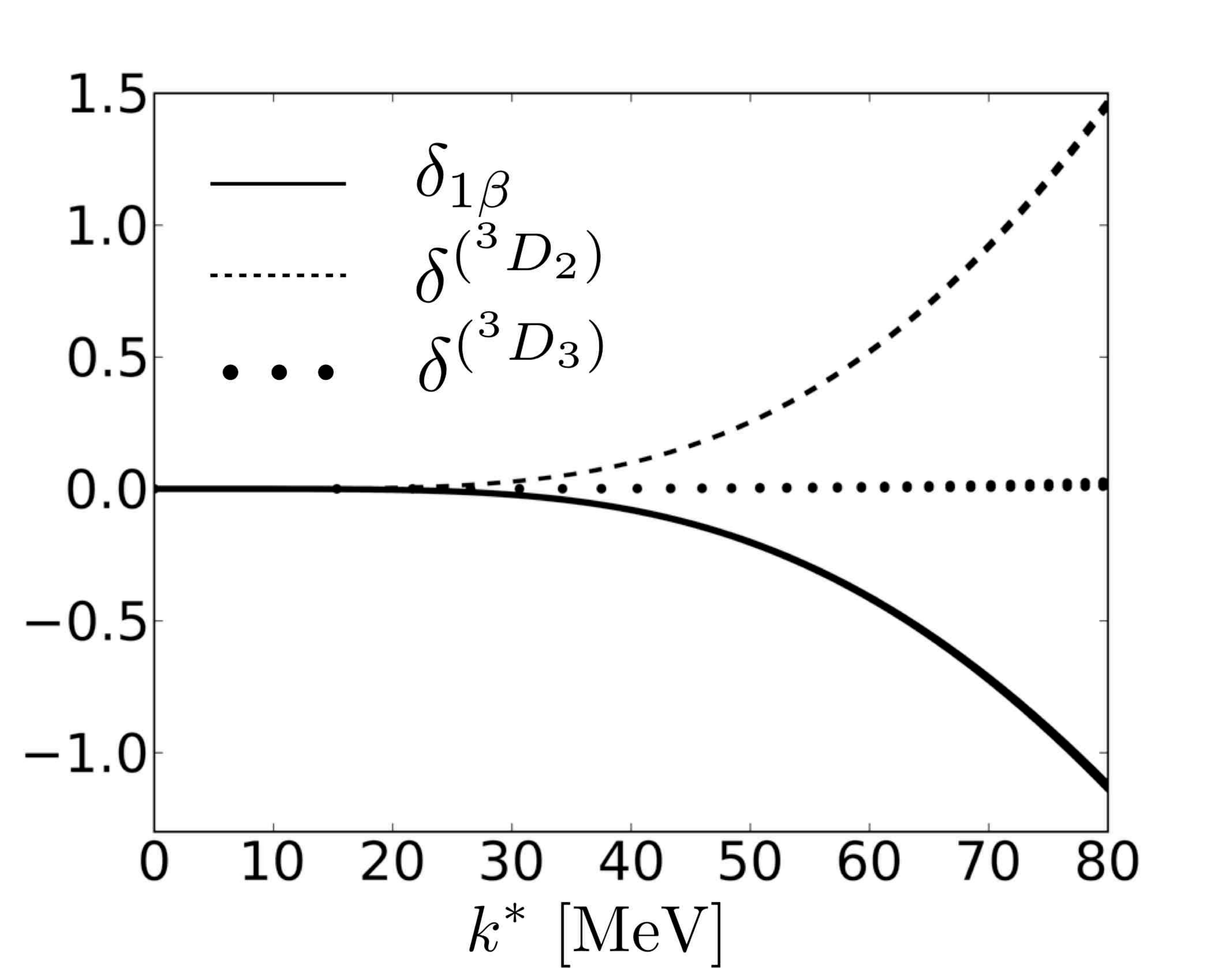}}
\subfigure[]{
\includegraphics[scale=0.145]{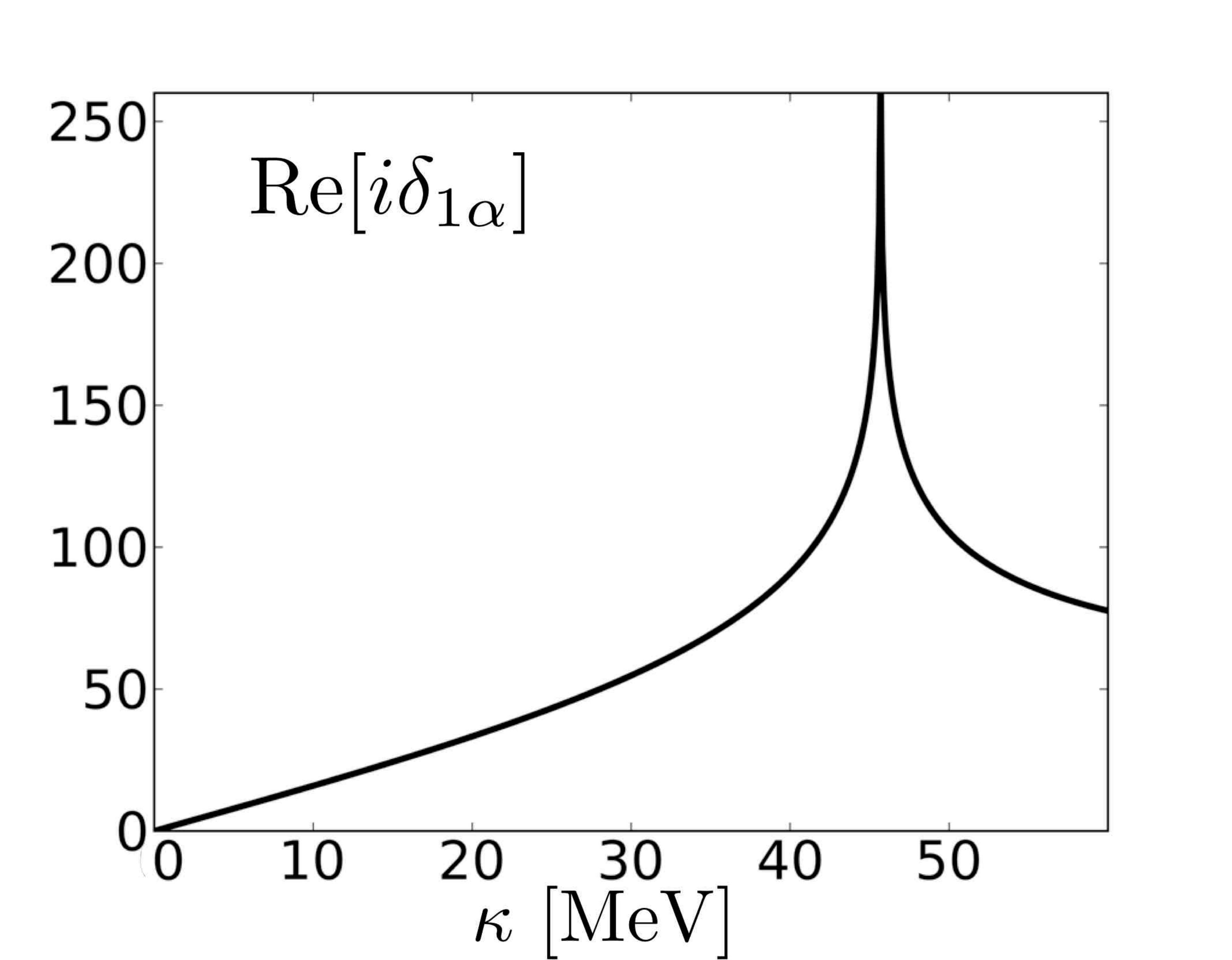}}
\subfigure[]{
\includegraphics[scale=0.145]{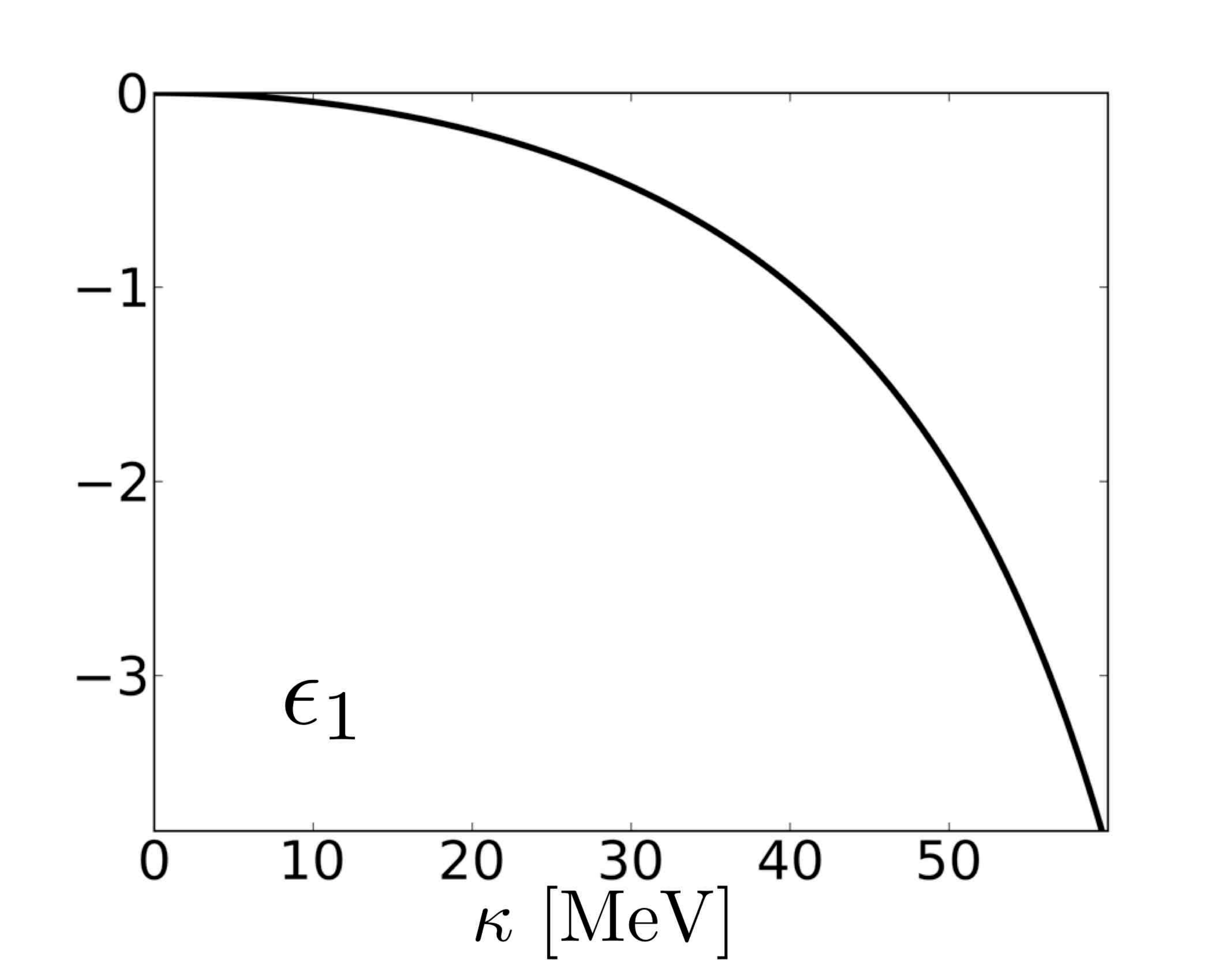}}
\subfigure[]{
\includegraphics[scale=0.145]{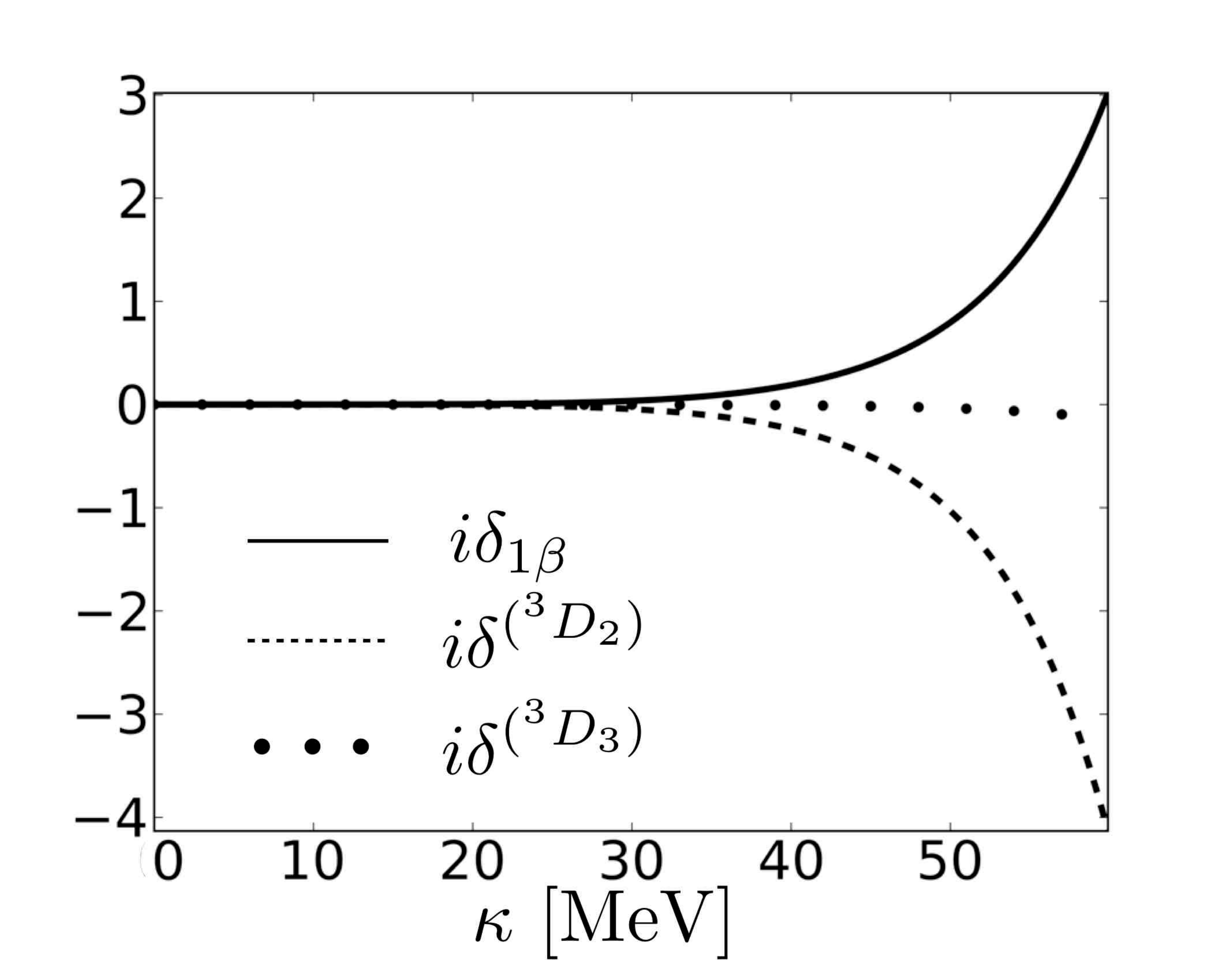}}
\caption{
Fits to the experimental values of 
(a) the $\alpha$-wave phase shift, 
(b) the mixing angle and 
(c) the $J=1$ $\beta$-wave and $J=2,3$ D-wave  phase shifts 
(in  degrees), in the Blatt-Biedenharn (BB) parameterization~\cite{Blatt:1952zza}, 
 as a function of
  momentum of each nucleon in the CM frame, $k^*$, 
based on six different phase shifts analyses~\protect\cite{PhysRevC.48.792,
    PhysRevC.49.2950, PhysRevC.54.2851, PhysRevC.54.2869}. 
(d) The $\alpha$-wave phase shift, 
(e) the mixing angle and 
(f)  the $J=1$ $\beta$-wave and $J=2,3$ D-wave phase shifts 
(in degrees) as a function of $\kappa=-ik^*$
obtained from the fit functions. 
}
\label{fig:Fits}
\end{center}
\end{figure}
Although the ultimate goal is to utilize the QCs in the analysis of the NN
spectra extracted from LQCD calculations, 
they can be used, in combination with the experimental NN scattering data, 
to predict the FV  spectra at the physical light-quark masses,
providing important guidance for future LQCD calculations. 
While for scattering states, 
the phase shifts and mixing angle from 
phenomenological  analyses of the experimental data~\cite{PhysRevC.48.792,
  PhysRevC.49.2950, PhysRevC.54.2851, PhysRevC.54.2869}
can be used in the QCs,
for bound states, however, it is necessary to use fit functions 
of the correct form to be continued to negative energies. 
The Blatt-Biedenharn (BB) parameterization~\cite{Blatt:1952zza,PhysRev.93.1387} is  chosen for the $J=1$ S-matrix, 
\begin{eqnarray}
S_{(J=1)}=\left( \begin{array}{cc}
\cos\epsilon_1&-\sin\epsilon_1\\
\sin\epsilon_1&\cos\epsilon_1\\
\end{array} \right)
\left( \begin{array}{cc}
e^{2i\delta_{1\alpha}}&0\\
0&e^{2i\delta_{1\beta}}\\
\end{array} \right)
\left( \begin{array}{cc}
\cos\epsilon_1&\sin\epsilon_1\\
-\sin\epsilon_1&\cos\epsilon_1\\
\end{array} \right),
\label{eq:BBSmatrix}
\end{eqnarray}
whose mixing angle, $\epsilon_1$, when evaluated at the deuteron binding
energy, is directly related to the asymptotic D/S
ratio in the deuteron wavefunction. 
$\delta_{1\alpha}$ and $\delta_{1\beta}$ are the scattering phase shifts
corresponding to two eigenstates of the S-matrix; 
the so called ``$\alpha$'' and ``$\beta$'' waves respectively. 
At low energies, the $\alpha$-wave is predominantly S-wave with a small
admixture of the D-wave, 
while the $\beta$-wave is predominantly D-wave with a small admixture of the
S-wave. 
The location of the deuteron pole is determined
by one condition 
on the $\alpha$-wave phase shift, $\cot \delta_{1\alpha}|_{k^*=i\kappa}=i$.  
In addition, $\epsilon_1$ in this parameterization is an analytic function of energy near
the deuteron pole 
(in contrast with $\bar{\epsilon}$ in the barred parameterization
\cite{Stapp:1956mz}).  
With a truncation of $l_{max}=2$ imposed upon the scattering amplitude matrix in Eq.~(\ref{QC}),
the scattering parameters required for the analysis of the FV spectra
are $\delta_{1\alpha}$, $\delta_{1\beta}$, $\delta^{(^3D_2)}$,
$\delta^{(^3D_3)}$ and $\epsilon_1$. 
Fits to six different phase-shift analyses 
(PWA93~\cite{PhysRevC.48.792}, Nijm93~\cite{PhysRevC.49.2950}, Nijm1~\cite{PhysRevC.49.2950}, 
Nijm2~\cite{PhysRevC.49.2950}, Reid93~\cite{PhysRevC.49.2950} and 
ESC96~\cite{PhysRevC.54.2851, PhysRevC.54.2869}) 
obtained from Ref.~\cite{NIJMEGEN} are shown in
Fig.~\ref{fig:Fits}(a-c).~\footnote{
The $\alpha$-wave was fit by  a pole term and a
polynomial, while the other parameters were fit with polynomials alone.  
The order of the polynomial for each parameter was determined by the goodness
of fit to phenomenological model data below the t-channel cut.
} 
In order to obtain the scattering parameters at negative energies,
the fit functions are
continued to imaginary momenta, $k^*\rightarrow i\kappa$. 
Fig.~\ref{fig:Fits}(d-f)  shows the phase shifts and the mixing angle as a
function of $\kappa$ 
below the t-channel cut
(which approximately corresponds to the 
positive-energy fitting range). 
$\epsilon_1$ is observed  to be 
positive for positive energies, and becomes
negative when continued to negative energies
(see Fig.~\ref{fig:Fits}). 
The slight difference between phenomenological models 
gives rise to a small  ``uncertainty band'' for each of the parameters.

\begin{center}
\begin{table}[ht!]
\label{tab:param2}
\begin{tabular}{|c|c|c|c|c|c|}
\hline
J &$O$&${D}_{4}$&${D}_{2}$& ${D}_{3}$
 \\\hline\hline
1
&$\mathbb{T}_1:(\mathcal{Y}_{11},\mathcal{Y}_{10},\mathcal{Y}_{1-1})$
&$\mathbb{A}_2:\mathcal{Y}_{10}$
&$\mathbb{B}_1:\mathcal{Y}_{10}$
& $\mathbb{A}_2:\mathcal{Y}_{10}$ \\
&
&$\mathbb{E}:\left(\overline{\mathcal{Y}}_{11},\widetilde{\mathcal{Y}}_{11}\right)$
&$\mathbb{B}_2:\overline{\mathcal{Y}}_{11},~\mathbb{B}_3:\widetilde{\mathcal{Y}}_{11}$
& $\mathbb{E}:\left(\overline{\mathcal{Y}}_{11},\widetilde{\mathcal{Y}}_{11}\right)$ \\
\hline
\end{tabular}
\caption{
Decomposition of the $J=1$ irrep of the rotational group in terms of
  the irreps of the cubic ($O$), 
tetragonal ($D_4$), orthorhombic ($D_2$) and trigonal ($D_3$) groups, see
Refs. \cite{Luscher:1990ux, Feng:2004ua, Dresselhaus}. 
The corresponding basis functions  of each irrep are also shown
in terms of the SO(3) functions
$\mathcal{Y}_{lm}$,
where 
$\overline{\mathcal{Y}}_{lm}\equiv \mathcal{Y}_{lm}+\mathcal{Y}_{l-m}$ and 
$\widetilde{\mathcal{Y}}_{lm}\equiv \mathcal{Y}_{lm}-\mathcal{Y}_{l-m}$.
}
\label{irreps}
\end{table}
\end{center}

\begin{figure}[!ht]
\begin{center}  
\subfigure[]{
\label{T1spec}
\includegraphics[scale=0.215]{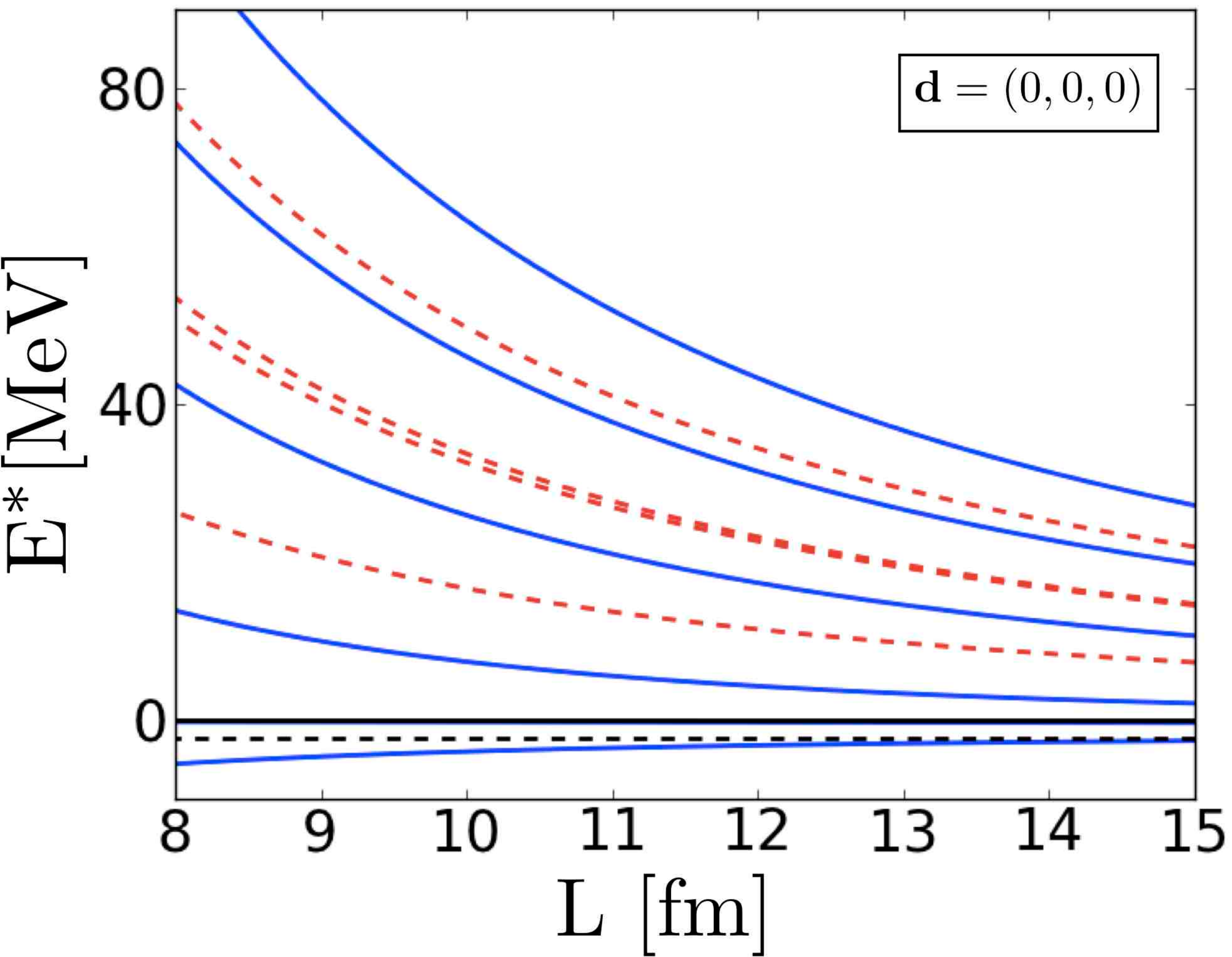}}
\subfigure[]{
\label{T1qtilde}
\includegraphics[scale=0.2175]{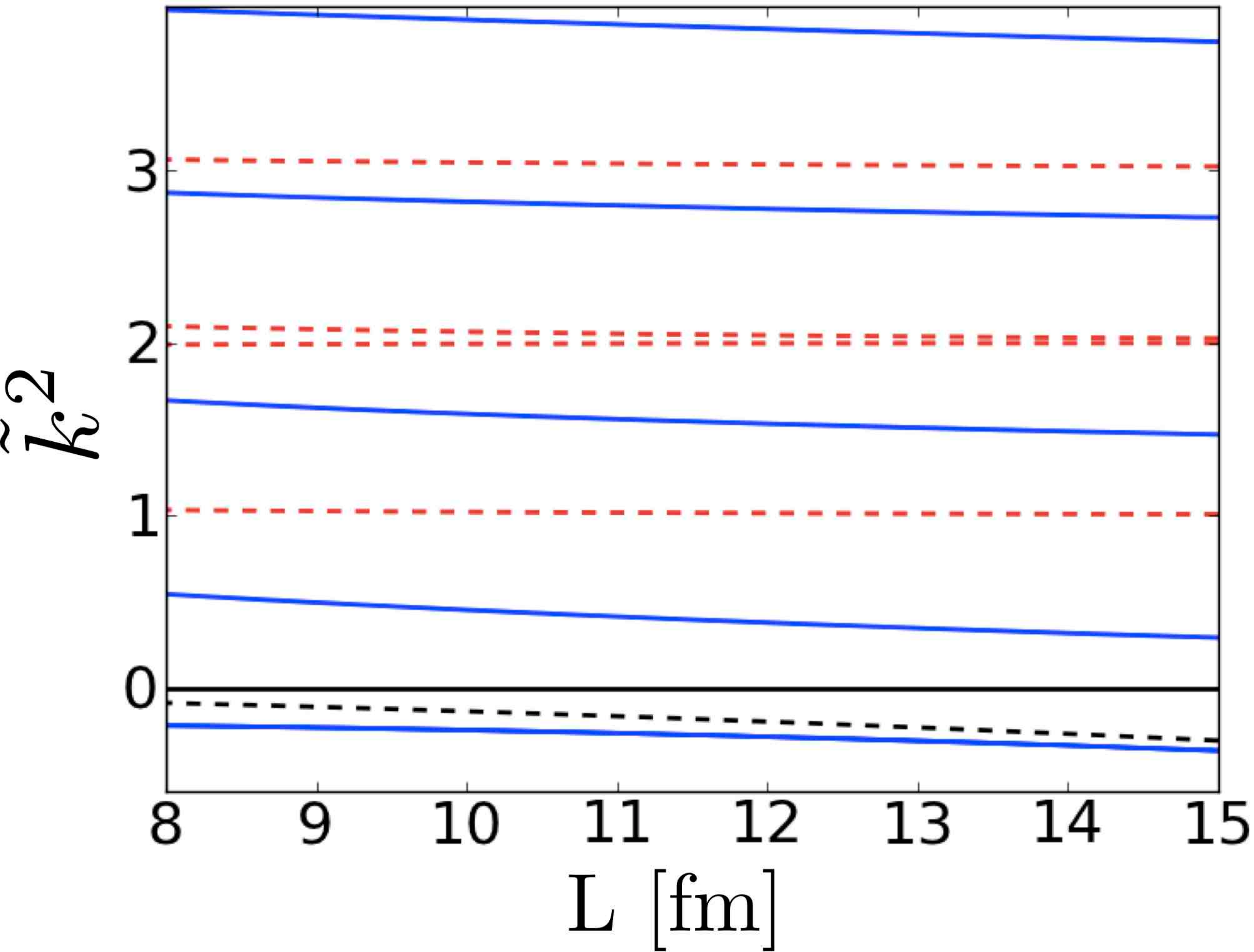}}
\caption
{(a) 
The nine lowest energy eigenvalues satisfying the QC for the
$\mathbb{T}_1$-irrep of the cubic group, 
Eq.~(\ref{I000T1}), and 
(b) the dimensionless quantity $\tilde{k}^2=ME^*{\rm{L}}^2/4\pi^2$ as a
function of ${\rm L}$.
The blue-solid lines correspond to states that are predominantly S-wave,
while the red-dashed lines represent states that are predominantly D-wave. 
The black-dashed line shows the infinite-volume binding energy of the
deuteron. 
}
\label{T1specfull}
\end{center}
\end{figure}

For the NN system at rest in the 
positive-parity isoscalar channel,
the only irrep of the cubic group that 
has overlap with the $J=1$ sector is $\mathbb{T}_1$,  see Table
\ref{tab:param2},
which also has overlap with the J=3 and higher channels. 
Using the scattering parameters of the $J=1$ and $J=3$  
channels, the nine lowest $\mathbb{T}_1$ energy levels (including the
bound-state level)
are shown in  Fig.~\ref{T1spec}  as a function of 
${\rm L}$. 
In the limit that $\epsilon_1$ vanishes, the $\mathbb{T}_1$ QC,
given in Eq.~(\ref{I000T1}), can be written as a product of two 
independent QCs. 
One of these QCs depends only on  
$\delta_{1\alpha}\rightarrow \delta^{(^3S_1)}$, while the other depends 
on $\delta_{1\beta}\rightarrow \delta^{(^3D_1)}$ and $\delta^{(^3D_3)}$. 
By comparing the $\mathbb{T}_1$ spectrum with that
obtained for $\epsilon_1=0$, the $\mathbb{T}_1$
states can be classified as predominantly S-wave 
or predominantly D-wave states. 
The dimensionless quantity
$\tilde{k}^2=ME^*{\rm{L}}^2/4\pi^2$ is shown as a 
function of volume in Fig. \ref{T1qtilde}, 
from which it is clear that the predominantly
D-wave energy levels remain close to 
the non-interacting energies, corresponding to
$\tilde{k}^2=1,2,3,4,5,6,8,\ldots$,
consistent
with the fact that both the mixing angle and the D-wave phase shifts are
small at low energies, as seen in Fig.~\ref{fig:Fits}. 
The states that are predominantly
S-wave are negatively shifted in energy 
compared with the non-interacting  states due to the attraction of the NN interactions.

\begin{figure}[ht!]
\begin{center}  
\subfigure[]{
\label{deut_cub-I}
\includegraphics[scale=0.220]{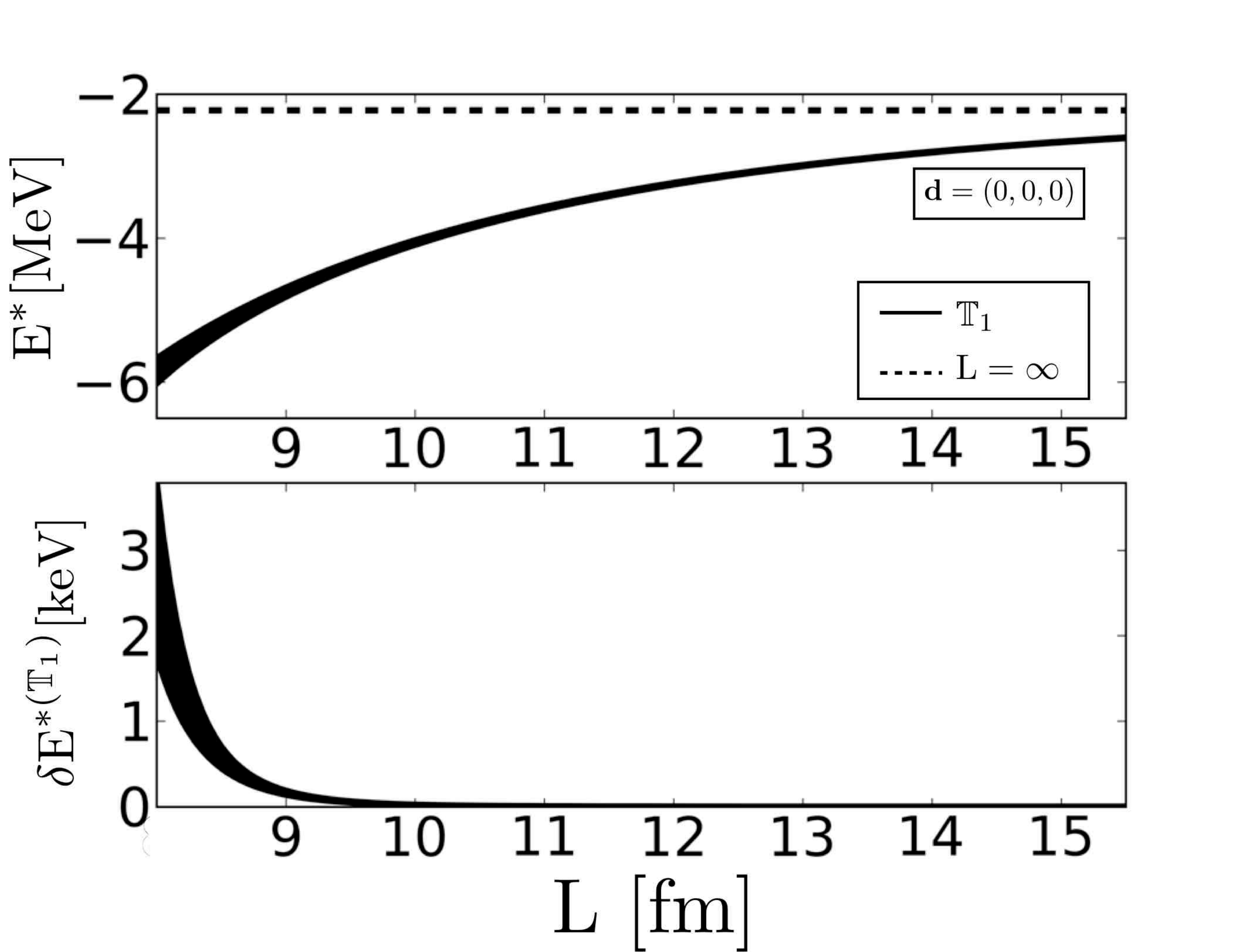}}
\subfigure[]{
\label{deut_trig}
\includegraphics[scale=0.215]{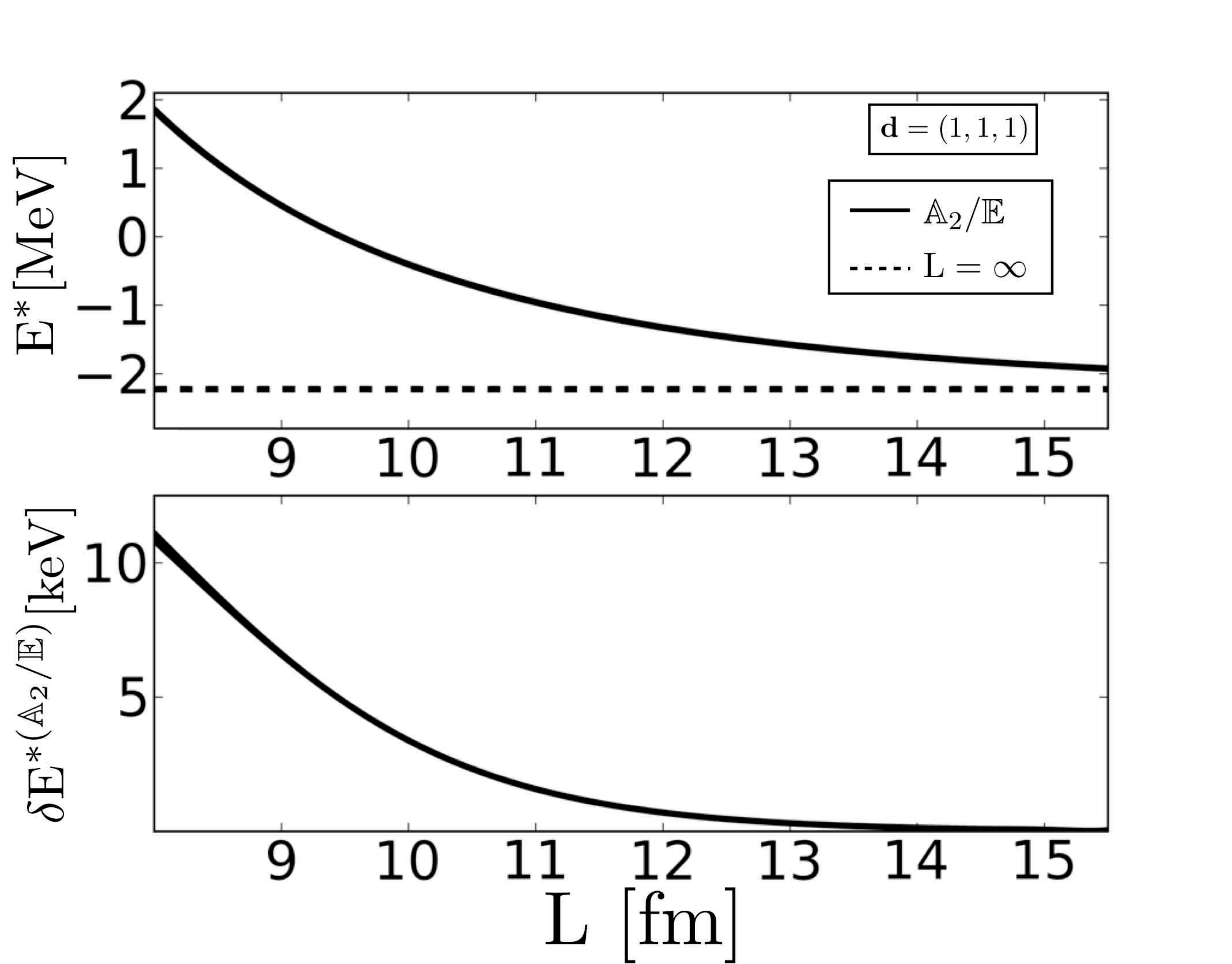}}
\caption{
(a) The upper panel shows the energy of two nucleons at rest in the 
positive-parity isoscalar channel  as a function of ${\rm L}$
extracted from the $\mathbb{T}_1$ QC given in  Eq.~(\ref{I000T1}). 
The uncertainty band is
associated with fits to different phenomenological analyses of the
experimental data,
and the dashed line denotes the infinite-volume deuteron binding
energy. 
The lower panel shows the contribution of the mixing angle to the 
energy,
$\delta {\rm E}^{*(\mathbb{T}_1)}= {\rm E}^{*(\mathbb{T}_1)}-{\rm E}^{*(\mathbb{T}_1)}(\epsilon_1=0)$.
(b) The same quantities as in (a) for the NN
system with
${\bf d}=(1,1,1)$ obtained from the  $\mathbb{A}_2/\mathbb{E}$ QCs, Eqs.~(\ref{I111A2},
\ref{I111E}).
}
\label{deut_cub}
\end{center}
\end{figure}

Focusing on the deuteron, it is important to quantify the effect of the
mixing between the S-wave and D-wave on the energy of the deuteron in the FV. 
The upper panel of Fig. \ref{deut_cub-I} provides 
a closer look at the binding energy of the deuteron as function of ${\rm L}$ 
extracted from the $\mathbb{T}_1$ QC given in Eq.~(\ref{I000T1}). 
While in larger volumes the uncertainties in the predictions due to the fits to experimental data 
are a few keV, 
in smaller volumes the uncertainties increase
because the fit functions are valid only below the t-channel cut 
and are not expected to describe the data above the cut. 
It is interesting  to examine the difference between the bound-state energy
obtained from the full $\mathbb{T}_1$ QC 
and that  obtained  with $\epsilon_1=0$, 
$\delta
{\rm E}^{*(\mathbb{T}_1)}={\rm E}^{*(\mathbb{T}_1)}-{\rm E}^{*(\mathbb{T}_1)}(\epsilon_1=0)$.
This quantity, shown in the lower panel of Fig.~\ref{deut_cub-I}, does not
exceed a few $\rm{keV}$ in smaller volumes, ${\rm L} \lesssim9~\rm{fm}$, 
and is significantly smaller in larger volumes,
demonstrating that the spectrum of $\mathbb{T}_1$ irrep is quite  insensitive
to the small mixing angle 
in the $\siii$-$\diii$ coupled channels.
Therefore, a determination of the mixing angle from the spectrum of two nucleons at rest
will be challenging for LQCD calculations. 
The spectra in the $\mathbb{A}_2/\mathbb{E}$ irreps of the trigonal group for
$\mathbf{d}=(1,1,1)$ 
exhibit the same feature, as shown in Fig. \ref{deut_trig}. 
By investigating the QCs in Eqs.~(\ref{I000T1}, \ref{I111A2}, \ref{I111E}), 
it is straightforward to show that the difference between the bound-state
energy 
extracted from the full QCs 
(including physical and FV-induced mixing between S-waves and D-waves) 
and from the uncoupled QC is proportional to ${\sin^2 \epsilon_1}$,
and is further suppressed by FV corrections and the 
small  $\beta$-wave and D-wave
phase shifts.

The boost vectors $\mathbf{d}=(0,0,1)$ and $(1,1,0)$
distinguish the $z$-axis from the 
other two axes, and result in an asymmetric volume as viewed
in the rest frame of the deuteron. 
In terms of the periodic images of the deuteron, 
images that are located in the $z$-direction with opposite signs compared with the
images in the $x$- and $y$-directions~\cite{Bour:2011ef,Davoudi:2011md} 
result in the quadrupole-type shape
modifications to the 
deuteron, as will be elaborated  on in Sec.~\ref{sec:wavefunction}. 
As a result, the energy of the deuteron, as well as its
shape-related quantities such as its quadrupole moment, 
will be affected more by the finite extent of the volume (compared with the
systems with $\mathbf{d}=(0,0,0)$ and $(1,1,1)$).

\begin{figure}[ht!]
\begin{center}  
\includegraphics[scale=0.30]{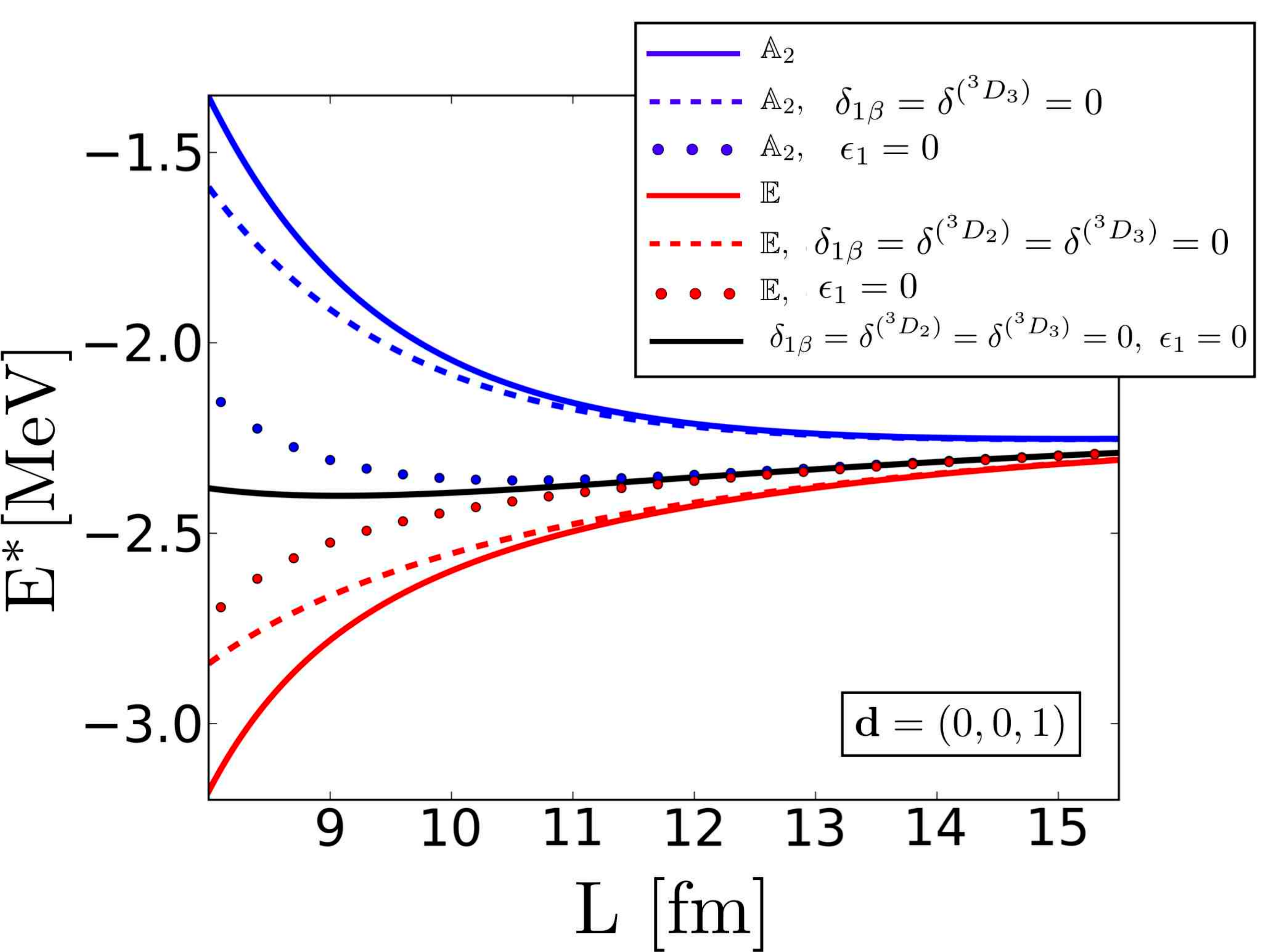}
\caption{The energy of two nucleons in the
 positive-parity
isoscalar channel with $\mathbf{d}=(0,0,1)$ as a function of
  ${\rm L}$, 
extracted from the $\mathbb{A}_2$ (red) and $\mathbb{E}$ (blue) QCs given in 
Eq.~(\ref{I001A2}) and Eqs.~(\ref{I001E}), respectively. 
The systematic uncertainties  associated with fitting 
different phenomenological analyses of the experimental data  are included. 
}
\label{deut_tet}
\end{center}
\end{figure}
As is clear from the QCs for $\mathbf{d}=(0,0,1)$ systems, given in Eqs.~(\ref{I001A2}, \ref{I001E}), 
there are two
irreps of the tetragonal group, $\mathbb{A}_2$ 
and $\mathbb{E}$,  that have overlap with the $J=1$ channel. 
These irreps represent states with $M_J=0$ for the $\mathbb{A}_2$ 
irrep and $M_J=\pm 1$ for the $\mathbb{E}$ irreps,
see Table~\ref{irreps}. 
The bound-state energies of  these two irreps  are shown in  Fig.~\ref{deut_tet}
as a function of ${\rm L}$. 
For comparison, the energy of the bound state with  ${\bf d}=(0,0,1)$
in the limit of vanishing mixing angle and D-wave phase shifts is also shown 
(black-solid curve) in Fig. \ref{deut_tet}. 
The energy of the bound states obtained in  both the $\mathbb{A}_2$ irrep
(blue-solid curve) and the $\mathbb{E}$ irrep (red-solid curve) 
deviate substantially from the energy of the purely S-wave bound state for modest
volumes, ${\rm L}\lesssim14~\rm{fm}$.~\footnote{LQCD calculations at the physical
  pion mass require volumes with ${\rm L} \gtrsim 9~\rm{fm}$ so that the
  systematic uncertainties associated with
the finite range of the nuclear forces  are below the percent level.} 
These deviations are such that the energy gap between the systems in the two
irreps is 
$\sim 80\%$ of the
infinite-volume deuteron binding energy 
at ${\rm L}=8~\rm{fm}$, decreasing to $\sim 5\%$  for ${\rm L}=14~\rm{fm}$. 
This gap is largely due to the mixing between S-wave and D-wave in
the infinite volume, as verified by evaluating the bound-state energy in the
$\mathbb{A}_2$ and $\mathbb{E}$ irrep in the limit where $\epsilon_1=0$ 
(the blue and red dotted curves in Fig. \ref{deut_tet}, respectively.)
Another feature of the $\textbf{d}=(0,0,1)$ FV bound-state energy is that
the contribution from the 
 $\beta$-wave and D-wave
states cannot be neglected for ${\rm L}\lesssim 10~\rm{fm}$. 
The blue (red) dashed curve in Fig.~\ref{deut_tet} 
results from the $\mathbb{A}_2$ ($\mathbb{E}$) QC in this limit. 
The D-wave states in the $J=2$ and $J=3$ channels mix with the $J=1$ $\alpha$-
and $\beta$-waves due to the reduced symmetry of the system, 
and as a result they, and the $\beta$-wave state, contribute to the energy of the
predominantly S-wave bound state in the FV.

\begin{figure}[!ht]
\begin{center}  
\includegraphics[scale=0.30]{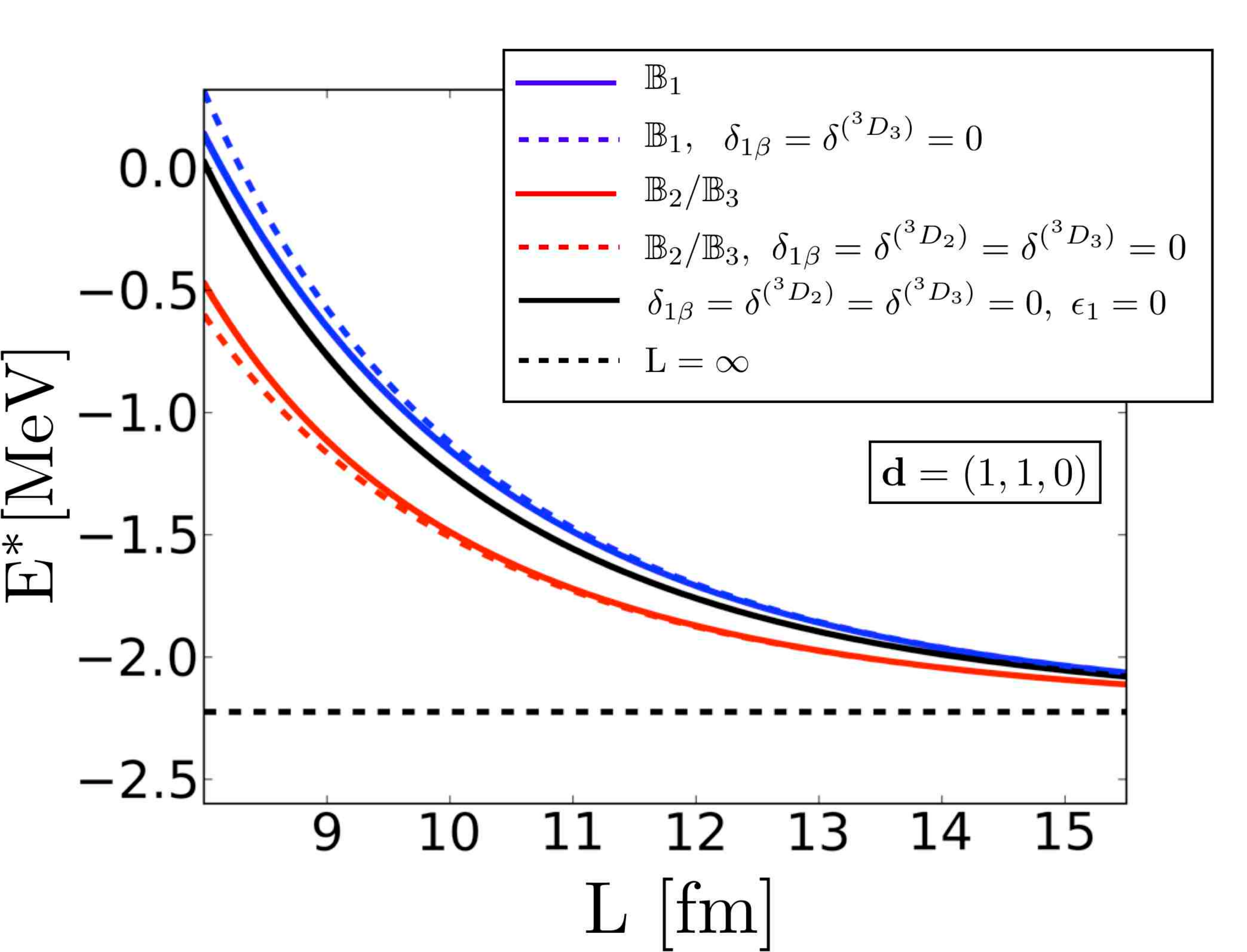}
\caption{
The energy of two nucleons in the
positive-parity
isoscalar channel with $\mathbf{d}=(1,1,0)$ as a function of
  ${\rm L}$, 
extracted from the $\mathbb{B}_1$ (red) and $\mathbb{B}_2/\mathbb{B}_3$ (blue)
QCs given in Eqs.~(\ref{I110B1}) and (\ref{I110B2}, \ref{I110B3}), 
respectively. 
The systematic uncertainties associated with fitting to different
phenomenological analyses of the experimental 
data are included. 
}
\label{deut_ort}
\end{center}
\end{figure}
The FV energy eigenvalues 
for the NN system in the positive-parity
isoscalar channel with ${\bf d}=(1,1,0)$ 
can be obtained from QCs in 
Eqs.~(\ref{I110B1}) and (\ref{I110B2}, \ref{I110B3}), corresponding to
$\mathbb{B}_1$ and $\mathbb{B}_2/\mathbb{B}_3$ irreps of 
the orthorhombic group, respectively. 
These irreps represent states with $M_J=0$ for the $\mathbb{B}_1$ irrep and 
$M_J=\pm 1$ for the $\mathbb{B}_2/\mathbb{B}_3$ irreps,
see Table~\ref{irreps}. 
The bound-state energies of these systems are shown in
Fig.~\ref{deut_ort}, and are found to  
deviate noticeably from the purely  S-wave limit (black-solid curve in
Fig.~\ref{deut_ort}), 
however the deviation is not as large as the 
case of ${\bf d}=(0,0,1)$. 
The energy gap between the systems in the two irreps is $\sim 30\%$ of the
 infinite-volume deuteron binding energy 
at ${\rm L}=8~\rm{fm}$, decreasing  to $\sim 5\%$  for ${\rm L}=14~\rm{fm}$. 
Eliminating the  $\beta$-wave and $J=2,3$ D-wave interactions, leads to the dashed curves in
Fig.~\ref{deut_ort}, indicating the 
negligible effect that they have on the bound-state energy in
these irreps.

To understand the large FV energy shifts from the purely  $\alpha$-wave
estimates for ${\bf d}=(0,0,1)$ and $(1,1,0)$ systems compared 
with $(0,0,0)$ and $(1,1,1)$ systems, 
it is instructive to examine the QCs given in Appendix~\ref{app: QC} in the limit
where the  $\beta$-wave and D-wave phase shifts vanish. 
This is a reasonable approximation  for volumes with ${\rm L} \gtrsim 10~\rm{fm}$,  as
illustrated in Fig.~\ref{deut_tet} and Fig.~ \ref{deut_ort}.
It is straightforward to show that in this limit, 
the QC of the system with $\mathbf{d}=(0,0,0)$ reduces to a purely $\alpha$-wave condition
\begin{eqnarray}
\mathbb{T}_1&:&\hspace{.1cm}k^*\cot\delta_{1\alpha}-4 \pi
c_{00}^{(0,0,0)}(k^{*2}; {\rm L})=0
\ \ \ .
\label{appr-T1}
\end{eqnarray}
The QCs for a system with $\mathbf{d}=(0,0,1)$ are
\begin{eqnarray}
\mathbb{A}_2:\hspace{.1cm}
&& k^*\cot\delta_{1\alpha}
-4 \pi  c_{00}^{(0,0,1)}(k^{*2}; {\rm L})
\ =\ 
-{1\over\sqrt{5}}
\frac{4\pi}{k^{*2}}\ 
c_{20}^{(0,0,1)}(k^{*2}; {\rm L})
\ 
(\sqrt{2}\sin2\epsilon_1-\sin^2\epsilon_1)
\ \ ,
\\
\mathbb{E}:\hspace{.1cm}
&& k^* \cot \delta_{1\alpha}
-4 \pi  c_{00}^{(0,0,1)}(k^{*2}; {\rm L})
\ =\ 
+{1\over 2\sqrt{5}}
\frac{4\pi}{k^{*2}}\ 
c_{20}^{(0,0,1)}(k^{*2}; {\rm L})
\ 
(\sqrt{2}\sin2\epsilon_1-\sin^2\epsilon_1)
\ \ ,
\label{appr-E}
\end{eqnarray}
which includes corrections to the $\alpha$-wave limit that scale with  $\sin \epsilon_1$
at LO. 
This is the origin of  the large deviations of these energy eigenvalues from the purely S-wave values. 
The same feature is seen in the systems with $\mathbf{d}=(1,1,0)$, where the QCs reduce to
\begin{eqnarray}
\mathbb{B}_1:
\hspace{.1cm}
&& k^*\cot\delta_{1\alpha}
-4 \pi  c_{00}^{(1,1,0)}(k^{*2}; {\rm L})
\ =\ 
-{1\over\sqrt{5}}
\frac{4\pi}{k^{*2}}\ 
c_{20}^{(1,1,0)}(k^{*2}; {\rm L})
\ 
(\sqrt{2}\sin2\epsilon_1-\sin^2\epsilon_1)
\ \ \ ,
\\
\mathbb{B}_2/\mathbb{B}_3:
\hspace{.1cm}
&& k^* \cot \delta_{1\alpha}
-4 \pi  c_{00}^{(1,1,0)}(k^{*2}; {\rm L})
\ =\ 
+{1\over 2\sqrt{5}}
\frac{4\pi}{k^{*2}}\ 
c_{20}^{(1,1,0)}(k^{*2}; {\rm L})
\ 
(\sqrt{2}\sin2\epsilon_1-\sin^2\epsilon_1)
\ \ \ .
\label{appr-B2}
\end{eqnarray}
Similarly, the QC with $\mathbf{d}=(1,1,1)$ in this limit is
\begin{eqnarray}
\mathbb{A}_2/\mathbb{E}&:&\hspace{.1cm}k^*\cot\delta_{1\alpha}-4 \pi
c_{00}^{(1,1,1)}(k^{*2}; {\rm L})=0
\ \ \ .
\label{appr-EA2}
\end{eqnarray}

The LO corrections to the QCs 
in Eqs.~(\ref{appr-T1}-\ref{appr-EA2}) 
are not only suppressed by
the 
$J=1$ $\beta$-wave and $J=2,3$ D-wave
phase shifts, but also by FV corrections 
that are further exponentially suppressed compared with the leading FV corrections. 
It is straightforward to show that the leading neglected terms in the QCs
presented above are 
$\sim \frac{1}{\rm L}e^{-2\kappa {\rm L}}\tan{\delta_{1\beta}}$ and 
$\frac{1}{\rm L}e^{-2\kappa {\rm L}}\tan{\delta_{D_{J=2,3}}}$,
while the
FV contributions to the approximate relations given in 
Eqs.~(\ref{appr-T1}-\ref{appr-EA2})  are  $\sim {1\over {\rm L}}e^{-\kappa {\rm L}}$. 
In Appendix \ref{app: clm}, the explicit volume dependence of $c^{\mathbf{d}}_{LM}$ functions
are given 
for the case of $k^{*2}=-\kappa^2<0$. 
These explicit forms are useful in obtaining the leading exponential corrections to the
QCs. 
We emphasize that the smaller volumes considered have $\kappa {\rm L} =2-2.5$, 
and therefore it is not a good approximation to replace the $c^{\mathbf{d}}_{LM}$
functions with their leading exponential terms,
and the complete form of these functions should be used in analyzing the FV spectra.

\begin{figure}[!ht]
\begin{center}  
\subfigure[]{
\label{spin-ave-001}
\includegraphics[scale=0.21]{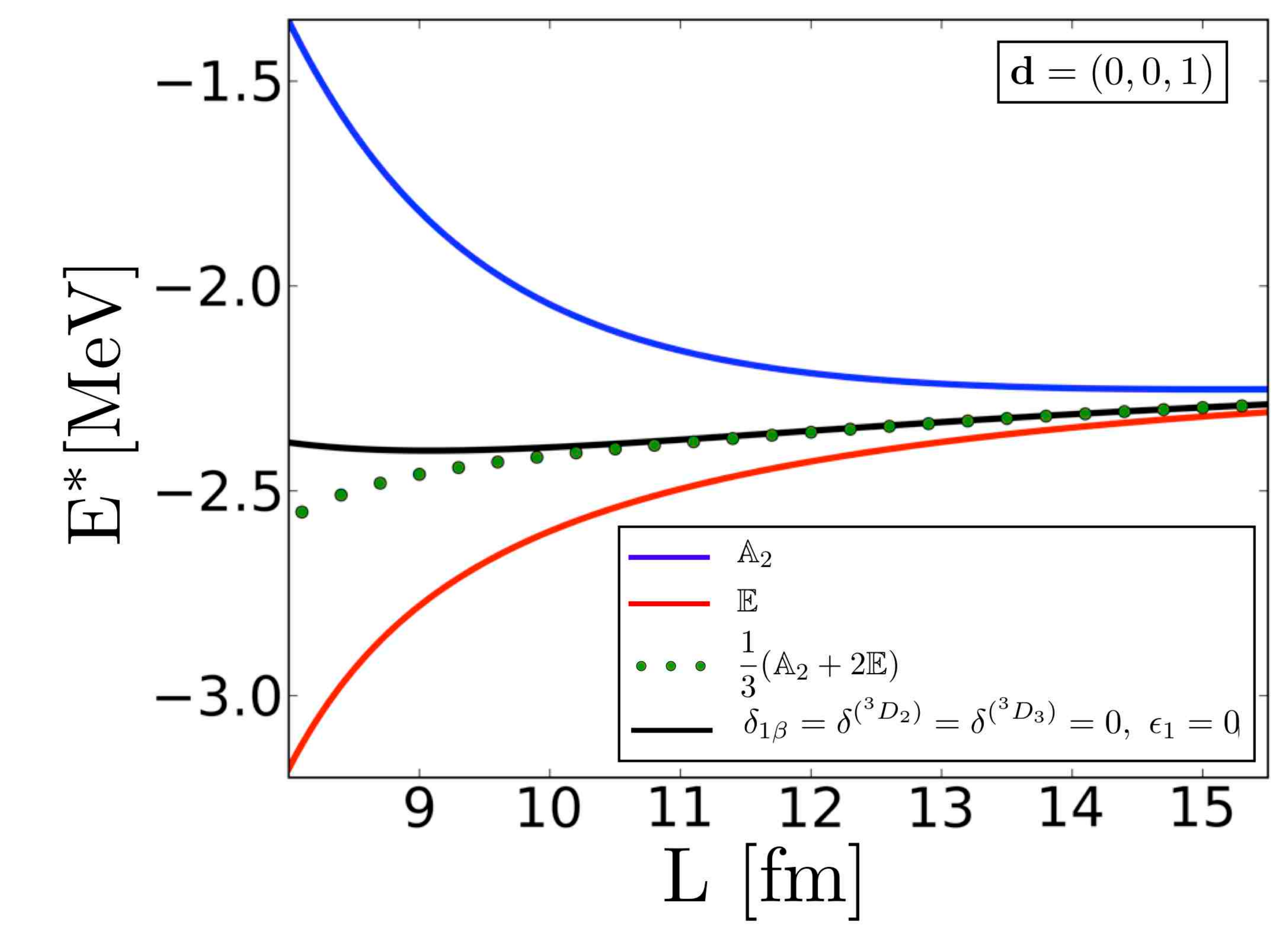}}
\subfigure[]{
\label{spin-ave-110}
\includegraphics[scale=0.21]{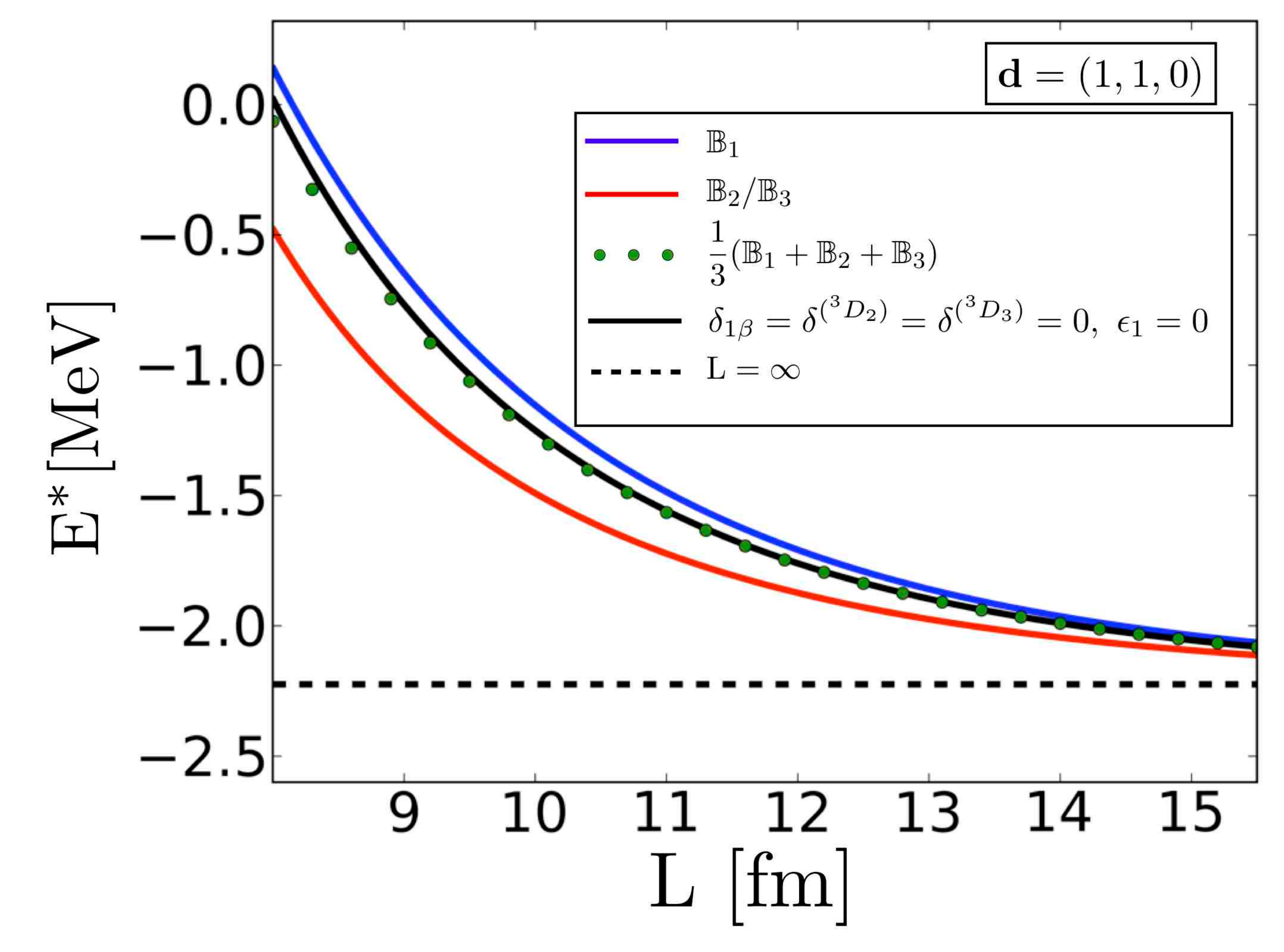}}
\caption{
(a) The dotted curve shows the $M_J$-averaged quantity
$\frac{1}{3}(E^{*(\mathbb{A}_2)}+2E^{*(\mathbb{E})})$ 
as function of ${\rm L}$, while the solid curves show the energy of the state  
in the $\mathbb{A}_2$ (blue) and $\mathbb{E}$ (red) irreps of the tetragonal
group, 
as well as that of the state with $\epsilon_1=0$ (black). 
(b) The dotted curve shows the $M_J$-averaged quantity
$\frac{1}{3}(E^{*(\mathbb{B}_1)}+E^{*(\mathbb{B}_2)}+E^{*(\mathbb{B}_3)})$
as function of ${\rm L}$, while the solid curves show the energy of the state 
in the $\mathbb{B}_1$ (blue) and $\mathbb{B}_2/\mathbb{B}_3$ (red) irreps of
the orthorhombic group, 
as well as that of the state with $\epsilon_1=0$ (black).}
\label{spin-ave}
\end{center}
\end{figure}
In the limit of vanishing 
$J=1$ $\beta$-wave and $J=2,3$ D-wave
phase shifts,
the QCs show that the energy shift of each pair of irreps of the systems 
with $\mathbf{d}=(0,0,1)$ and $(1,1,0)$ differ in sign. 
It is also the case that the $M_J$-averaged energies are approximately the
same as the purely S-wave case. 
In fact, as illustrated in Fig. \ref{spin-ave-001}, the energy level
corresponding to 
$\frac{1}{3}(E^{*(\mathbb{A}_2)}+2E^{*(\mathbb{E})})$ quickly
converges to the S-wave energy with ${\bf d}=(0,0,1)$. 
Similarly, the $M_J$-averaged quantity
$\frac{1}{3}(E^{*(\mathbb{B}_1)}+E^{*(\mathbb{B}_2)}+E^{*(\mathbb{B}_3)})$
almost coincides with the S-wave state with ${\bf d}=(1,1,0)$,
Fig. \ref{spin-ave-110}. 
This is to be expected, as $M_J$-averaging is equivalent to averaging
over the orientations of the image systems, suppressing  the anisotropy induced by
the boost phases in the FV corrections, Eqs.~(\ref{c00-exp})-(\ref{c40-exp}).
These expressions also demonstrate that, unlike the case of degenerate, scalar
coupled-channels systems \cite{Berkowitz:2012xq,Oset:2012bf}, 
the NN spectra (with spin degrees of freedom) depend on the sign of
$\epsilon_1$. 
Of course, this sensitivity to the sign of $\epsilon_1$ can be deduced
from the full QCs in Eqs.~(\ref{I000T1}-\ref{I111E}). 
Upon fixing the phase  convention of the angular momentum states, 
both the magnitude and sign of the mixing angle  can be extracted from 
FV calculations, 
as will be discussed in more detail in Section~\ref{sec:extraction}.

\section{Extracting the Scattering Parameters From Synthetic Data
\label{sec:extraction}
}
\noindent
Given the features of the energy spectra associated with different boosts, 
it is interesting  to consider how well the scattering parameters
can be extracted from future LQCD calculations at the physical pion mass.
With the truncations we have imposed, 
the full QCs for the FV  states that have overlap with the
$\siii$-$\diii$ 
coupled channels depend on four scattering phase shifts and the $J=1$ mixing angle.
As discussed in Sec.~\ref{sec:DeutFV}, for bound states these are equivalent to
QCs that depend solely on $\delta_{1\alpha}$
and $\epsilon_1$  up to corrections of
$\sim \frac{1}{\rm L}e^{-2\kappa {\rm L}}\tan{\delta_{1\beta}}$ and 
$\frac{1}{\rm L}e^{-2\kappa {\rm L}}\tan{\delta_{D_{J=2,3}}}$,
as given in Eqs.~(\ref{appr-T1}-\ref{appr-EA2}). 
By considering the boosts with $|\textbf{d}|\leq\sqrt{3}$, six
independent bound-state energies that asymptote to 
the physical deuteron energy can be obtained. 
For a single volume, these give six different constraints on
$\delta_{1\alpha}$ and $\epsilon_1$ for energies in the 
vicinity of the deuteron pole. 
Therefore, by parameterizing the momentum dependence of these two
parameters, and requiring them 
to simultaneously satisfy Eqs.~(\ref{appr-T1}-\ref{appr-EA2}), 
their low-energy behavior can be extracted.

Using the fact that the $\alpha$-wave is dominantly S-wave with $\epsilon_1$
and $\delta^{(\diii)}$
small, we use the  effective range expansion (ERE) of the inverse 
S-wave scattering amplitude,
which is valid below the t-channel cut,
to parameterize~\cite{PhysRev.93.1387}
\begin{eqnarray}
\label{eq:ERE}
k^*\cot\delta_{1\alpha}&=&-\frac{1}{a^{(^3S_1)}}+\frac{1}{2}r^{(^3S_1)}k^{*2}+ ...
\ \ ,
\\
\epsilon_1&=&h_1~k^{*2}+...
\ \ \ .
\end{eqnarray}
Therefore, up to $\mathcal{O}(k^{*2})$, 
the three  parameters, denoted by $a^{(^3S_1)}$, $r^{(^3S_1)}$ and $h_1$,
 can be over-constrained by the bound-state spectra in 
a single volume. 
To illustrate this point, we fit the six independent energies 
to ``synthetic data''
using the approximated QCs, Eqs.~(\ref{appr-T1}-\ref{appr-EA2}). 
The precision with which  $\{a^{(^3S_1)},r^{(^3S_1)},h_1\}$ can be extracted 
depends on the precision and correlation of the energies
determined in LQCD calculations. 
With this in mind, we consider four possible scenarios,
corresponding  to the energies being extracted 
from a given LQCD calculation
with $1\%$ and $10\%$ precision, and with uncertainties that are
uncorrelated or fully correlated with each other. 
It is likely that the energies of these irreps will be determined in 
LQCD calculations on the same ensembles of gauge-field configurations, 
and consequently they are likely to
be highly correlated - a feature that has been
exploited 
extensively in the past when determining energy differences.

\begin{figure}[t]
\begin{center} 
\subfigure[]{

\label{a_corr}
\includegraphics[scale=0.20]{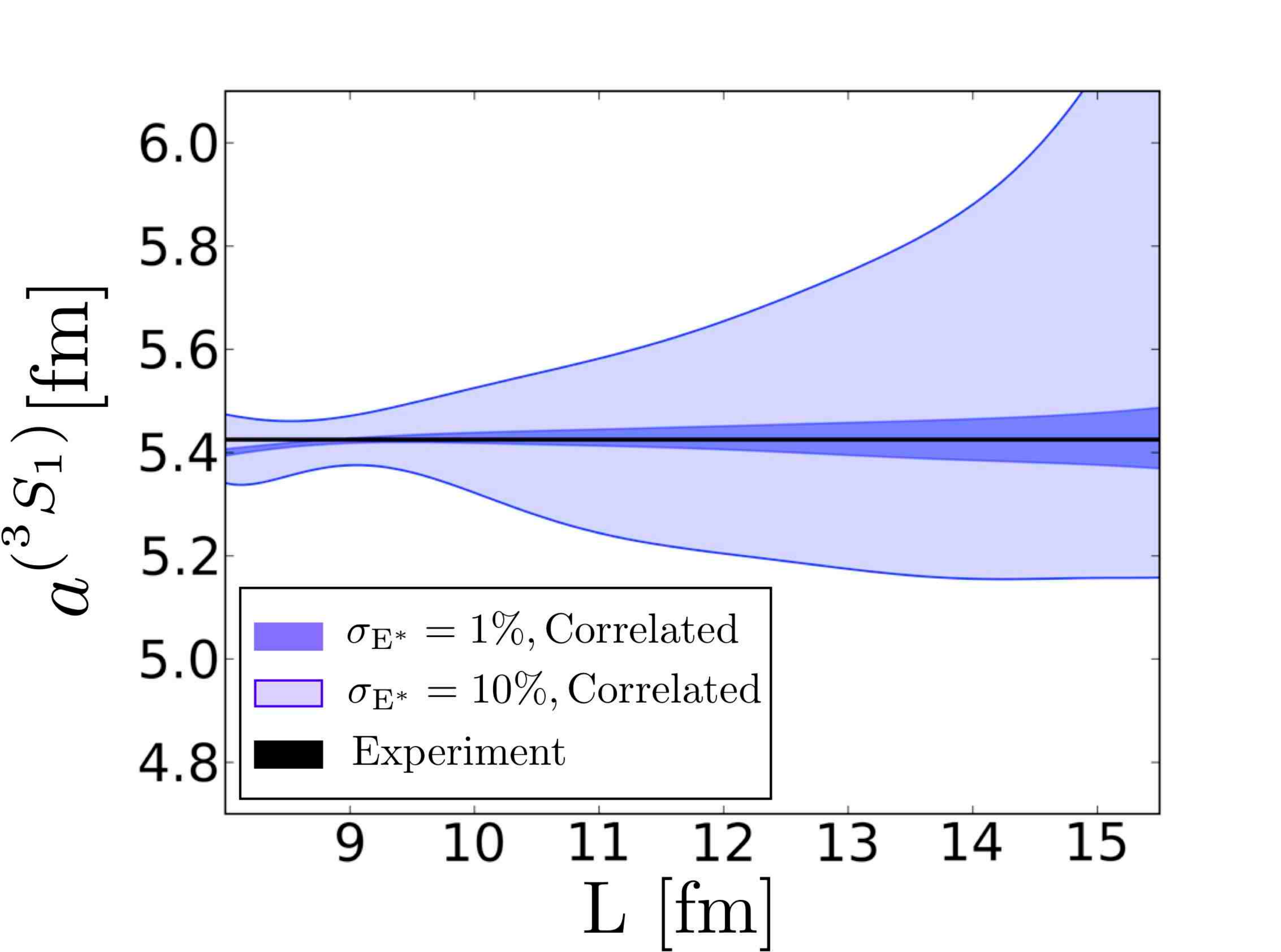}}
\subfigure[]{
\label{a_uncorr}
\includegraphics[scale=0.20]{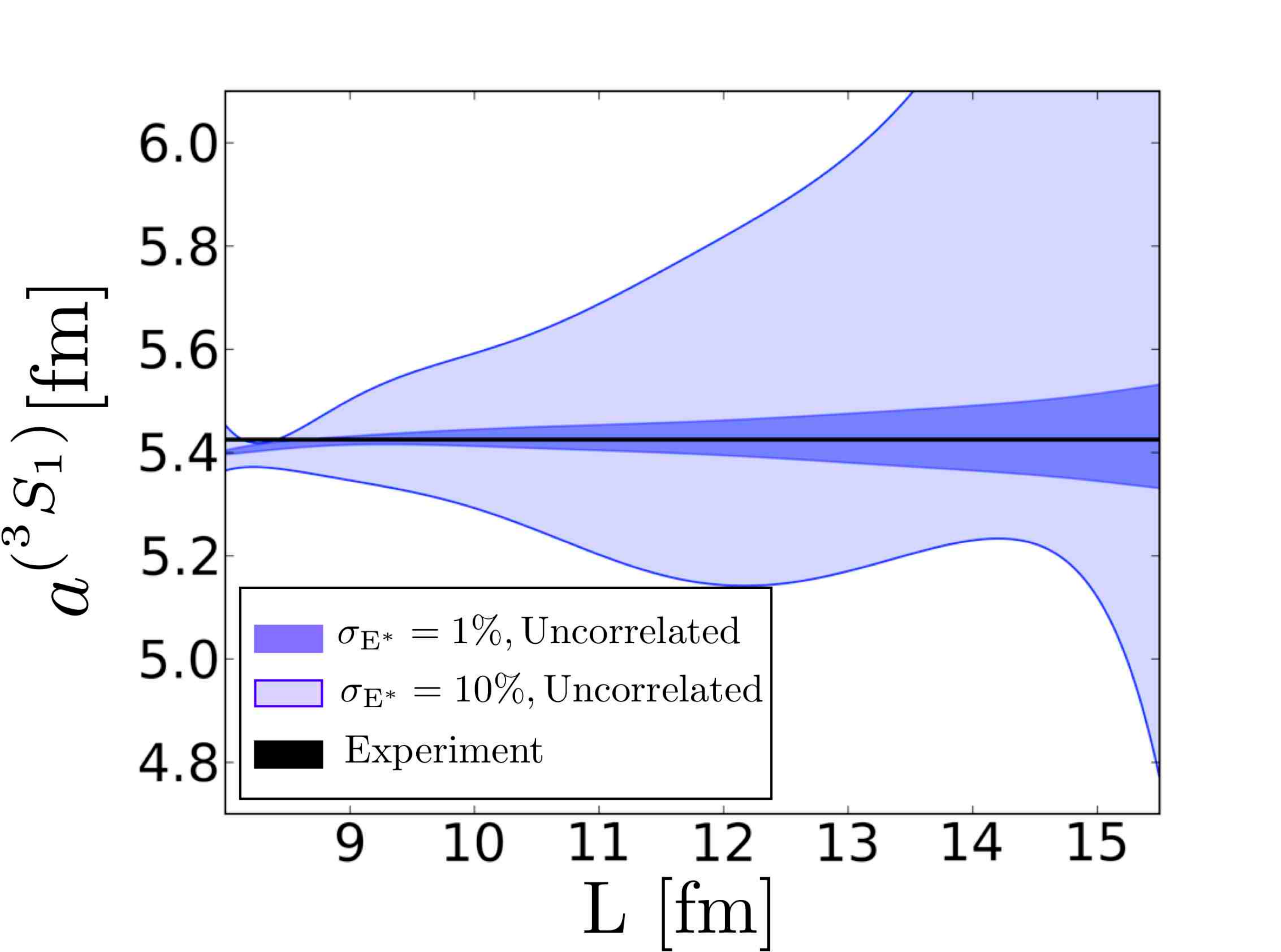}}
\subfigure[]{
\label{r_corr}
\includegraphics[scale=0.20]{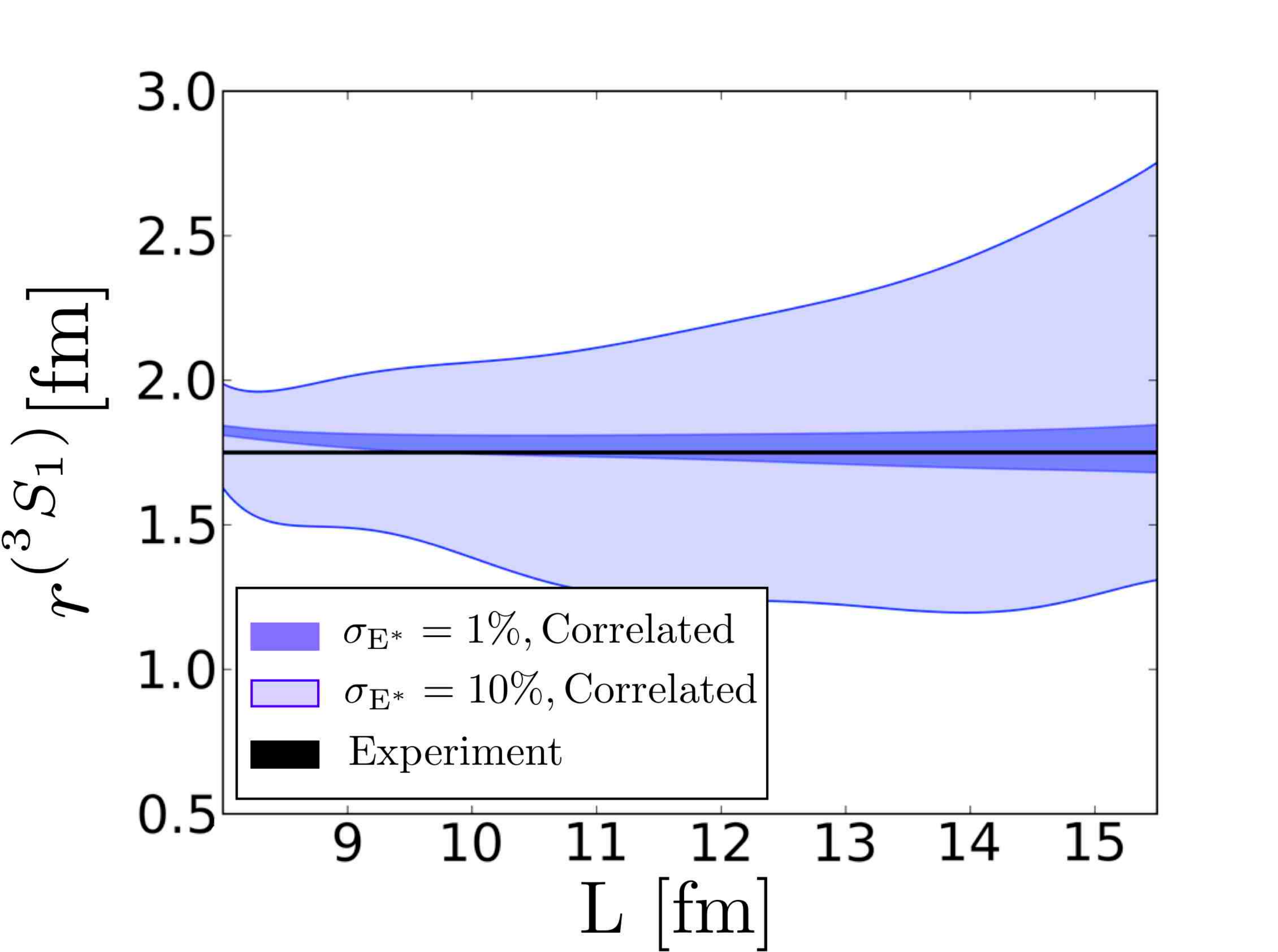}}
\subfigure[]{
\label{r_uncorr}
\includegraphics[scale=0.20]{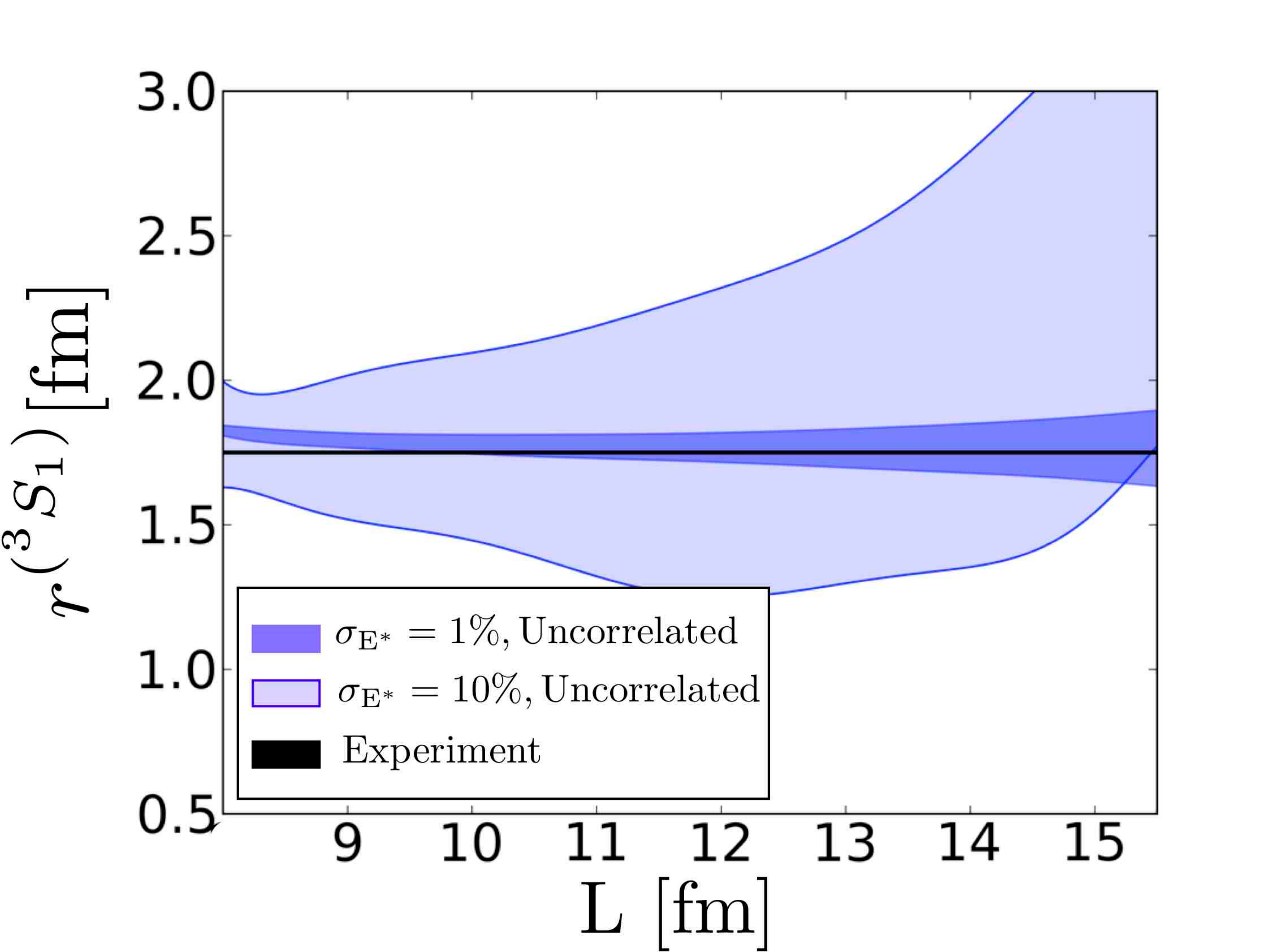}} 
\caption{
The  values of $\{a^{(^3S_1)}, r^{(^3S_1)}\}$ obtained by
  fitting the six independent bound-state energies with $|\textbf{d}|\leq\sqrt{3}$
  (depicted in Figs.~\ref{deut_cub}, \ref{deut_tet}, \ref{deut_ort}),
generated from synthetic LQCD calculations,
using the approximate QCs in Eqs.~(\ref{appr-T1}-\ref{appr-EA2}), as discussed
in the text. 
The black lines denote
the experimental value of these quantities determined by fitting the scattering
parameters obtained from Ref.~\cite{NIJMEGEN}. 
The dark (light) inner (outer) band is the 
$1\sigma$ band corresponding to the energies being  determined with 
1\% (10\%) precision. 
}
\label{fig:fakedata1}
\end{center}
\end{figure}

\begin{figure}[t]
\begin{center}  
\subfigure[]{
\label{Bd_corr}
\includegraphics[scale=0.20]{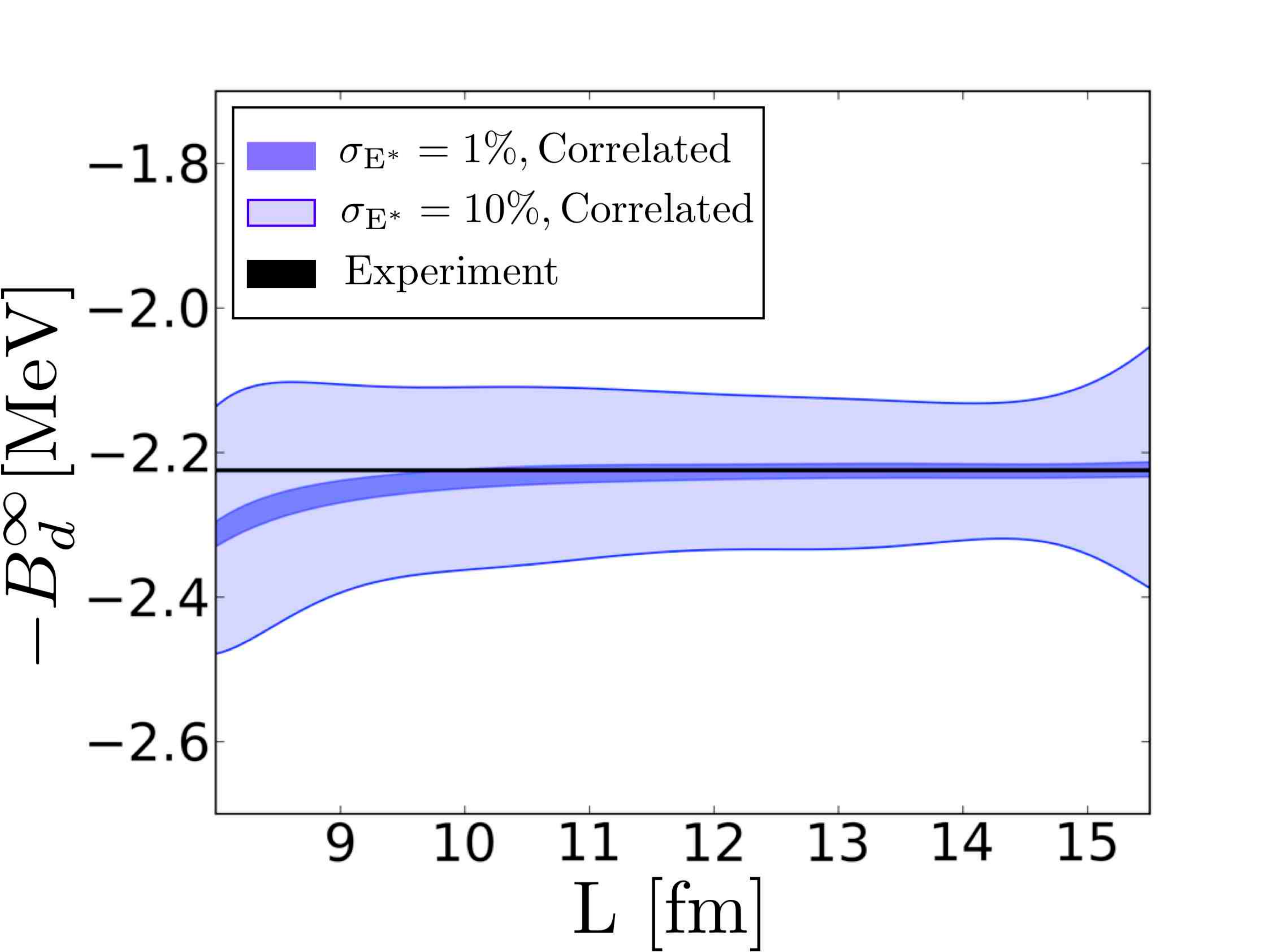}}
\subfigure[]{
\label{Bd_uncorr}
\includegraphics[scale=0.20]{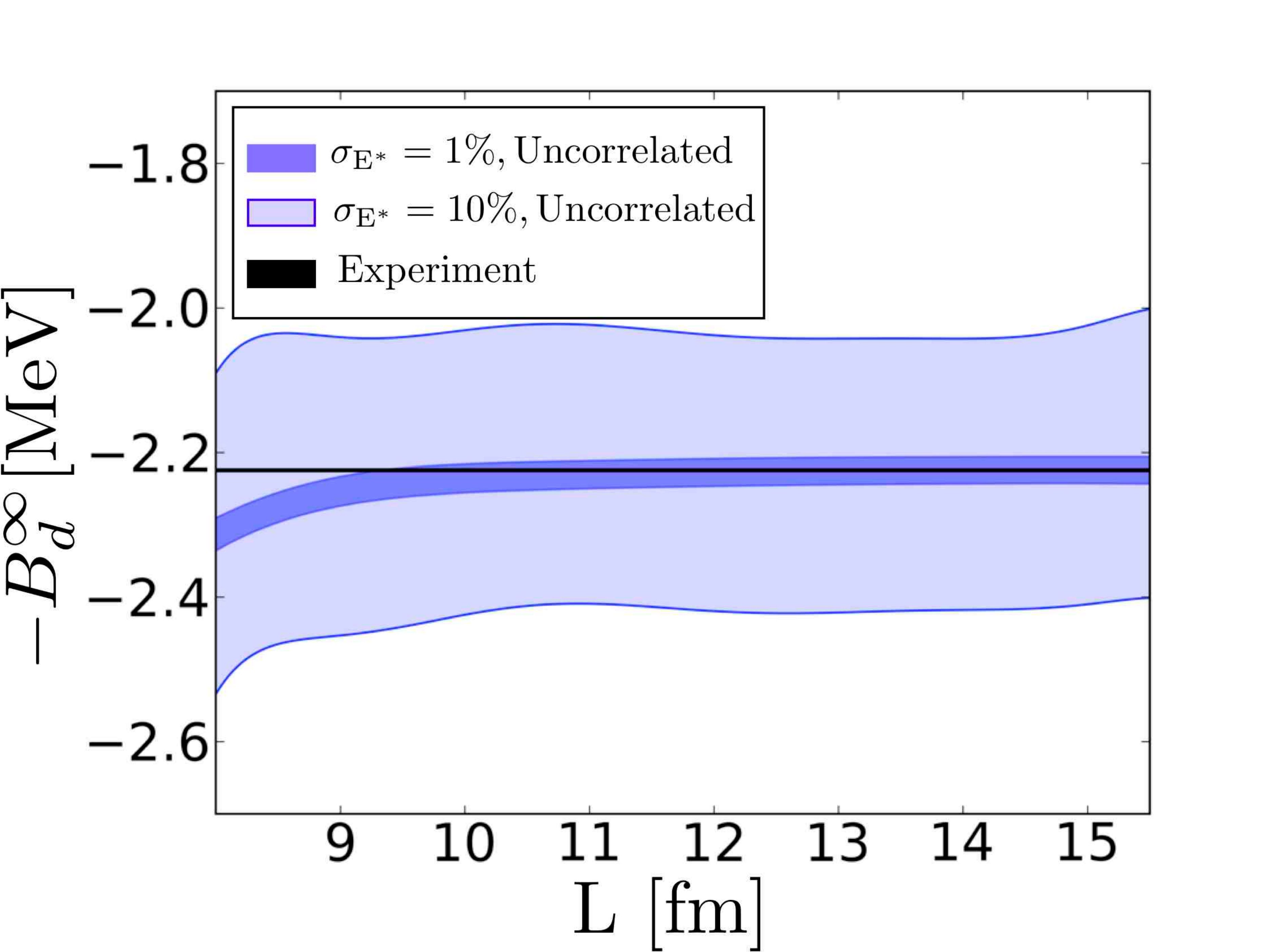}}
\subfigure[]{
\label{e1_corr}
\includegraphics[scale=0.20]{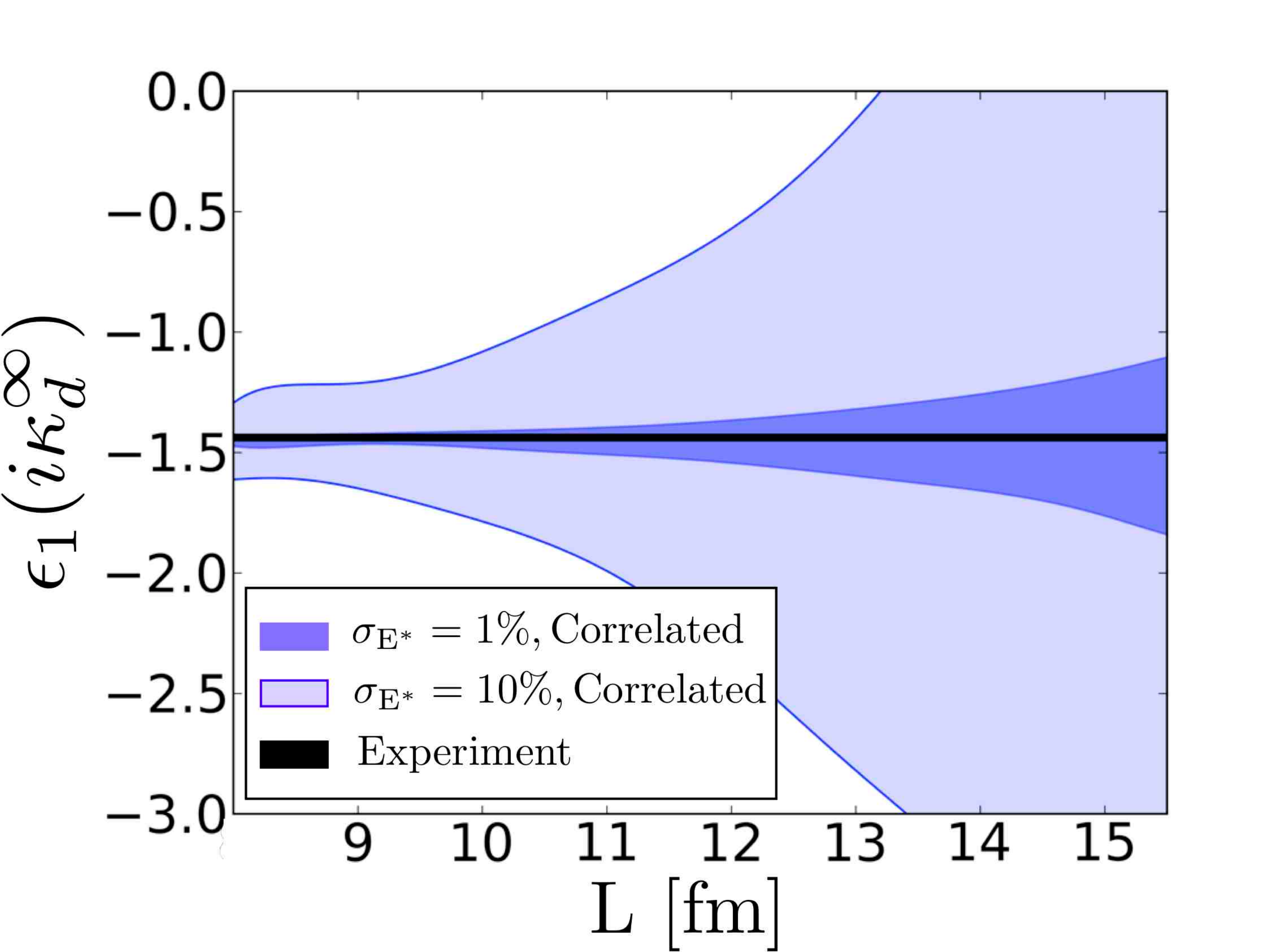}}
\subfigure[]{
\label{e1_uncorr}
\includegraphics[scale=0.20]{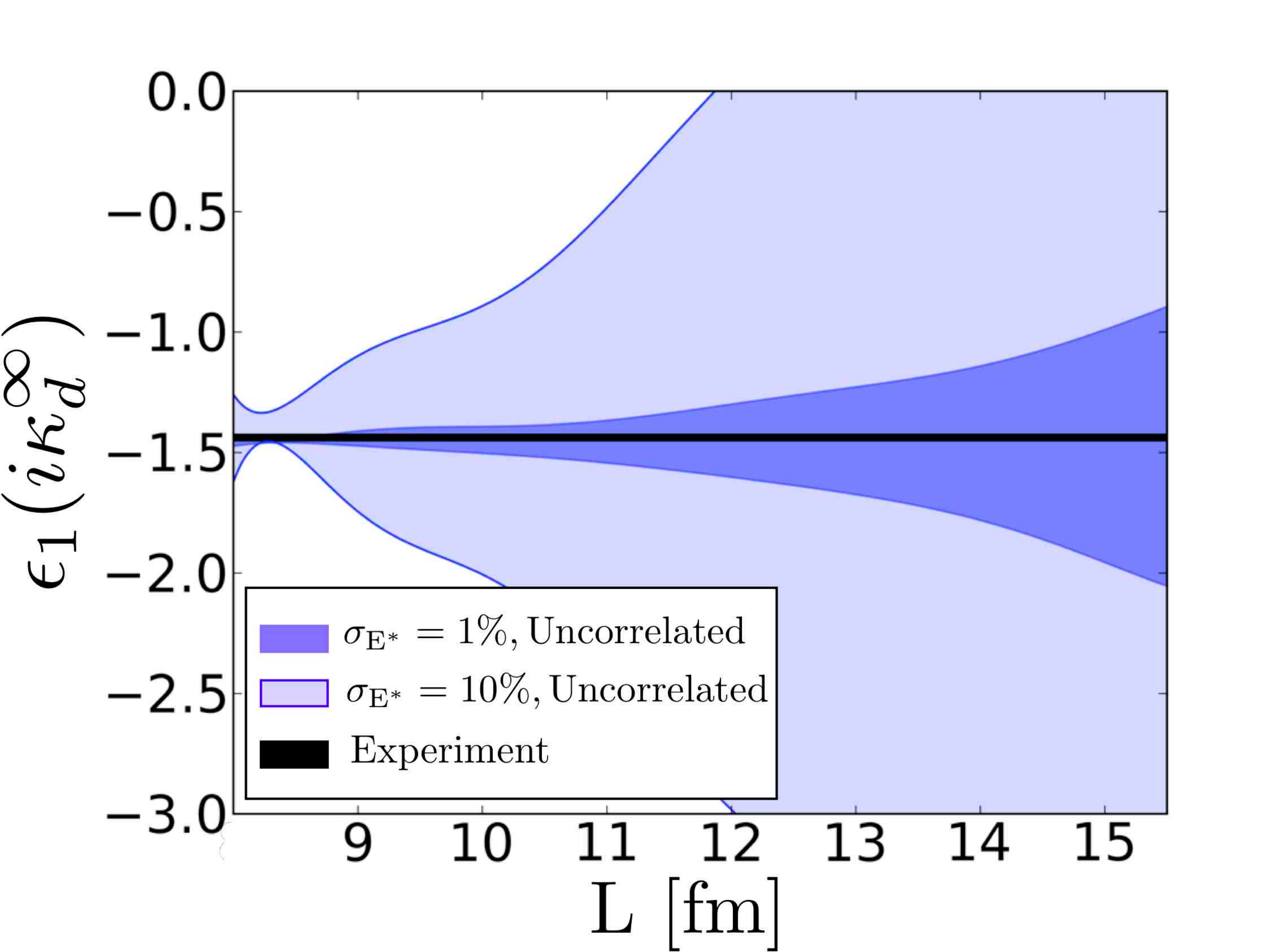}}
\caption{
The values of
  $\{-B_d^{\infty},\epsilon_1(\mathit{i}\kappa_d^{\infty})\}$ obtained by fitting the
  six independent 
bound-state energies with $|\textbf{d}|\leq\sqrt{3}$ (depicted in
Figs.~\ref{deut_cub}, \ref{deut_tet}, \ref{deut_ort}),
generated from synthetic LQCD calculations,
using the approximate QCs in
Eqs.~(\ref{appr-T1}-\ref{appr-EA2}). 
$\epsilon_1$ is in degrees and ${B_d^\infty}(\kappa_d^\infty)$ 
denotes the infinite-volume deuteron binding energy (momentum). 
The black lines denote
the experimental value of these quantities determined by fitting the scattering
parameters obtained from Ref.~\cite{NIJMEGEN}. 
The dark (light) inner (outer) band is the 
$1\sigma$ band corresponding to the energies being  determined with 
1\% (10\%) precision. 
}
\label{fig:fakedata2}
\end{center}
\end{figure}

Using the QCs,
the ground-state energy in each irrep is determined for a given lattice volume.
The level of precision of such a future LQCD calculation is introduced  by selecting a
modified energy for each ground state from a Gaussian distribution with the
true energy for its mean and the precision level multiplied by the mean
for its standard deviation.
This generates one set of uncorrelated ``synthetic LQCD calculations''.
To generate fully correlated ``synthetic LQCD calculations'', the same
fluctuation (appropriately scaled) 
is chosen for each energy.~\footnote{Partially-correlated 
``synthetic LQCD calculations'' 
can be generated by forming a weighted average of the uncorrelated and
fully-correlated calculations.}
These synthetic data are then taken to be the results of a possible future LQCD
calculation and analyzed accordingly to extract the scattering parameters.
The values of $\{a^{(^3S_1)},
  r^{(^3S_1)},-B_d^{\infty},\epsilon_1(\mathit{i}\kappa_d^{\infty})\}$ 
extracted from an analysis of the synthetic data
are shown in Figs.~\ref{fig:fakedata1},~\ref{fig:fakedata2} for both
correlated and uncorrelated energies. 
Since for ${\rm L}\lesssim 10~\rm{fm}$ the contribution of the D-wave phase shifts to the
bound-state spectrum is not negligible, the mean values of the scattering
parameters extracted using the approximated 
QCs deviate from their experimental values. 
This is most noticeable when the binding energies are determined at the 1\%
level of precision, 
where the S-matrix parameters and predicted $B_d^\infty$ can deviate by
$\sim 3\sigma$ from the experimental values for this range of volumes. 
For $10~\rm{fm}$$ < {\rm L} < $$14~\rm{fm}$, one can see that these quantities can be extracted
with high accuracy using this method,
but it is important to note that the precision with which $\{a^{(^3S_1)},
  r^{(^3S_1)},\epsilon_1(\mathit{i}\kappa_d^{\infty)}\}$ 
can be extracted decreases as a function of increasing volume. 
The reason is that the bound-state energy in each irrep asymptotes to the physical deuteron binding energy
in the infinite-volume limit.
In this limit, sensitivity to $\epsilon_1$ is lost and the $\alpha$-wave phase shift
is determined  at a single energy, the deuteron pole. 
Therefore, for sufficiently large volumes one cannot  independently  resolve $a^{(^3S_1)}$
and $r^{(^3S_1)}$. 
This analysis of synthetic data reinforces the fact that the FV spectrum not
only depends on the magnitude of $\epsilon_1$ but also its sign. 
As discussed in Sect.~\ref{sec:DeutFV}, this sensitivity can be deduced
from the full QCs in Eqs.(\ref{I000T1}-\ref{I111E}), 
but it is most evident from the approximated
QCs in Eqs.~(\ref{appr-T1}-\ref{appr-EA2}).

In performing this analysis, we have benefited from two important pieces of
{\it apriori} knowledge at the physical light-quark masses. 
First is that in the volumes of interest, the bound-state energy in each
irrep falls within the radius of convergence of the 
ERE, $|{\rm E}^*|<m_\pi^2/4M$. 
For unphysical light-quark masses, the S-matrix elements could in principle
change in such a way that this need not be the case
and pionful EFTs would be required 
to extract the scattering parameters from the FV spectrum. 
Second is that the D-wave phase shifts are naturally small. 
Again, since the dependence of these phase shifts on the light-quark masses 
can only be estimated, further investigation would be required.
To improve upon this analysis,  
the J=1 $\beta$-wave and  $J=2,3$ D-wave phase shifts
would have to be extracted from the scattering states.
As is evident from Fig.~\ref{T1specfull}, states that have a strong dependence
on the D-wave phase shifts will, in general, 
lie above the t-channel cut. 
In principle, one could attempt to extract them by fitting the FV bound-state
energies for ${\rm L}\leq 10~{\rm fm}$ with the full QCs. 
In practice, this will be 
challenging as eight scattering parameters appear in the ERE at
the order at which the 
$J=1$ $\beta$-wave and $J=2,3$ D-wave
phase shifts first contribute.
This is also formally problematic since for small volumes, $m_\pi {\rm L} \lsim
2\pi $, 
finite range effects are no longer negligible. 
Although these finite range effects have been estimated for two nucleons in a  S-wave~\cite{Sato:2007ms}, 
they remain to be examined for the general NN  system.

\section{The Finite-Volume Deuteron Wavefunction and the Asymptotic D/S
  Ratio \label{sec:wavefunction}}
\noindent
The S-matrix dictates the asymptotic behavior of the NN wavefunction, 
and as a result the infra-red (IR) distortions of the wavefunction inflicted by
the boundaries of the lattice volume have a direct connection to the parameters
of the scattering matrix, as exploited by L\"uscher. 
Outside the range of the nuclear forces,   the FV wavefunction of the NN system 
is obtained from the solution of the
Helmholtz equation in a cubic volume with the 
PBCs~\cite{Luscher:1986pf,Luscher:1990ux, Rummukainen:1995vs, Ishizuka:2009bx}. 
By choosing the amplitude of the $l=0$ and $l=2$ components of the 
FV wavefunction to recover the asymptotic D/S ratio of the infinite-volume
deuteron,
it is straightforward to show~\cite{Luscher:1986pf,Luscher:1990ux} 
that the unnormalized FV deuteron wavefunctions
associated with the approximate QCs in Eqs.~(\ref{appr-T1}-\ref{appr-EA2})
are
\begin{eqnarray}
\psi^{V,{\bf d}}_{1,M_J} (\mathbf{r};\kappa)
\ =\ 
\psi^{\infty}_{1,M_J} (\mathbf{r};\kappa)
\ +\ 
\sum_{\mathbf{n} \neq \mathbf{0}} e^{i \pi \mathbf{n} \cdot \mathbf{d}} \ 
\psi^{\infty}_{1,M_J} (\mathbf{r}+\mathbf{n}{\rm L};\kappa)
\ \ \ ,
\label{psi-V}
\end{eqnarray}
with $r = |{\bf r}|>R$,
where  $\mathbf{r}$ denotes the relative displacement 
of the two nucleons, 
and $R >>  1/m_\pi$ is the approximate range of the nuclear interactions.
The subscripts on the wavefunction refer to the $J=1,~M_J=0,\pm1$ 
quantum numbers of the state and 
$\mathbf{n}$ is an integer triplet. 
In order for Eq.~(\ref{psi-V}) to be an energy eigenstate of the Hamiltonian, 
$E^*=-{\kappa^2}/{M}$ has to be an energy eigenvalue of the NN  system in
the finite volume,
obtained from the  QCs in Eqs.~(\ref{appr-T1}-\ref{appr-EA2}). 
$\psi^{\infty}_{1,M_J}(\mathbf{r})$ is the asymptotic infinite-volume wavefunction of the deuteron,
\begin{eqnarray}
\psi^{\infty}_{1,M_J} (\mathbf{r};\kappa)
 \ =\ 
\mathcal{A}_S 
\ \left(\ 
\frac{e^{-\kappa r}}{r}
 \mathcal{Y}_{1M_J;01}(\hat{\mathbf{r}})
\ +\ 
\eta~\frac{e^{-\kappa r}}{r} (1+\frac{3}{\kappa
  r}+\frac{3}{\kappa^2r^2}) 
\mathcal{Y}_{1M_J;21}(\hat{\mathbf{r}})
\ \right)
\ \ \ .
\label{psi-inf}
\end{eqnarray}
with $\mathcal{Y}_{JM_J;L1}$ being the well-known spin-orbital functions,
\begin{eqnarray}
\mathcal{Y}_{JM_J;L1}(\hat{\mathbf{r}})
\ =\ 
\sum_{M_L,M_S}\left\langle L M_L 1 M_S|J M_J\right\rangle  \
Y_{L M_L}(\hat{\mathbf{r}})\ \mathcal{\chi}_{1 M_S}
\ \ \ ,
\label{Y-def}
\end{eqnarray}
where $\mathcal{\chi}_{1 M_S}$ is the spin wavefunction of the deuteron. 
$\eta$ is the deuteron asymptotic D/S ratio which is
related to the mixing angle via
$\eta=-\tan{\epsilon_1}|_{k^*=i\kappa^\infty_d}$~\cite{Blatt:1952zza}. 
As is well known from the effective range theory~\cite{PhysRev.76.38,
  PhysRev.77.647}, 
the short-distance contribution to the \textit{outer} quantities of the
deuteron, 
such as  the quadrupole moment,
can be approximately taken into account by requiring the
normalization  of the asymptotic wavefunction of the deuteron,
obtained from the residue of the S-matrix at the
deuteron pole, 
to be approximately 
$|\mathcal{A}_S|^2\approx{2\kappa}/({1-\kappa  r^{(^3S_1)}})$. 
Corrections to this normalization arise at 
$\mathcal{O}\left({\kappa^3}/{R}^2,\kappa\eta^2\right)$,
at the same order the 
$J=1$ $\beta$-wave and $J=2,3$ D-waves
contribute.
In writing the FV wavefunction in Eq.~(\ref{psi-V}) 
contributions from  these
waves have been neglected.

\begin{figure}[h]
\begin{center}  
\subfigure[]{
\label{WF-T1-L10}
\includegraphics[scale=0.3]{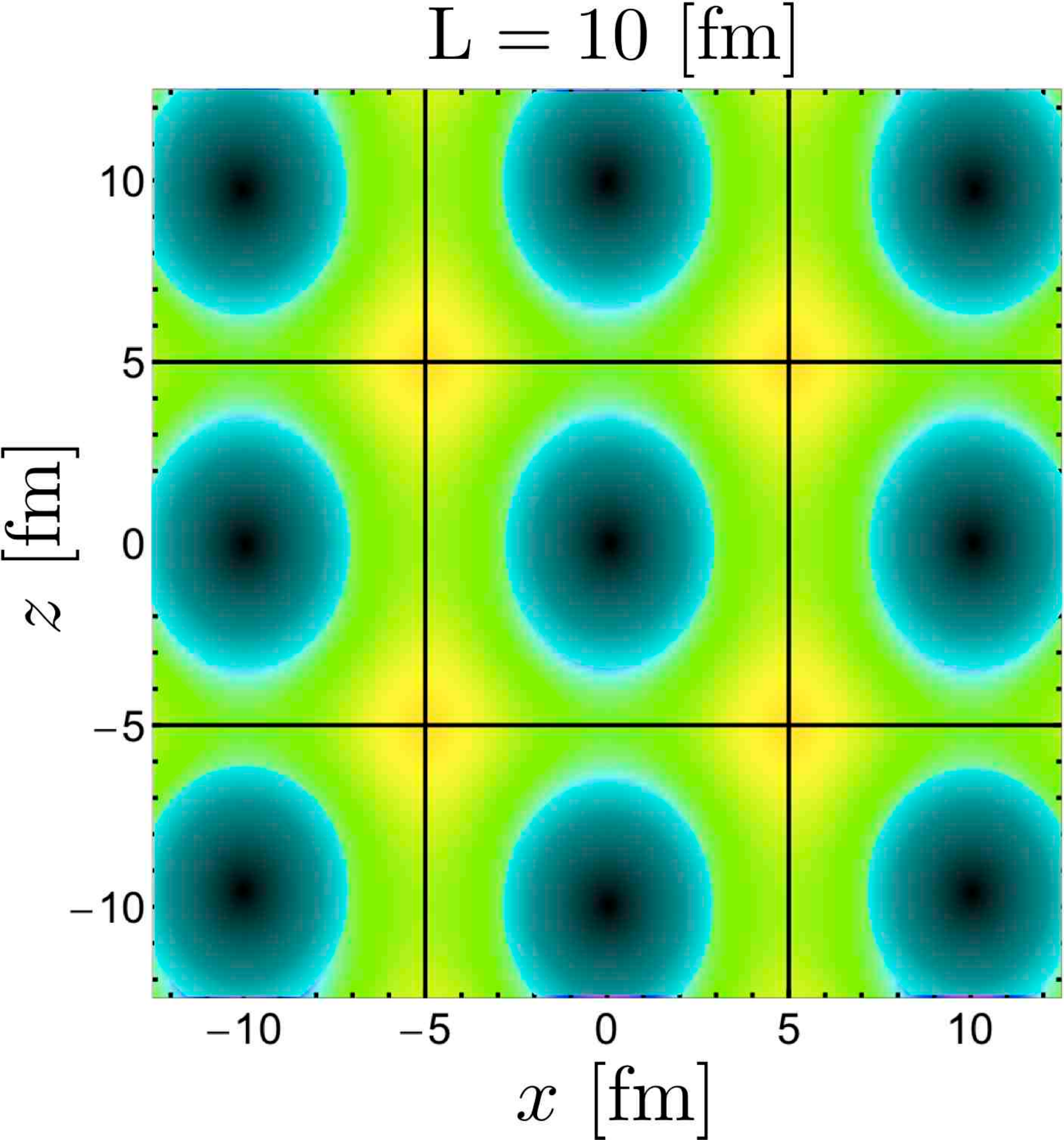}}
\subfigure[]{
\label{WF-T1-L15}
\includegraphics[scale=0.3]{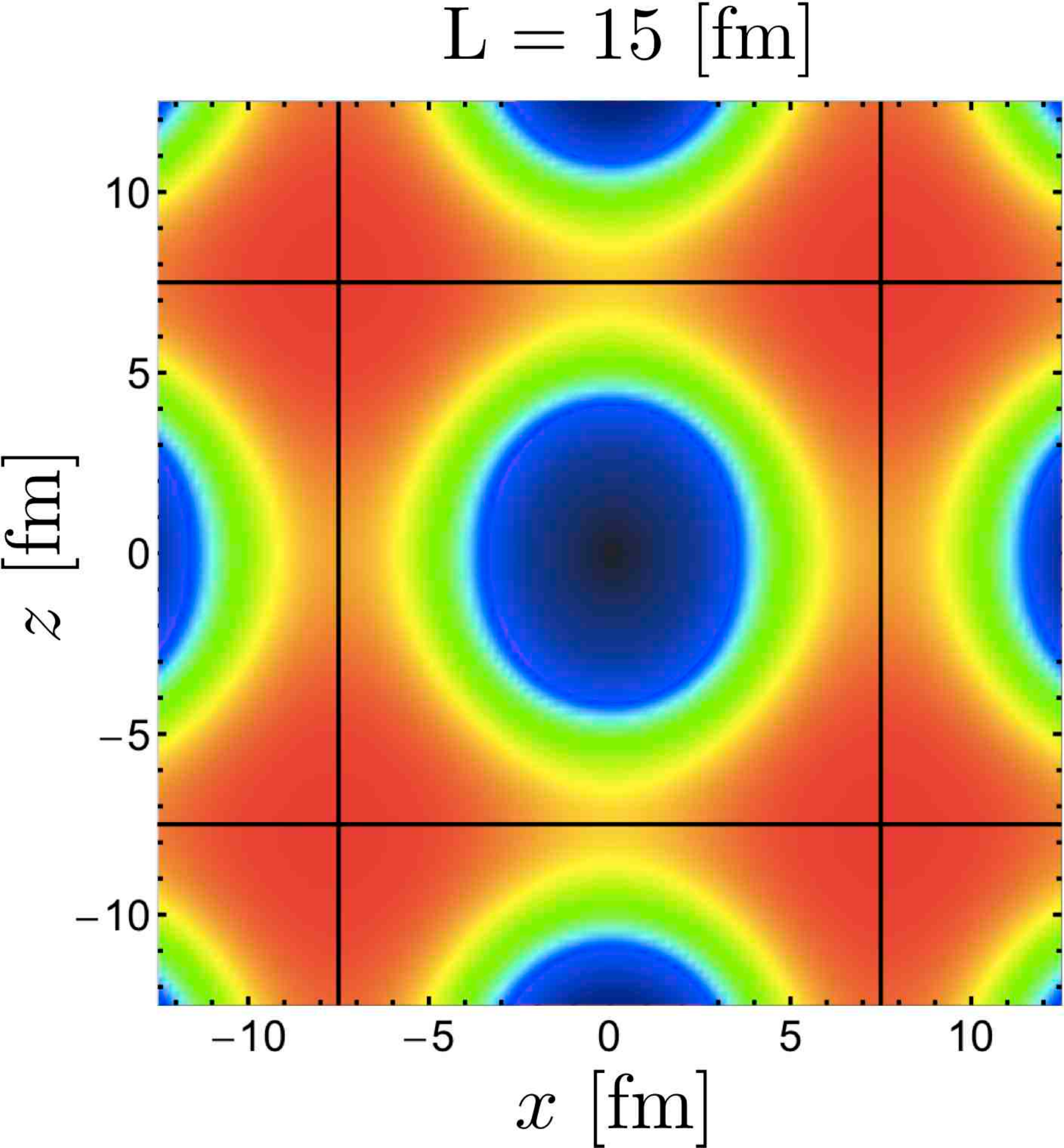}}
\subfigure[]{
\label{WF-T1-20}
\includegraphics[scale=0.3]{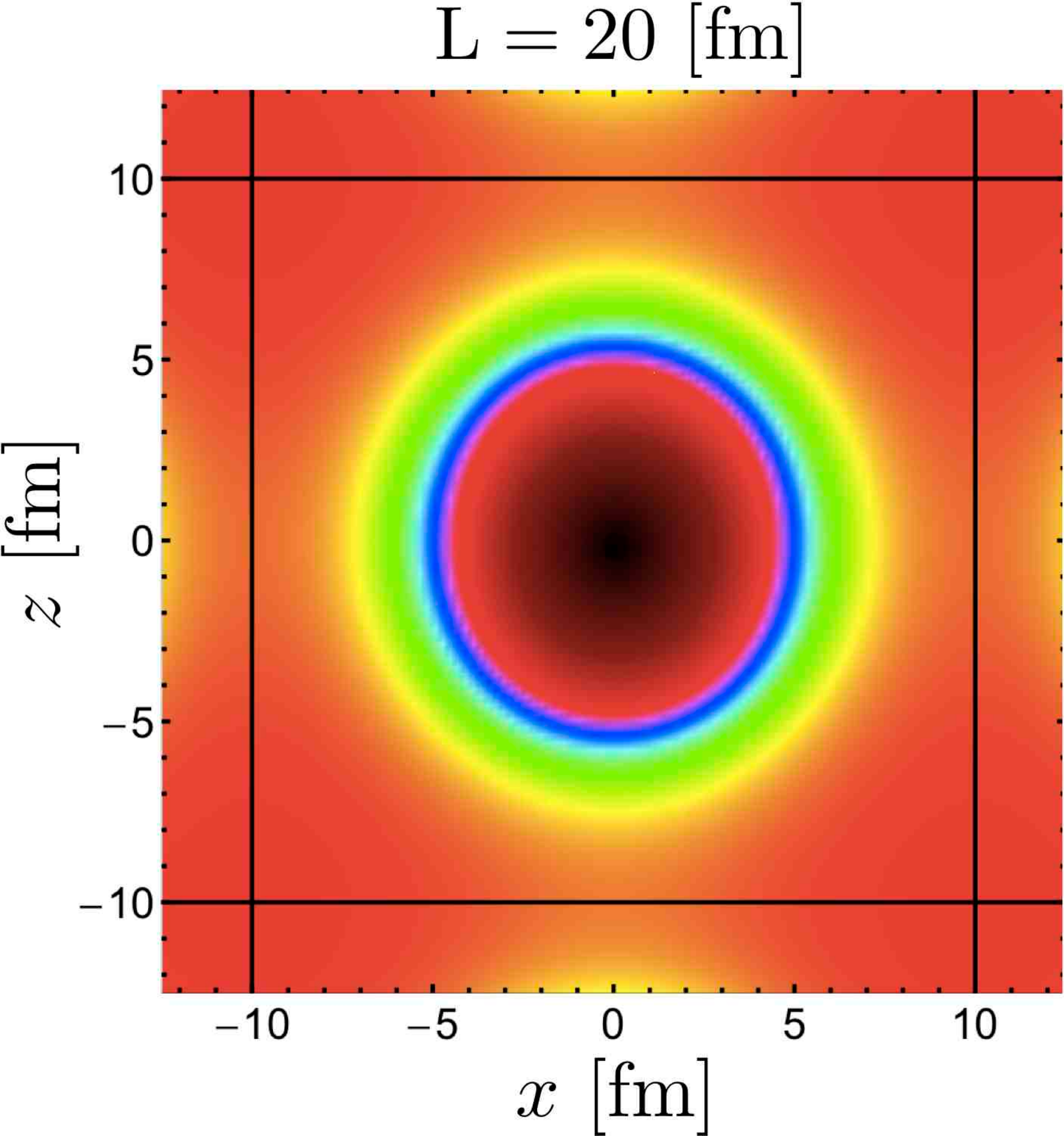}}
\subfigure[]{
\label{WF-T1-30}
\includegraphics[scale=0.3]{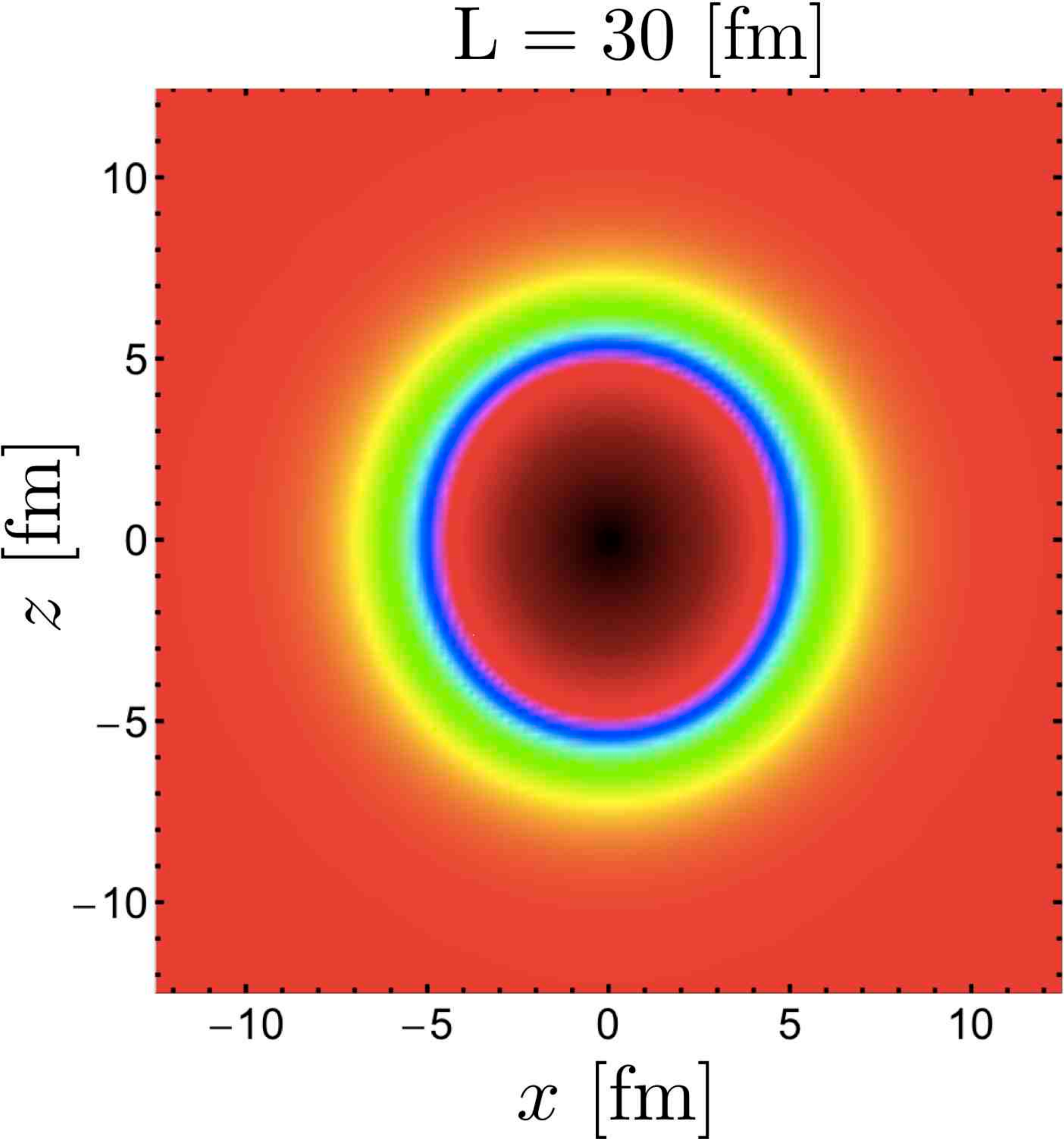}}
\caption{The mass density  in the $xz$-plane from the $\mathbb{T}_1$ FV deuteron wavefunction  at rest
  for 
${\rm L}=10,15,20$, and $30~{\rm fm}$.
}
\label{WF-T1}
\end{center}
\end{figure}
An important feature of the FV wavefunction in Eq. (\ref{psi-V}) is the
contribution from partial waves 
other than $l=0$ and $l=2$, 
which results from the cubic distribution of the periodic images. 
While there are also FV corrections to the $l=0$ component of the wavefunction, 
the FV corrections to the $l=2$ component are enhanced  for systems with
${\bf d}=(0,0,1)$ and $(1,1,0)$. 
By forming appropriate linear combinations of the $\psi^{V,{\bf d}}_{1,M_J}$ 
that transform according to a given irrep of the cubic, tetragonal, 
orthorhombic and trigonal point groups (see Table \ref{irreps}),
wavefunctions for the systems with ${\bf d}=(0,0,0)$, $(0,0,1)$, $(1,1,0)$ and
$(1,1,1)$
can be obtained.
The 
mass density in the $xz$-plane from 
the FV wavefunction  of the deuteron 
at rest in the volume, obtained from the $\mathbb{T}_1$ irrep of the
  cubic group 
is shown in Fig.~\ref{WF-T1}
for ${\rm L}=10,15,20$, and $30~{\rm fm}$,
and for the boosted systems in  Figs.~\ref{WF-A2}-\ref{WF-A2E} of Appendix \ref{sec:Wavefunc}. 
As the interior region of the wavefunctions cannot be deduced from its asymptotic
behavior alone, it is ``masked'' 
in  Fig.~\ref{WF-T1} and Figs.~\ref{WF-A2}-\ref{WF-A2E} by a 
shaded disk.
Although the deuteron wavefunction exhibits its slight prolate shape 
(with respect to its spin axis) 
at large volumes, 
it is substantially deformed in smaller volumes, 
such that the deuteron can no longer be thought as a compact bound state within
the lattice volume. 
When the system is at rest, the FV deuteron 
is more prolate than the infinite-volume deuteron. 
When the deuteron is boosted along the $z$-axis with ${\bf d}=(0,0,1)$, 
the distortion of the wavefunction 
is large, 
and in fact, for a significant  
range of volumes (${\rm L} \lesssim 30~\rm{fm}$), the FV effects
give rise to an oblate (as opposed to prolate) deuteron in the 
$\mathbb{E}$ irrep, 
Fig. \ref{WF-E}, and a more prolate shape in the 
$\mathbb{A}_2$ irrep, 
Fig. \ref{WF-A2}. 
For ${\bf d}=(1,1,0)$, 
the system remains prolate for the deuteron in the 
$\mathbb{B}_2/\mathbb{B}_3$ irreps,
Fig. \ref{WF-B2B3}, 
while it becomes  
oblate in the $\mathbb{B}_1$ irrep, Fig. \ref{WF-B1}, for volumes up to ${\rm L} \sim
30~\rm{fm}$. 

\begin{figure}[h]
\begin{center}  
\subfigure[]{
\label{NDS-T1}
\includegraphics[scale=0.25]{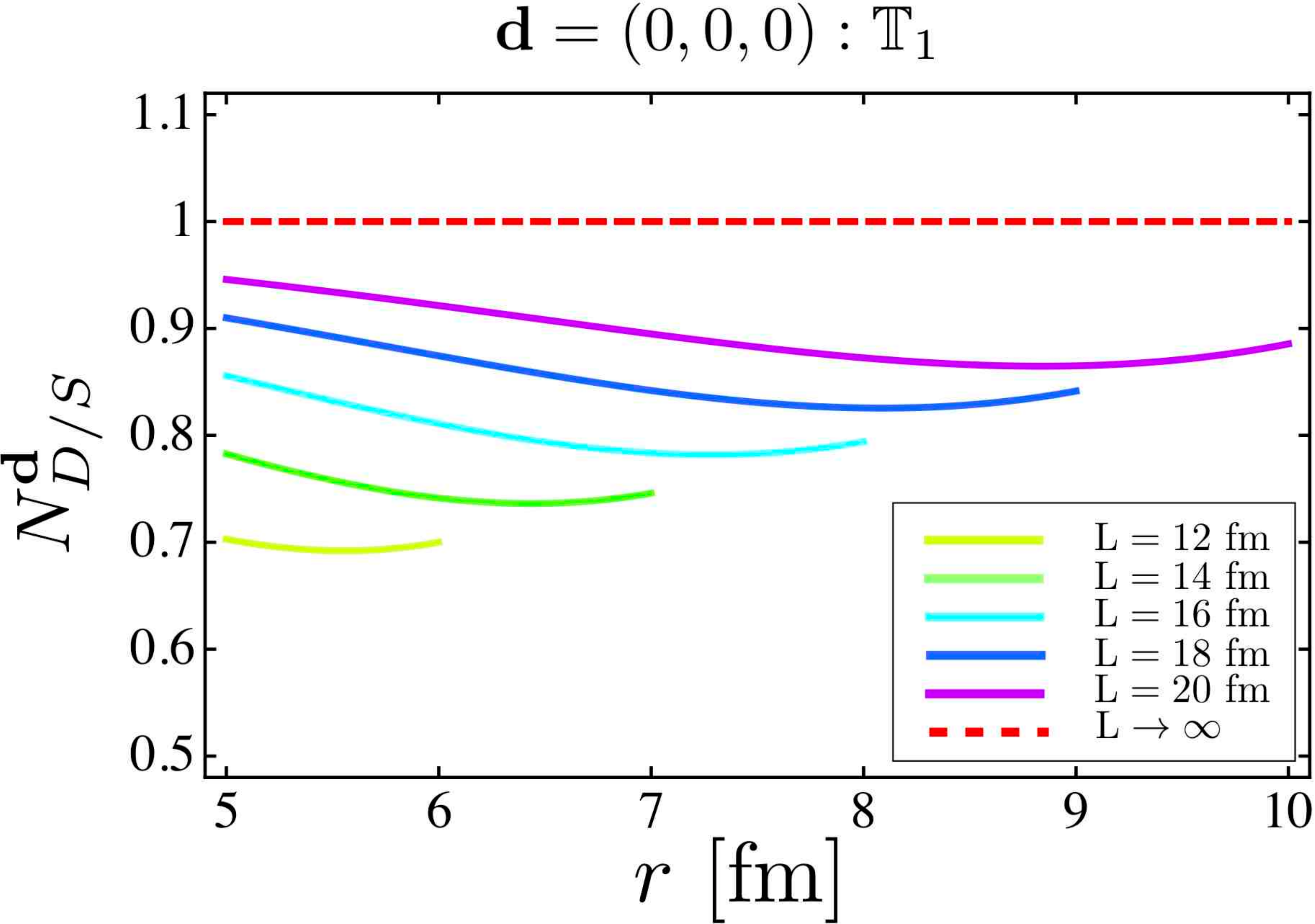}}
\subfigure[]{
\label{NDS-E}
\includegraphics[scale=0.25]{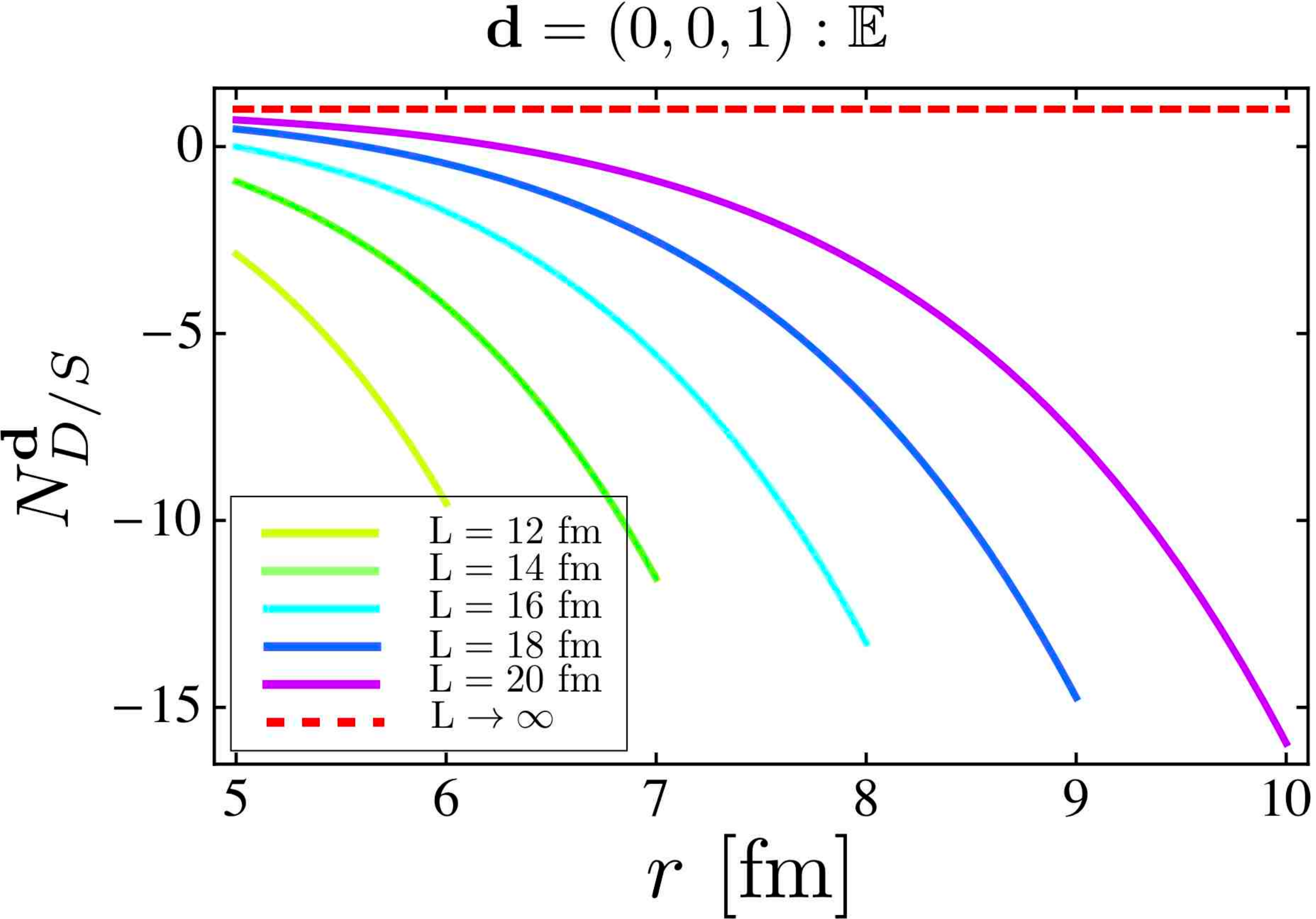}}
\subfigure[]{
\label{NDS-B2B3}
\includegraphics[scale=0.25]{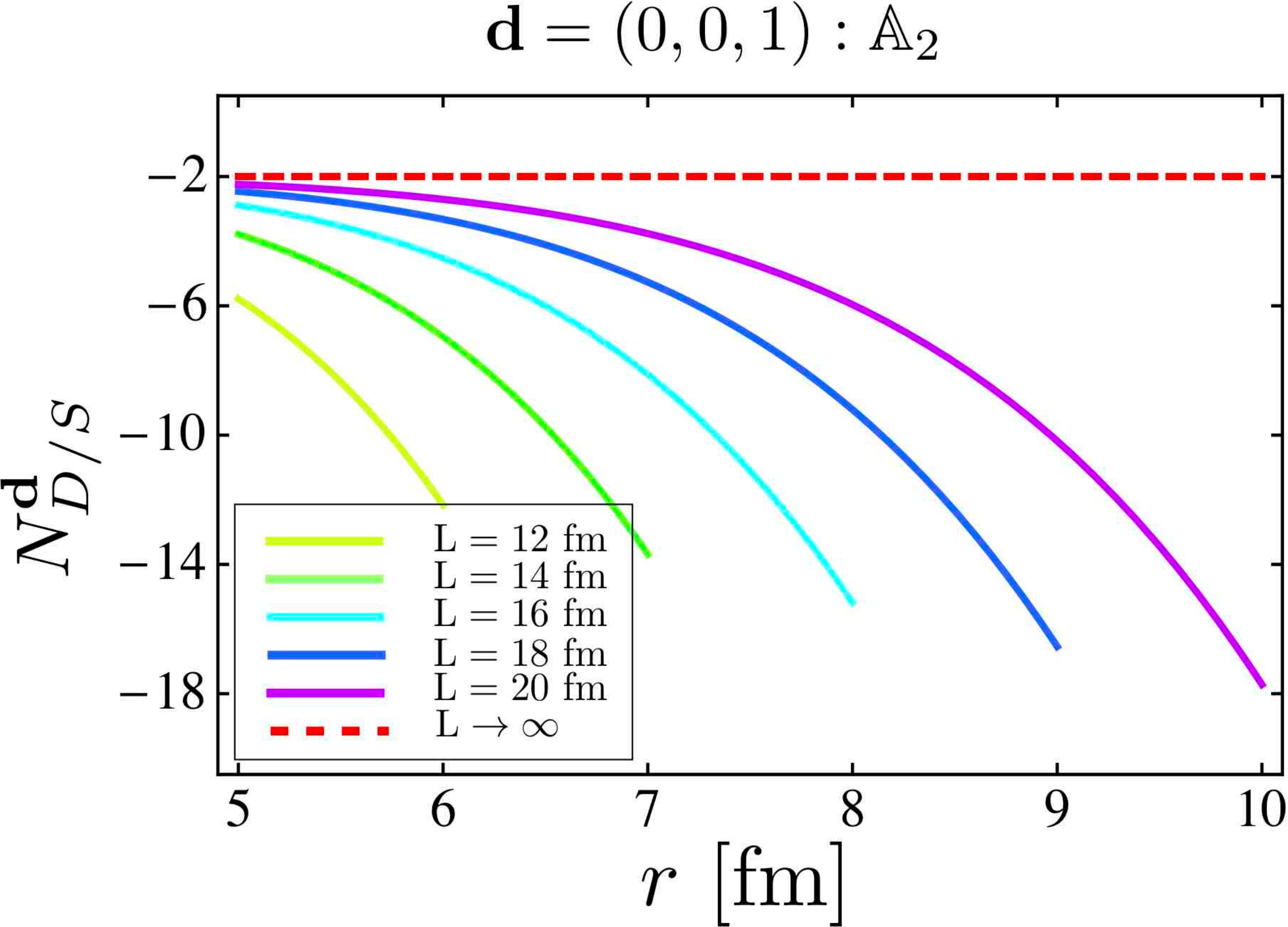}}
\subfigure[]{
\label{NDS-A2E}
\includegraphics[scale=0.25]{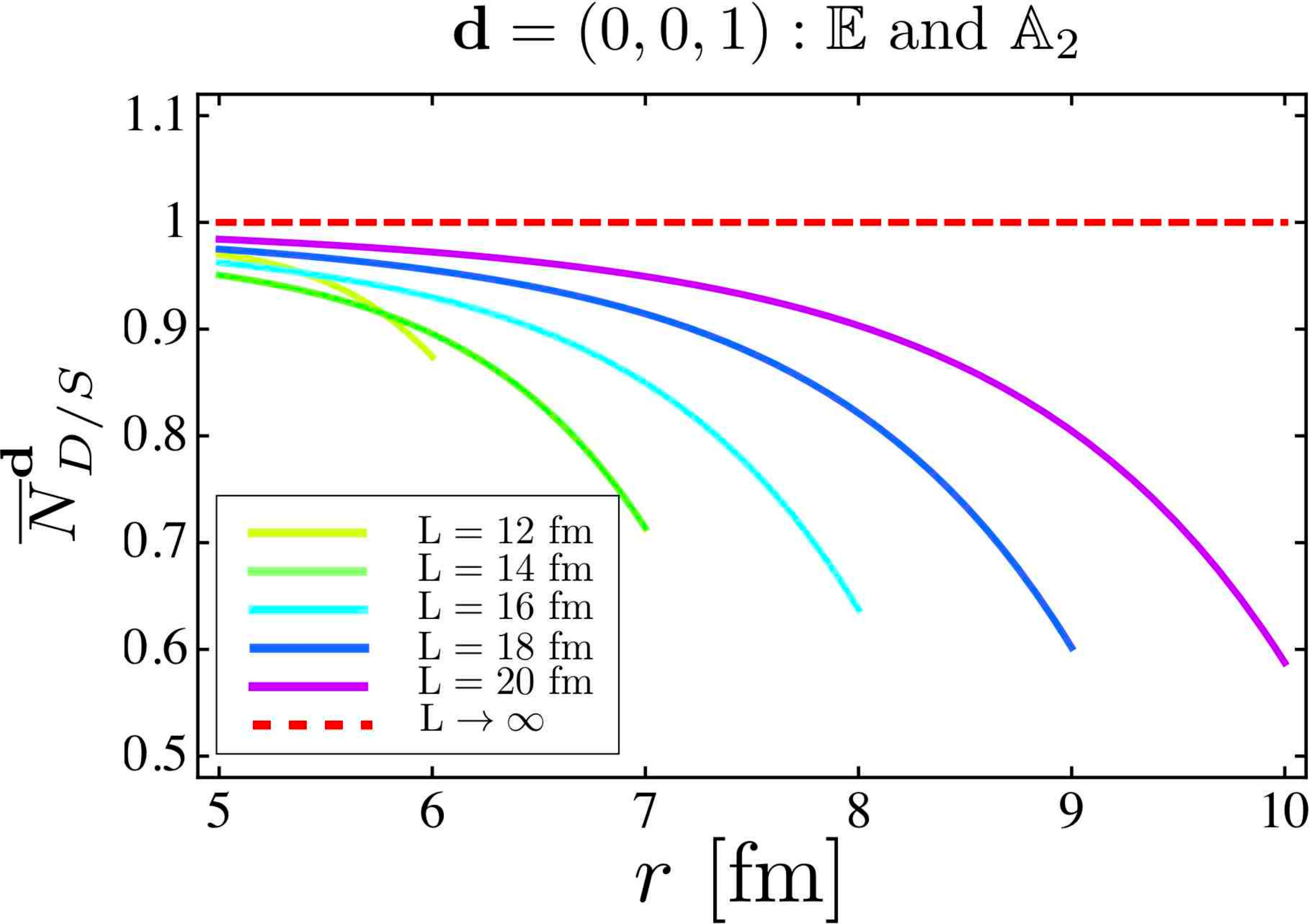}}
\caption{
The normalized D/S ratio of the deuteron wavefunction with $M_J=M_S=1$, defined in 
Eq.~(\protect\ref{ratio}) and Eq.~(\protect\ref{ratiob})
in the 
(a) $\mathbb{T}_1$, 
(b) $\mathbb{E}$ and 
(c) with $M_J=M_S=0$ in the $\mathbb{A}_2$ irrep,
along with 
(d) the difference of the D/S ratios in  the  $\mathbb{E}$ and  $\mathbb{A}_2$ irreps, defined
in Eq.~(\protect\ref{eq:A2Ediff}). 
The red-dashed lines show 
the infinite-volume value.}
\label{fig:NDS}
\end{center}
\end{figure}

Although the normalization factor $\mathcal{A}_S$ corrects  for
the fact that the complete wavefunction is not given by the asymptotic form
given in Eq.~(\ref{psi-inf}) for  $|{\bf r}|\lesssim r^{(^3S_1)}/2$ in infinite volume, it gives rise to a
normalization ambiguity in the FV. 
On the other hand, the asymptotic D/S ratio is protected by the
S-matrix, and 
can be directly extracted from the long-distance tail of the lattice
wavefunctions.~\footnote{
The energy-dependent ``potentials'' generated by HALQCD
  and used to compute scattering parameters, including $\epsilon_1$ (at
unphysical light-quark masses)~\cite{Murano:2013xxa},
are expected to reproduce the predictions of QCD only at the energy-eigenvalues of
their LQCD calculations.  
Hence, if they had found a bound deuteron,
their prediction for $\epsilon_1$ would be expected to be correct at the
calculated deuteron binding energy.
} 
It is evident from Eq.~(\ref{psi-inf}) that the ratio
\begin{eqnarray}
N_{D/S}^{{\bf d}; M_J, M_S}(r;\kappa) \equiv
\frac{\psi_{D;M_J,M_S}^{V,{\bf d}}(r;\kappa)}{\eta~\chi(r;\kappa)~\psi_{S;M_J,M_S}^{V,{\bf d}}(r;\kappa)}
\ \ \ ,
\label{ratio}
\end{eqnarray}
with $\chi(r;\kappa)=\sqrt{\frac{1}{10}}(1+\frac{3}{\kappa
  r}+\frac{3}{\kappa^2r^2})$, 
is unity  for the $M_J=M_S=1$ component of the infinite-volume deuteron wavefunction
(and is equal to $-2$ for the $M_J=M_S=0$ component),
where
$\psi_{S;M_J,M_S}^{V,{\bf d}}$  and $\psi_{D;M_J,M_S}^{V,{\bf d}}$ 
are 
\begin{eqnarray}
\psi_{S;M_J,M_S}^{V,{\bf d}}(r;\kappa)
& = &  
\int d\Omega_{\hat{\mathbf{r}}} \ 
\psi^{V,{\bf d}}_{1,M_J}(\mathbf{r},\kappa)
\big|_{M_S}\ 
\ Y_{00}(\hat{\mathbf{r}})
\ ,
\nonumber\\
\psi_{D;M_J,M_S}^{V,{\bf d}}(r;\kappa)
& = & 
\int d\Omega_{\hat{\mathbf{r}}}\  
\psi^{V,{\bf d}}_{1,M_J}(\mathbf{r},\kappa) \big|_{M_S}\ 
\ Y_{20}(\hat{\mathbf{r}})
\ ,
\label{ratiob}
\end{eqnarray}
with $r\le {\rm L}/2$.
By evaluating the FV wavefunction 
in different irreps with $|\mathbf{d}|\leq \sqrt{3}$, 
this ratio can be determined in the FV, as is shown in 
Fig.~\ref{fig:NDS}. 
Not only does it exhibit strong dependence on the volume, 
but also varies dramatically as a function of $r$. 
This is due to the fact that the periodic images give rise to exponentially
growing contributions 
to the FV wavefunction in $r$. 
For the FV deuteron at rest and with $\mathbf{d}=(1,1,1)$, 
a sufficiently small $r$ gives rise to a $N_{D/S}^{{\bf d};M_J,M_S}$ that is not severely
distorted by volume effects even in small volumes.
In contrast,  
this ratio deviates significantly from its infinite-volume value for systems
with $\mathbf{d}=(0,0,1)$ and $(1,1,0)$ even in large volumes (${\rm L} \lesssim
20~\rm{fm}$). 
This feature is understood by noting that 
while  the leading correction to $N_{D/S}^{{\bf d};M_J,M_S}$ 
is $\sim \eta~e^{-\kappa {\rm L}}$ for systems with ${\bf d}=(0,0,0)$ and
$(1,1,1)$, 
they are $\sim e^{-\kappa {\rm L}}$ for systems
with ${\bf d}=(0,0,1)$ and $(1,1,0)$. 
The periodic images
of the wavefunction with 
the latter boosts are quadrupole distributed, and consequently modify the
$l=2$ component of the 
wavefunction by contributions that are not suppressed by $\eta$.
However, for these systems, there are two irreps that receive similar FV
corrections to their ratios,
which can be largely removed
by forming differences, 
e.g. for the system with ${\bf d}=(0,0,1)$,
\begin{eqnarray}
\overline{N}_{D/S}^{(0,0,1)}
& = & 
{1\over 3}\left(
N_{D/S}^{(0,0,1);1,1}
\ -\ 
N_{D/S}^{(0,0,1);0,0}
 \right)
\ =\ 
{1\over 3}\left(
N_{D/S}^{(0,0,1);\mathbb{E}}
\ -\ 
N_{D/S}^{(0,0,1);\mathbb{A}_2}
 \right)
\ \ \ ,
\label{eq:A2Ediff}
\end{eqnarray}
as shown in 
Fig.~\ref{fig:NDS}.  A similar improvement is found for systems with 
${\bf d}=(1,1,0)$.
It is also worth noting that the contributions to the wavefunction from higher
partial waves, $l\geq2$, can be added to $\psi^{V,{\bf d}}$ with 
coefficients that depend on their corresponding phase shifts, and therefore are
small  under the assumption of low-energy scattering \cite{Luscher:1990ux,
Rummukainen:1995vs, Ishizuka:2009bx}. 
However, the partial-wave decomposition of the FV wavefunction in Eq.~(\ref{psi-V})
contains contributions with $l\geq2$. 
In the limit where the corresponding phase shifts vanish, the wavefunction, in contrast to the spectrum,  
remains  sensitive to these contributions,
resulting in  the larger FV modifications  of
quantities   compared with their spectral analogues.

An extraction of $\eta$ is possible by taking sufficiently large volumes such that a large NN
separation can be 
achieved without approaching the boundaries of the volume. 
While 
the wavefunctions corresponding to the deuteron at rest or with
$\mathbf{d}=(1,1,1)$ 
provide an opportunity to extract $\eta$
with an accuracy of $\sim 15-20\%$ in volumes of  ${\rm L}\sim 14~\rm{fm}$,
combinations of the ratios obtained from the two irreps 
in both the systems with $\mathbf{d}=(0,0,1)$ and $(1,1,0)$
will provide for a $\sim 10\%$ determination  in volumes of  ${\rm L}\sim 12~\rm{fm}$, 
as shown in  Fig. \ref{fig:NDS}.
As it is possible that the uncertainties in the
extraction of $\eta$ can be systematically reduced,
those due to the neglect of the
$J=1$ $\beta$-wave and $J=2,3$ D-wave
phase shifts, as well as higher order terms in the ERE, 
deserve further investigation.

\section{Summary and Conclusion 
\label{sec:conclusion}}
\noindent
A Lattice QCD calculation of the deuteron and its properties would be a
theoretical milestone on the path toward  calculating quantities of importance in low-energy nuclear physics 
from quantum chromodynamics without uncontrolled approximations or assumptions.
While there is no formal impediment to calculating the deuteron binding energy
to arbitrary precision when sufficient computational resources become
available, determining its  properties and interactions presents a
challenge that has largely remained unexplored~\cite{Detmold:2004qn,Meyer:2012wk,Briceno:2013lba}.
As LQCD calculations are performed in a finite Euclidean spacetime volume with
certain boundary conditions imposed upon the fields, 
calculating the properties of the deuteron requires establishing  a rigorous 
connection between FV correlation functions  and the S-matrix. 
Using the NN formalism developed in Ref.~\cite{Briceno:2013lba}, 
we have explored  the 
FV energy 
spectra of states that
have an overlap with the $\siii$-$\diii$ coupled-channels system in which
 the deuteron resides.
Although the full FV QCs associated with the 
$^3S_1$-$^3D_1$ coupled channels depend on interactions in all 
positive-parity isoscalar channels, 
a low-energy expansion
depends only on four scattering
phases and one mixing angle.
Further, for the deuteron, these truncated QCs can be further simplified to
depend only upon one phase shift and the mixing angle,
with corrections suppressed by 
$\sim\frac{1}{\rm L}e^{-2\kappa {\rm L}} \tan{\delta_i}$
where $\delta_i$ denotes the 
$J=1$ $\beta$-wave and $J=2,3$ D-wave
phase shifts
which are all small at the deuteron binding energy.
We have demonstrated that the infinite-volume deuteron binding energy and
leading scattering parameters,
including the mixing angle, $\epsilon_1$, 
that dictate the low-energy behavior of the scattering amplitudes,
can  be (in principle) determined with precision
from the bound-state spectra of deuterons, both at rest and in motion,
in a single modest volume, with ${\rm L}=10$-$14~\rm{fm}$.
Calculations in a second lattice volume would reduce
the systematic uncertainties introduced by truncating the QCs.

We have investigated the feasibility of extracting $\epsilon_1$ from the
asymptotic D/S ratio of the deuteron FV wavefunction using the
periodic images associated with the $\alpha$-wavefunction. 
As the amplitude of the $J=1$ $\beta$-wave and the $J=2,3$ D-wave components of
the wavefunction are not constrained by the infinite-volume deuteron wavefunction, 
the analysis is limited by an imposed truncation of the ERE, 
which is at the same level of approximation as the approximate QCs. 
The systematic uncertainties introduced by this
truncation 
are currently unknown, but will be 
suppressed by the small phase shifts in those channels in addition to being 
exponentially suppressed with ${\rm L}$.
This is in contrast to the extraction from the FV spectra where the
systematic uncertainties have been determined to be small. 
With this approximation, it is estimated  that 
volumes with ${\rm L} \gsim 12~\rm{fm}$ are required to extract $\epsilon_1$  
with $\sim 10\%$ level of accuracy from the asymptotic form of the
wavefunctions.


\noindent
\subsection*{Acknowledgment}
RB, ZD and MJS were supported in part by the DOE grant No. DE-FG02-97ER41014.
ZD and MJS were also supported in part by DOE grant No. DE-FG02-00ER41132.
The work of TL was performed under the auspices of the U.S. Department of
Energy by Lawrence Livermore National Laboratory under Contract DE-AC52-07NA27344. 

\bibliography{bibi}

\newpage

\appendix

\section{Quantization Conditions \label{app: QC}}
\noindent
The NN FV QCs in the positive-parity isoscalar channels
that have an overlap with  $\siii$-$\diii$ coupled channels
are listed in this appendix for a number of CM boosts. 
With the notation introduced in Ref.~\cite{Briceno:2013lba},
the QC for the irrep $\Gamma_i$ can be written as
\begin{eqnarray}
\det\left({\mathbb{M}}^{(\Gamma_i)}+\frac{iMk^{*}}{4\pi
  }-\mathcal{F}^{(\Gamma_i),{\textbf{d}}}\right)=0
\ \ \ ,
\label{QC-simplified}
\end{eqnarray} 
where
\begin{eqnarray}
\mathcal{F}^{(\Gamma_i),{\textbf{d}}}(k^{*2}; {\rm L} )
& = &
{M}\sum_{l,m}\frac{1}{k^{*l}}~{\mathbb{F}
}_{lm}^{(\Gamma_i)}~{c_{lm}^{\textbf{d}}(k^{*2};{\rm L})}
\ \ \ ,
\nonumber\\
{\mathbb{M}}^{(\Gamma_i)}
& = & \left( \mathcal{M}^{-1}\right)_{\Gamma_i}
\ \ \ ,
\label{def-F}
\end{eqnarray}
where 
the functions ${c_{lm}^{\textbf{d}}(k^{*2};{\rm L})}$ are defined in
Eq.~(\ref{clm}), $M$ is the nucleon mass and $k^*$ is the on-shell momentum of each nucleon in the CM frame.
It is straightforward to decompose $\mathcal{M}^{-1}$ into 
$\left( \mathcal{M}^{-1}\right)_{\Gamma_i}$
using the eigenvectors of the FV functions.
The matrices $\mathbb{F}$ and $\mathbb{M}$ are given in the following subsections.

For notational convenience,
$\mathcal{M}_{1,S(D)}$ denotes the scattering amplitude in the channel with total
angular momentum $J=1$ 
and orbital angular momentum $L=0~(L=2)$, 
$\mathcal{M}_{1,SD}$ is the amplitude between $S$ and $D$ partial
waves in the $J=1$ channel, 
and $\text{det} \mathcal{M}_1$ is the determinant of the $J=1$ sector of the scattering-amplitude matrix,
\begin{eqnarray}
\det\mathcal{M}_{1}=\det \left( \begin{array}{cc}
\mathcal{M}_{1,S}&\mathcal{M}_{1,SD}\\
\mathcal{M}_{1,DS}&\mathcal{M}_{1,D}\\
\end{array} \right)
\ \ \ .
\end{eqnarray}
%

\subsubsection{$\mathbf{d}=(0,0,0)$}
\begin{align}
& \mathbb{T}_1: \hspace{.5cm}
\mathbb{F}_{00}^{(\mathbb{T}_1)}=\textbf{I}_{3},\hspace{.5cm}
\mathbb{F}_{40}^{(\mathbb{T}_1)}=
\left(
\begin{array}{ccc}
 0 & 0 & 0 \\
 0 & 0 & \frac{2 \sqrt{6}}{7} \\
 0 & \frac{2 \sqrt{6}}{7} & \frac{2}{7} \\
\end{array}
\right),\hspace{.5cm}
{\mathbb{M}}^{(\mathbb{T}_1)}=\left(
\begin{array}{ccc}
 \frac{\mathcal{M}_{1,D}}{{\det\mathcal{M}_{1}}} & -\frac{\mathcal{M}_{1,SD}}{\det\mathcal{M}_{1}} & 0 \\
 -\frac{\mathcal{M}_{1,SD}}{\det\mathcal{M}_{1}} & \frac{\mathcal{M}_{1,S}}{\det\mathcal{M}_{1}} & 0 \\
 0 & 0 &\mathcal{M}_{3,D}^{-1} \\
\end{array}
\right). \label{I000T1}
\end{align}
%

\subsubsection{$\mathbf{d}=(0,0,1)$}
\begin{align}
& \mathbb{A}_2:\hspace{.5cm}
\mathbb{F}_{00}^{(\mathbb{A}_2)}=\textbf{I}_{3}
\ \ ,\hspace{.5cm}
\mathbb{F}_{20}^{(\mathbb{A}_2)}=
\left(
\begin{array}{ccc}
 \frac{2}{\sqrt{5}} & 0 & -\frac{9}{7 \sqrt{5}} \\
 0 & -\frac{1}{\sqrt{5}} & \frac{6 }{7}\sqrt{\frac{2}{5}} \\
 -\frac{9}{7 \sqrt{5}} & \frac{6 }{7}\sqrt{\frac{2}{5}} & \frac{8}{7 \sqrt{5}} \\
\end{array}
\right),\hspace{.5cm}
\mathbb{F}_{40}^{(\mathbb{A}_2)}=\left(
\begin{array}{ccc}
 0 & 0 & -\frac{4}{7} \\
 0 & 0 & -\frac{2 \sqrt{2}}{7}  \\
 -\frac{4}{7} & -\frac{2 \sqrt{2}}{7} & \frac{2}{7} \\
\end{array}
\right),
\nonumber\\
&\hspace{1cm}
{\mathbb{M}}^{(\mathbb{A}_2)}= \left(
\begin{array}{ccc}
 \frac{2 \mathcal{M}_{1,S}+2 \sqrt{2}\mathcal{M}_{1,SD}+\mathcal{M}_{1,D}}{3
   \det\mathcal{M}_{1}} 
& \frac{\sqrt{2} \mathcal{M}_{1,S}-\mathcal{M}_{1,SD}-\sqrt{2}\mathcal{M}_{1,D}}{3 \det\mathcal{M}_{1}} & 0 \\
 \frac{\sqrt{2}
   \mathcal{M}_{1,S}-\mathcal{M}_{1,SD}-\sqrt{2}\mathcal{M}_{1,D}}{3
   \det\mathcal{M}_{1}} 
& \frac{\mathcal{M}_{1,S}-2 \sqrt{2}\mathcal{M}_{1,SD}+2\mathcal{M}_{1,D}}{3 \det\mathcal{M}_{1}} & 0 \\
 0 & 0 & \mathcal{M}_{3,D}^{-1}  \\
\end{array}
\right).
\label{I001A2}
\end{align}
\begin{align}
& E:\hspace{.5cm}\mathbb{F}_{00}^{(\mathbb{E})}=\textbf{I}_{5}
\ \ \ ,\hspace{.5cm}
\mathbb{F}_{20}^{(\mathbb{E})}=
\left(
\begin{array}{ccccc}
 \frac{1}{2 \sqrt{5}} & 0 & -\frac{\sqrt{3}}{2} & 0 & \frac{4 \sqrt{\frac{3}{5}}}{7} \\
 0 & -\frac{1}{\sqrt{5}} & 0 & 0 & -\frac{3}{7}   \sqrt{\frac{6}{5}} \\
 -\frac{\sqrt{3}}{2} & 0 & \frac{\sqrt{5}}{14} & 0 & -\frac{2}{7} \\
 0 & 0 & 0 & -\frac{2 }{7} \sqrt{5}  & 0 \\
 \frac{4 \sqrt{\frac{3}{5}}}{7} & -\frac{3}{7}   \sqrt{\frac{6}{5}} & -\frac{2}{7} & 0 & \frac{6}{7 \sqrt{5}} \\
\end{array}
\right),
\nonumber\\
&\hspace{1cm}
\mathbb{F}_{40}^{(\mathbb{E})}=\left(
\begin{array}{ccccc}
 0 & 0 & 0 & 0 & \frac{\sqrt{3}}{7} \\
 0 & 0 & 0 & 0 & \frac{\sqrt{6}}{7} \\
 0 & 0 & \frac{8}{21} & 0 & -\frac{5 \sqrt{5}}{21} \\
 0 & 0 & 0 & \frac{1}{7} & 0 \\
 \frac{\sqrt{3}}{7} & \frac{\sqrt{6}}{7} & -\frac{5 \sqrt{5}}{21} & 0 & \frac{1}{21}
\end{array}
\right),
\hspace{.5cm}
\mathbb{F}_{44}^{(\mathbb{E})}=
\left(
\begin{array}{ccccc}
 0 & 0 & 0 & \sqrt{\frac{2}{7}} & 0 \\
 0 & 0 & 0 & \frac{2}{\sqrt{7}} & 0 \\
 0 & 0 & 0 & \sqrt{\frac{10}{21}} & 0 \\
 \sqrt{\frac{2}{7}} & \frac{2}{\sqrt{7}} & \sqrt{\frac{10}{21}} & 0 & \sqrt{\frac{2}{21}} \\
 0 & 0 & 0 & \sqrt{\frac{2}{21}} & 0
\end{array}
\right),
\nonumber\\
&\hspace{1cm}{\mathbb{M}}^{(\mathbb{E})}=
 \left(
\begin{array}{ccccc}
 \frac{\mathcal{M}_{1,S}-2 \sqrt{2}\mathcal{M}_{1,SD}+2\mathcal{M}_{1,D}}{3
   \det\mathcal{M}_{1}} 
& \frac{\sqrt{2} \mathcal{M}_{1,S}-\mathcal{M}_{1,SD}-\sqrt{2}\mathcal{M}_{1,D}}{3 \det\mathcal{M}_{1}} & 0 & 0 & 0 \\
 \frac{\sqrt{2}
   \mathcal{M}_{1,S}-\mathcal{M}_{1,SD}-\sqrt{2}\mathcal{M}_{1,D}}{3
   \det\mathcal{M}_{1}} 
& \frac{2 \mathcal{M}_{1,S}+2 \sqrt{2}\mathcal{M}_{1,SD}+\mathcal{M}_{1,D}}{3 \det\mathcal{M}_{1}} & 0 & 0 & 0 \\
 0 & 0 & \mathcal{M}_{2,D}^{-1} & 0 & 0 \\
 0 & 0 & 0 & \mathcal{M}_{3,D}^{-1} & 0 \\
 0 & 0 & 0 & 0 & \mathcal{M}_{3,D}^{-1} \\
\end{array}
\right).
\label{I001E}
\end{align}
%

\subsubsection{$\mathbf{d}=(1,1,0)$}
\begin{align}
& \mathbb{B}_1: \hspace{.5cm}
\mathbb{F}_{00}^{(\mathbb{B}_1)}=\textbf{I}_{5}\ \ \ ,\hspace{.5cm}
\mathbb{F}_{20}^{(\mathbb{B}_1)}=\left(
\begin{array}{ccccc}
 \frac{2}{\sqrt{5}} & 0 & 0 & -\frac{9}{7 \sqrt{5}} & 0 \\
 0 & -\frac{1}{\sqrt{5}} & 0 & \frac{6}{7}\sqrt{\frac{2}{5}} & 0 \\
 0 & 0 & -\frac{\sqrt{5}}{7} & 0 & -\frac{\sqrt{10}}{7} \\
 -\frac{9}{7 \sqrt{5}} & \frac{6}{7}\sqrt{\frac{2}{5}} & 0 & \frac{8}{7 \sqrt{5}} & 0 \\
 0 & 0 & -\frac{\sqrt{10}}{7} & 0 & 0 \\
\end{array}
\right),
\nonumber\\
&\hspace{1cm}\mathbb{F}_{40}^{(\mathbb{B}_1)}=\left(
\begin{array}{ccccc}
 0 & 0 & 0 & -\frac{4}{7} & 0 \\
 0 & 0 & 0 & -\frac{2 \sqrt{2}}{7}   & 0 \\
 0 & 0 & -\frac{2}{21} & 0 & \frac{5 \sqrt{2}}{21} \\
 -\frac{4}{7} & -\frac{2 \sqrt{2}}{7}   & 0 & \frac{2}{7} & 0 \\
 0 & 0 & \frac{5 \sqrt{2}}{21} & 0 & -\frac{1}{3} \\
\end{array}
\right), \hspace{.5cm} \mathbb{F}_{44}^{(\mathbb{B}_1)}=\left(
\begin{array}{ccccc}
 0 & 0 & 0 & 0 & 0 \\
 0 & 0 & 0 & 0 & 0 \\
 0 & 0 & -\frac{2}{3}\sqrt{\frac{10}{7}} & 0 & -\frac{2}{3}  \sqrt{\frac{5}{7}} \\
 0 & 0 & 0 & 0 & 0 \\
 0 & 0 & -\frac{2}{3}  \sqrt{\frac{5}{7}} & 0 & -\frac{\sqrt{\frac{10}{7}}}{3} \\
\end{array}
\right),
\nonumber\\
&\hspace{1cm}{\mathbb{M}}^{(\mathbb{B}_1)}=
\left(
\begin{array}{ccccc}
 \frac{2 \mathcal{M}_{1,S}+2
   \sqrt{2}\mathcal{M}_{1,SD}+\mathcal{M}_{1,D}}{3\det\mathcal{M}_{1}} 
& \frac{\sqrt{2} \mathcal{M}_{1,S}-\mathcal{M}_{1,SD}-\sqrt{2}\mathcal{M}_{1,D}}{3 \det\mathcal{M}_{1}} & 0 & 0 & 0 \\
 \frac{\sqrt{2}
   \mathcal{M}_{1,S}-\mathcal{M}_{1,SD}-\sqrt{2}\mathcal{M}_{1,D}}{3
   \det\mathcal{M}_{1}} 
& \frac{\mathcal{M}_{1,S}-2 \sqrt{2}\mathcal{M}_{1,SD}+2\mathcal{M}_{1,D}}{3 \det\mathcal{M}_{1}} & 0 & 0 & 0 \\
 0 & 0 & \mathcal{M}_{2,D}^{-1} & 0 & 0 \\
 0 & 0 & 0 & \mathcal{M}_{3,D}^{-1} & 0 \\
 0 & 0 & 0 & 0 & \mathcal{M}_{3,D}^{-1} \\
\end{array}
\right).
\label{I110B1}
\end{align}
\begin{align}
& \mathbb{B}_2:\hspace{.5cm}\mathbb{F}_{00}^{(\mathbb{B}_2)}=\textbf{I}_{5}\ \
\ ,\hspace{.5cm}
\mathbb{F}_{20}^{(\mathbb{B}_2)}=\left(
\begin{array}{ccccc}
 -\frac{1}{\sqrt{5}} & 0 & 0 & 0 & -\frac{3}{7}  \sqrt{\frac{6}{5}} \\
 0 & \frac{1}{2 \sqrt{5}} & -\frac{\sqrt{3}}{2} & 0 & \frac{4 \sqrt{\frac{3}{5}}}{7} \\
 0 & -\frac{\sqrt{3}}{2} & \frac{\sqrt{5}}{14} & 0 & -\frac{2}{7} \\
 0 & 0 & 0 & -\frac{2\sqrt{5}}{7}   & 0 \\
 -\frac{3}{7}  \sqrt{\frac{6}{5}} & \frac{4 \sqrt{\frac{3}{5}}}{7} & -\frac{2}{7} & 0 & \frac{6}{7 \sqrt{5}} \\
\end{array}
\right),
\nonumber\\
&\hspace{1cm}
\mathbb{F}_{40}^{(\mathbb{B}_2)}=\left(
\begin{array}{ccccc}
 0 & 0 & 0 & 0 & \frac{\sqrt{6}}{7} \\
 0 & 0 & 0 & 0 & \frac{\sqrt{3}}{7} \\
 0 & 0 & \frac{8}{21} & 0 & -\frac{5 \sqrt{5}}{21}  \\
 0 & 0 & 0 & \frac{1}{7} & 0 \\
 \frac{\sqrt{6}}{7} & \frac{\sqrt{3}}{7} & -\frac{5 \sqrt{5}}{21}  & 0 & \frac{1}{21} \\
\end{array}
\right), \hspace{0.5cm}
\mathbb{F}_{44}^{(\mathbb{B}_2)}=\left(
\begin{array}{ccccc}
 0 & 0 & 0 & \frac{2 i}{\sqrt{7}} & 0 \\
 0 & 0 & 0 & i \sqrt{\frac{2}{7}} & 0 \\
 0 & 0 & 0 & i \sqrt{\frac{10}{21}} & 0 \\
 -\frac{2 i}{\sqrt{7}} & -i \sqrt{\frac{2}{7}} & -i \sqrt{\frac{10}{21}} & 0 & -i \sqrt{\frac{2}{21}} \\
 0 & 0 & 0 & i \sqrt{\frac{2}{21}} & 0 \\
\end{array}
\right),
\nonumber\\
&\hspace{1cm}{\mathbb{M}}^{(\mathbb{B}_2)}=\left(
\begin{array}{ccccc}
 \frac{2 \mathcal{M}_{1,S}+2
   \sqrt{2}\mathcal{M}_{1,SD}+\mathcal{M}_{1,D}}{3\det\mathcal{M}_{1}} 
& \frac{\sqrt{2} \mathcal{M}_{1,S}-\mathcal{M}_{1,SD}-\sqrt{2}\mathcal{M}_{1,D}}{3 \det\mathcal{M}_{1}} & 0 & 0 & 0 \\
 \frac{\sqrt{2}
   \mathcal{M}_{1,S}-\mathcal{M}_{1,SD}-\sqrt{2}\mathcal{M}_{1,D}}{3
   \det\mathcal{M}_{1}} 
& \frac{\mathcal{M}_{1,S}-2 \sqrt{2}\mathcal{M}_{1,SD}+2\mathcal{M}_{1,D}}{3 \det\mathcal{M}_{1}} & 0 & 0 & 0 \\
 0 & 0 & \mathcal{M}_{2,D}^{-1} & 0 & 0 \\
 0 & 0 & 0 & \mathcal{M}_{3,D}^{-1} & 0 \\
 0 & 0 & 0 & 0 & \mathcal{M}_{3,D}^{-1} \\
\end{array}
\right).
\label{I110B2}
\end{align}
\begin{align}
&\mathbb{B}_3:\hspace{.5cm}
\mathbb{F}_{00}^{(\mathbb{B}_3)}=\textbf{I}_{5}\ \ \ ,\hspace{.5cm}
\mathbb{F}_{20}^{(\mathbb{B}_3)}=\left(
\begin{array}{ccccc}
 \frac{1}{2 \sqrt{5}} & 0 & -\frac{\sqrt{3}}{2} & \frac{4 \sqrt{\frac{3}{5}}}{7} & 0 \\
 0 & -\frac{1}{\sqrt{5}} & 0 & -\frac{3}{7} \sqrt{\frac{6}{5}} & 0 \\
 -\frac{\sqrt{3}}{2} & 0 & \frac{\sqrt{5}}{14} & -\frac{2}{7} & 0 \\
 \frac{4 \sqrt{\frac{3}{5}}}{7} & -\frac{3}{7} \sqrt{\frac{6}{5}} & -\frac{2}{7} & \frac{6}{7 \sqrt{5}} & 0 \\
 0 & 0 & 0 & 0 & -\frac{2 \sqrt{5}}{7}  \\
\end{array}
\right),
\nonumber\\
&\hspace{1cm}
\mathbb{F}_{40}^{(\mathbb{B}_3)}=\left(
\begin{array}{ccccc}
 0 & 0 & 0 & \frac{\sqrt{3}}{7} & 0 \\
 0 & 0 & 0 & \frac{\sqrt{6}}{7} & 0 \\
 0 & 0 & \frac{8}{21} & -\frac{5 \sqrt{5}}{21}  & 0 \\
 \frac{\sqrt{3}}{7} & \frac{\sqrt{6}}{7} & -\frac{5 \sqrt{5}}{21}  & \frac{1}{21} & 0 \\
 0 & 0 & 0 & 0 & \frac{1}{7} \\
\end{array}
\right), \hspace{0.5cm}
\mathbb{F}_{44}^{(\mathbb{B}_3)}=\left(
\begin{array}{ccccc}
 0 & 0 & 0 & 0 & -i \sqrt{\frac{2}{7}} \\
 0 & 0 & 0 & 0 & -\frac{2 i}{\sqrt{7}} \\
 0 & 0 & 0 & 0 & -i \sqrt{\frac{10}{21}} \\
 0 & 0 & 0 & 0 & -i \sqrt{\frac{2}{21}} \\
 i \sqrt{\frac{2}{7}} & \frac{2 i}{\sqrt{7}} & i \sqrt{\frac{10}{21}} & i \sqrt{\frac{2}{21}} & 0 \\
\end{array}
\right),
\nonumber\\
&\hspace{1cm}{\mathbb{M}}^{(\mathbb{B}_3)}=
\left(
\begin{array}{ccccc}
 \frac{\mathcal{M}_{1,S}-2 \sqrt{2}\mathcal{M}_{1,SD}+2\mathcal{M}_{1,D}}{3
   \det\mathcal{M}_{1}} 
& \frac{\sqrt{2} \mathcal{M}_{1,S}-\mathcal{M}_{1,SD}-\sqrt{2}\mathcal{M}_{1,D}}{3 \det\mathcal{M}_{1}} & 0 & 0 & 0 \\
 \frac{\sqrt{2}
   \mathcal{M}_{1,S}-\mathcal{M}_{1,SD}-\sqrt{2}\mathcal{M}_{1,D}}{3
   \det\mathcal{M}_{1}} 
& \frac{2 \mathcal{M}_{1,S}+2 \sqrt{2}\mathcal{M}_{1,SD}+\mathcal{M}_{1,D}}{3 \det\mathcal{M}_{1}} & 0 & 0 & 0 \\
 0 & 0 & \mathcal{M}_{2,D}^{-1} & 0 & 0 \\
 0 & 0 & 0 & \mathcal{M}_{3,D}^{-1} & 0 \\
 0 & 0 & 0 & 0 & \mathcal{M}_{3,D}^{-1} \\
\end{array}
\right).
\label{I110B3}
\end{align}
%

\subsubsection{$\mathbf{d}=(1,1,1)$}
\begin{align}
& \mathbb{A}_2:\hspace{.4cm}
\mathbb{F}_{00}^{(\mathbb{A}_2)}=\textbf{I}_{4}\ \ \ ,\hspace{.4cm}
\mathbb{F}_{40}^{(\mathbb{A}_2)}=
\left(
\begin{array}{cccc}
 0 & 0 & 0 & 0 \\
 0 & 0 & \frac{2 \sqrt{6}}{7} & 0 \\
 0 & \frac{2 \sqrt{6}}{7} & \frac{2}{7} & 0 \\
 0 & 0 & 0 & -\frac{4}{7}
\end{array}
\right), \hspace{0.4cm}
{\mathbb{M}}^{(\mathbb{A}_2)}=\left(
\begin{array}{cccc}
 \frac{\mathcal{M}_{1,D}}{{\det\mathcal{M}_{1}}} & -\frac{\mathcal{M}_{1,SD}}{\det\mathcal{M}_{1}} & 0 & 0 \\
 -\frac{\mathcal{M}_{1,SD}}{\det\mathcal{M}_{1}} & \frac{\mathcal{M}_{1,S}}{\det\mathcal{M}_{1}} & 0 & 0 \\
 0 & 0 &\mathcal{M}_{3,D}^{-1} & 0 \\
 0 & 0 & 0 & \mathcal{M}_{3,D}^{-1} \\
\end{array}
\right).
\label{I111A2}
\\
& \mathbb{E}:\hspace{.5cm}\mathbb{F}_{00}^{(\mathbb{E})}=\textbf{I}_{6}\ \ \ ,\hspace{.5cm}
\mathbb{F}_{40}^{(\mathbb{E})}=
\left(
\begin{array}{cccccc}
 0 & 0 & 0 & 0 & 0 & 0 \\
 0 & 0 & \frac{2 \sqrt{6}}{7} & 0 & 0 & 0 \\
 0 & \frac{2 \sqrt{6}}{7} & \frac{2}{7} & 0 & 0 & 0 \\
 0 & 0 & 0 & \frac{8}{21} & -\frac{10 \sqrt{2}}{21} & 0 \\
 0 & 0 & 0 & -\frac{10 \sqrt{2}}{21} & -\frac{2}{21} & 0 \\
 0 & 0 & 0 & 0 & 0 & -\frac{4}{7}
\end{array}
\right),
\nonumber\\
&\hspace{1cm}{\mathbb{M}}^{(\mathbb{E})}=
 \left(
\begin{array}{cccccc}
  \frac{\mathcal{M}_{1,D}}{{\det\mathcal{M}_{1}}} & -\frac{\mathcal{M}_{1,SD}}{\det\mathcal{M}_{1}} & 0 & 0 & 0 & 0 \\
-\frac{\mathcal{M}_{1,SD}}{\det\mathcal{M}_{1}} & \frac{\mathcal{M}_{1,S}}{\det\mathcal{M}_{1}} & 0 & 0 & 0 & 0 \\
 0 & 0 & \mathcal{M}_{3,D}^{-1} & 0 & 0 & 0 \\
 0 & 0 & 0 & \mathcal{M}_{2,D}^{-1} & 0 & 0 \\
 0 & 0 & 0 & 0 & \mathcal{M}_{3,D}^{-1} & 0 \\
 0 & 0 & 0 & 0 & 0 & \mathcal{M}_{2,D}^{-1} \\
\end{array}
\right).
\label{I111E}
\end{align}

\newpage
%
\section{The Finite-Volume $c^{\mathbf{d}}_{LM}$ Functions \label{app: clm}}
\begin{figure}[h]
\begin{center}  
\subfigure[]{
\includegraphics[scale=0.25]{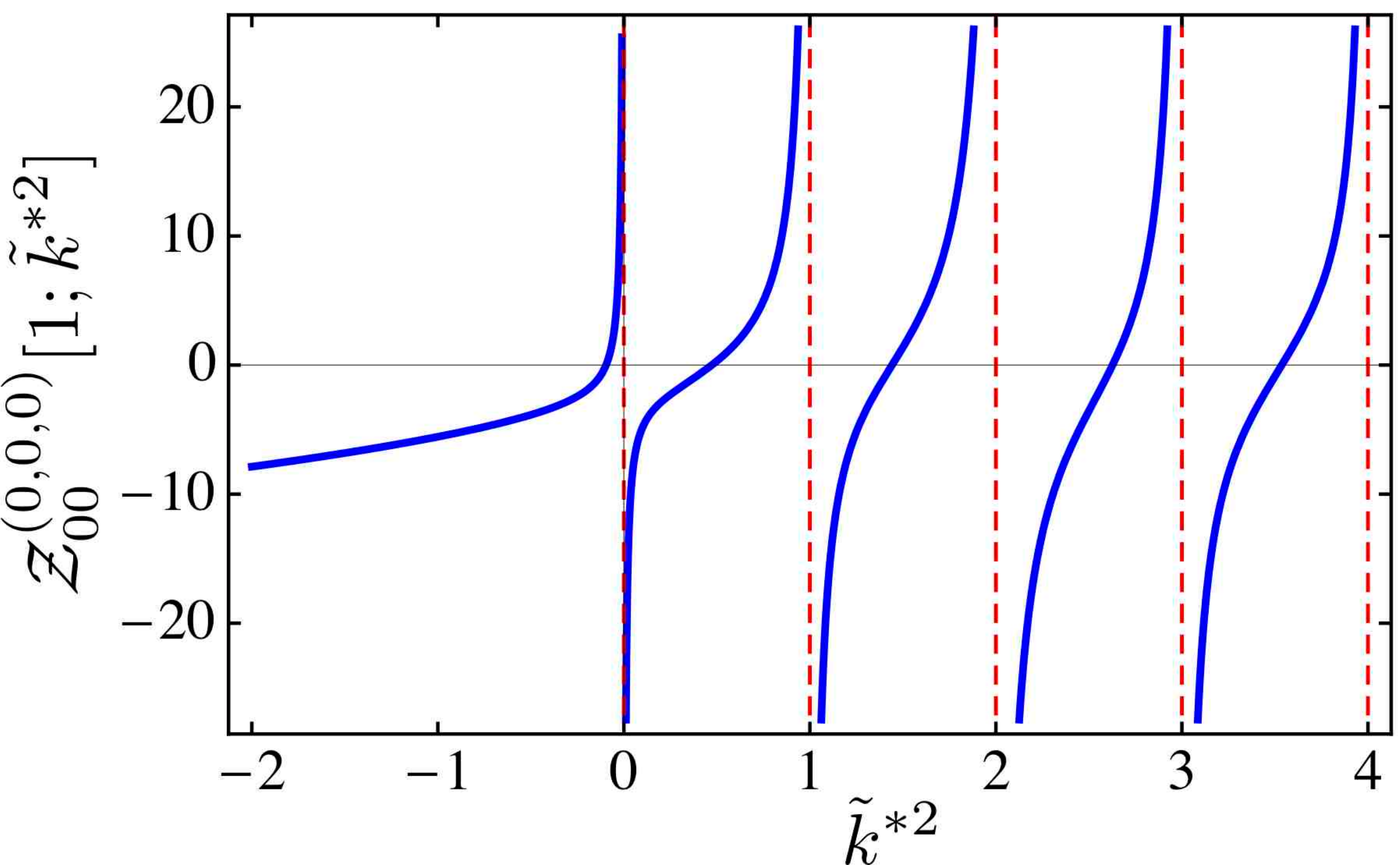}}
\subfigure[]{
\includegraphics[scale=0.25]{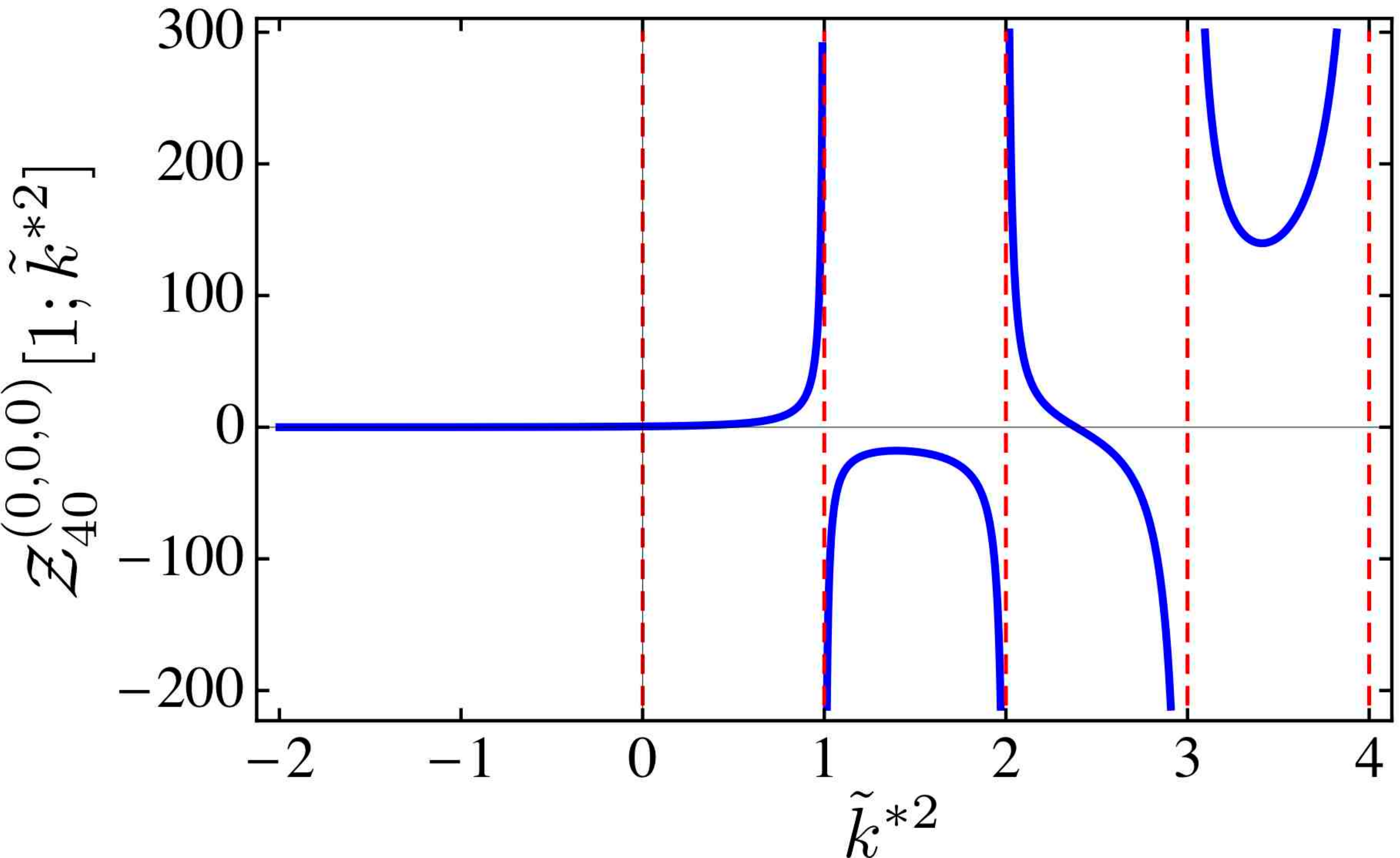}}
\caption{(a) $\mathcal{Z}_{00}^{\mathbf{d}}$ and (b)
  $\mathcal{Z}_{40}^{\mathbf{d}}$ for $\mathbf{d}=(0,0,0)$ as a function of
  $\tilde{k}^{*2}=k^{*2} {\rm L}^2/4\pi^2$.}
\label{fig:Z-func}
\end{center}
\end{figure}

\noindent
The FV NN energy spectra are determined by
the $c_{LM}^{\mathbf{d}}(k^{*2}; {\rm L})$ functions that are defined in
Eq.~(\ref{clm}). 
They are smooth analytic functions of $k^{*2}$ for negative
values of $k^{*2}$, but have poles at $k^{*2}=\frac{4\pi^2}{\rm L^2}(\mathbf{n}-\mathbf{d}/2)^2$,
where $\mathbf{n}$ is an integer triplet, corresponding to the
energy of two non-interacting 
nucleons in a cubic volume with the
PBCs. 
In obtaining  the spectra  in the
positive-parity isoscalar channels from the $\mathbb{T}_1$ irrep of the cubic
group that are shown in 
Fig.~\ref{T1specfull}, 
the $c_{00}^{(0,0,0)}(k^{*2}; {\rm L})
=
\mathcal{Z}_{00}^{(0,0,0)}[1;\tilde{k}^{*2}]/(2\pi^{3/2} {\rm L} )$ 
and $c_{40}^{(0,0,0)}(k^{*2}; {\rm
  L})=\mathcal{Z}_{40}^{(0,0,0)}[1;\tilde{k}^{*2}]/(8\pi^{5/2} {\rm L}^5)$ 
functions 
have been determined. 
The corresponding $\mathcal{Z}$ functions are shown 
in Fig.~\ref{fig:Z-func} as a function of $\tilde{k}^{*2}$,
see Ref.~\cite{Luu:2011ep}.  

When $k^{*2}=-\kappa^2\leq0$, the exponential volume dependence of the
$c_{LM}^{\mathbf{d}}$
can be made explicit 
by performing a Poisson resummation of Eq.~(\ref{clm}),
\begin{align}
& c^\mathbf{d}_{00}(-\kappa^2; {\rm L})=-\frac{\kappa}{4\pi}+\sum_{\textbf{n}\neq\textbf{0}}{e^{i \pi \textbf{n}\cdot\textbf{d}}}
\ \frac{e^{-n\kappa {\rm L}}}{4\pi n {\rm L}}
\ \ \ ,
\label{c00-exp}
\\
& c^\mathbf{d}_{20}(-\kappa^2;{\rm L})=
-\kappa^2\sqrt{4\pi}\sum_{\textbf{n}\neq\textbf{0}}{e^{i \pi
    \textbf{n}\cdot\textbf{d}}}
\ Y_{20}(\hat{\mathbf{n}})\left(1+\frac{3}{n\kappa
    {\rm L}} +\frac{3}{n^2\kappa^{2}{\rm L}^2}\right)
\frac{e^{-n\kappa {\rm L}}}{4\pi n{\rm L}}
\ \ \ ,
\label{c20-exp}
\\
& c^\mathbf{d}_{40(\pm4)}(-\kappa^2;{\rm L})=
\kappa^4\sqrt{4\pi}\sum_{\textbf{n}\neq\textbf{0}}{e^{i \pi
    \textbf{n}\cdot\textbf{d}}}
\ Y_{40(\pm4)}(\hat{\mathbf{n}})\left(1+\frac{10}{n\kappa {\rm L}} 
+\frac{45}{n^2\kappa^{2}{\rm L}^2}+\frac{105}{n^3\kappa^{3}{\rm L}^3}+\frac{105}{n^4\kappa^{4}{\rm L}^4}\right)
\frac{e^{-n\kappa {\rm L}}}{4\pi n{\rm L}}
\ \ \ ,
\label{c40-exp}
\end{align}
where $\mathbf{n}$ is an integer triplet and $n=|{\bf n}|$. 
The expansions of 
$c_{20}^{\mathbf{d}}$ and $c_{40(\pm 4)}^{\mathbf{d}}$ start at
$\sim {1\over{\rm L}}e^{-\kappa {\rm L}}$, 
while $c_{00}^{\mathbf{d}}$ has a leading term that does not vanish in the
infinite-volume limit. 
It is also evident from these relations that $c_{20}^{\mathbf{d}}$ is
non-vanishing only for 
$\mathbf{d}=(0,0,1)$ and $(1,1,0)$, 
which gives rise to the $\mathcal{O}(\sin \epsilon_1)$ contributions to the
corresponding QCs given in Sec. \ref{sec:DeutFV}.

\begin{figure}[!ht]
\begin{center}  
\subfigure[]{
\label{dZ00P000}
\includegraphics[scale=0.260]{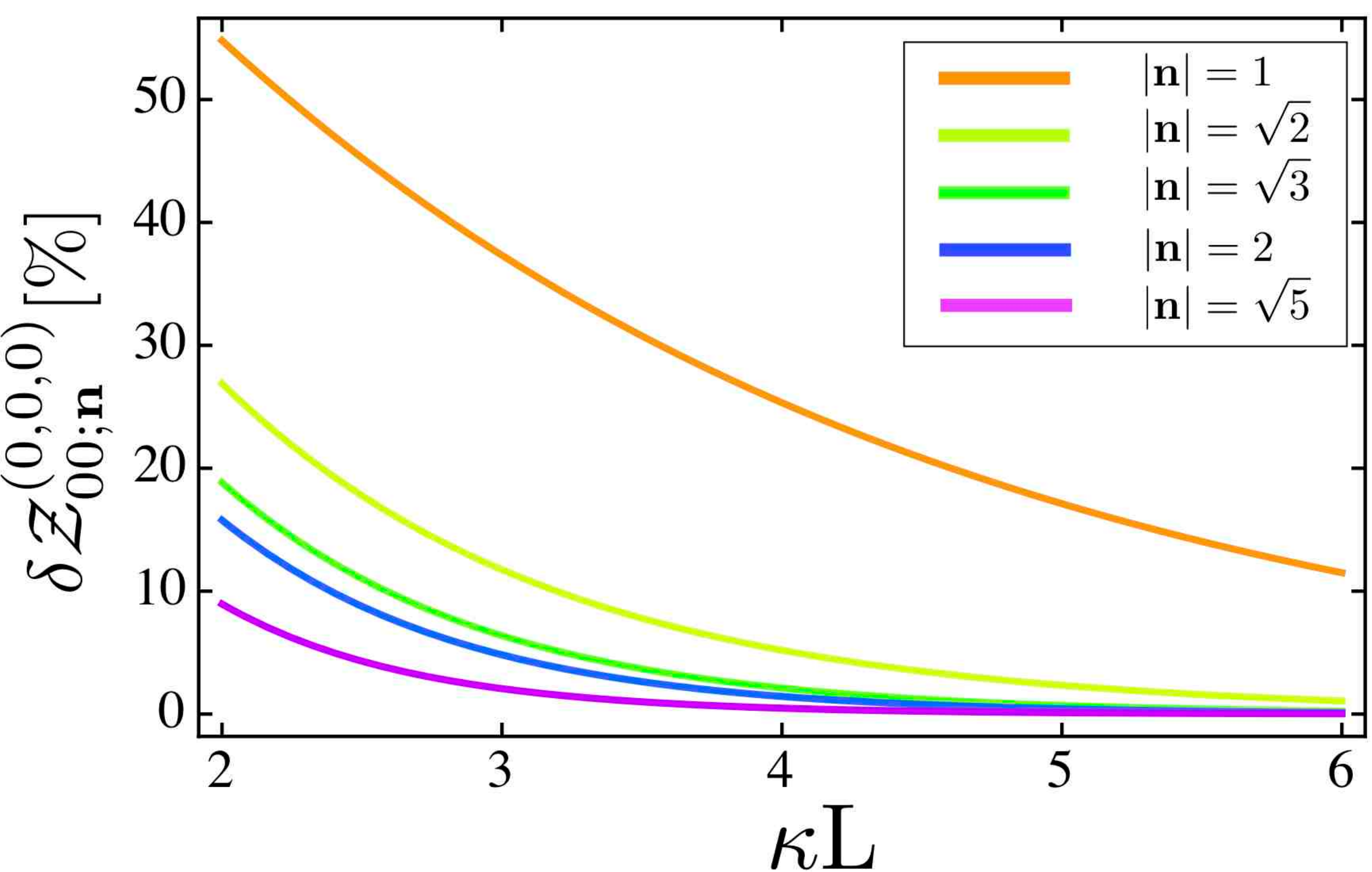}}
\subfigure[]{
\label{dZ00P001}
\includegraphics[scale=0.260]{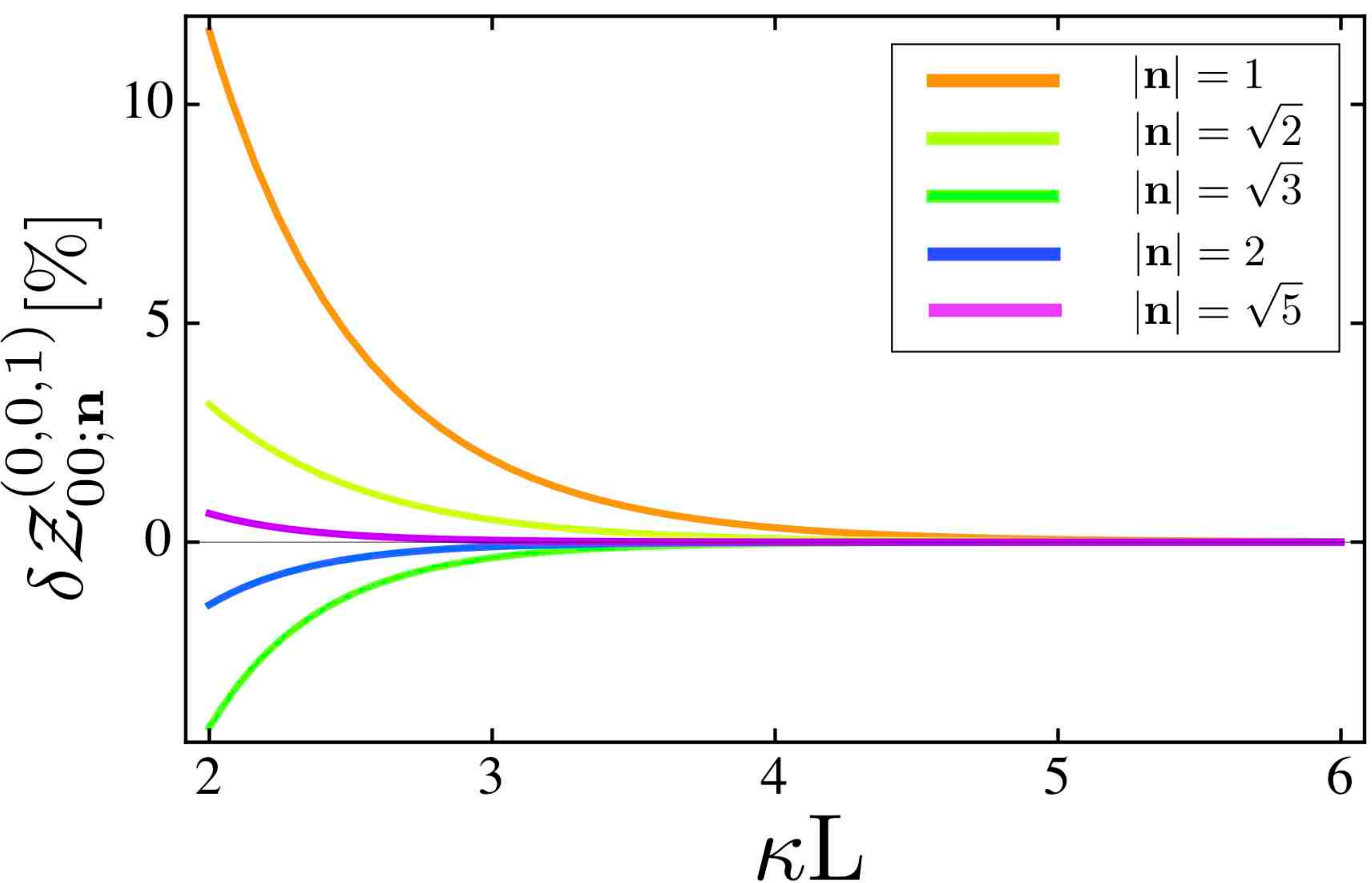}}
\subfigure[]{
\label{dZ00P110}
\includegraphics[scale=0.260]{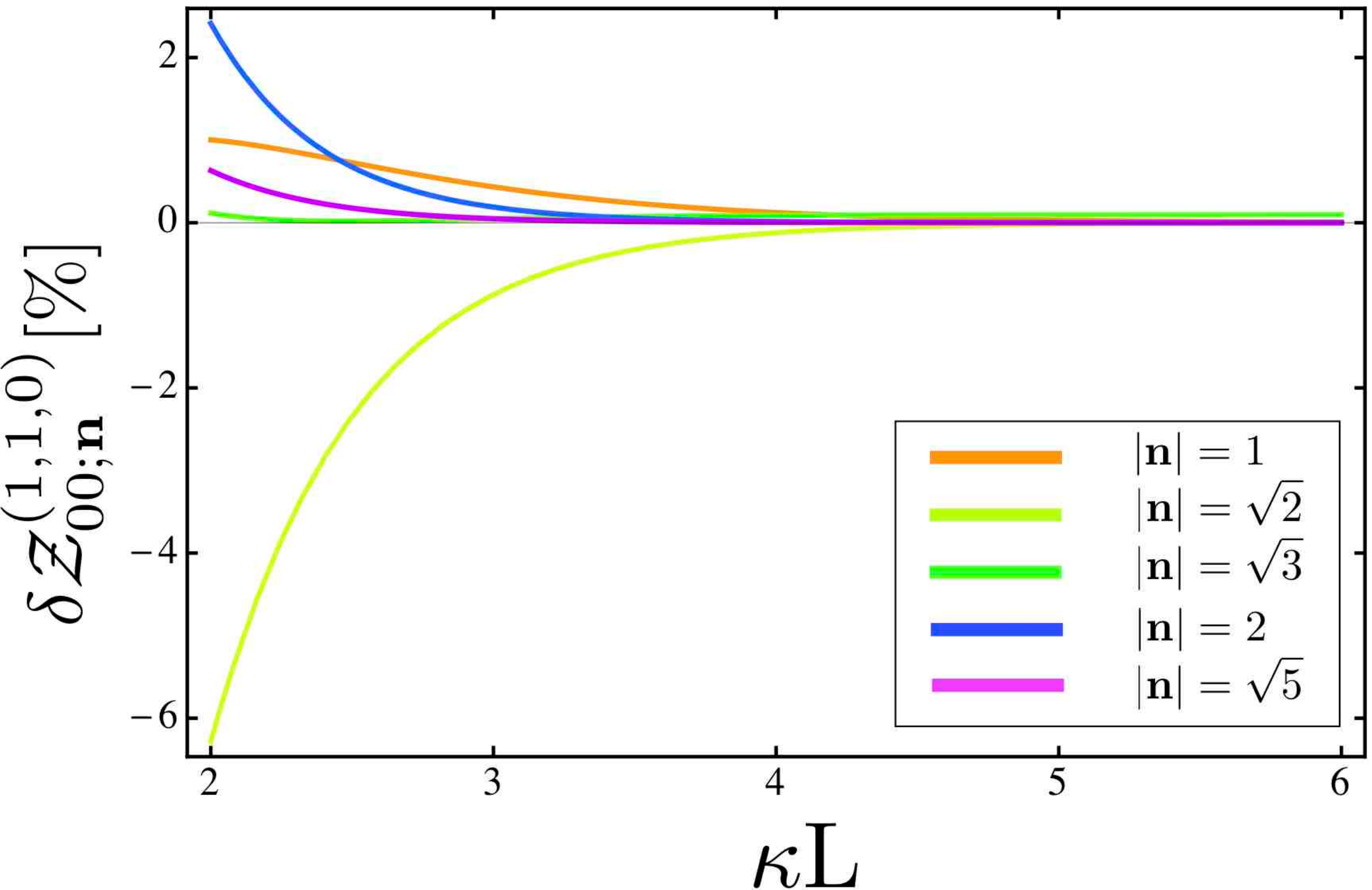}}
\subfigure[]{
\label{dZ00P111}
\includegraphics[scale=0.260]{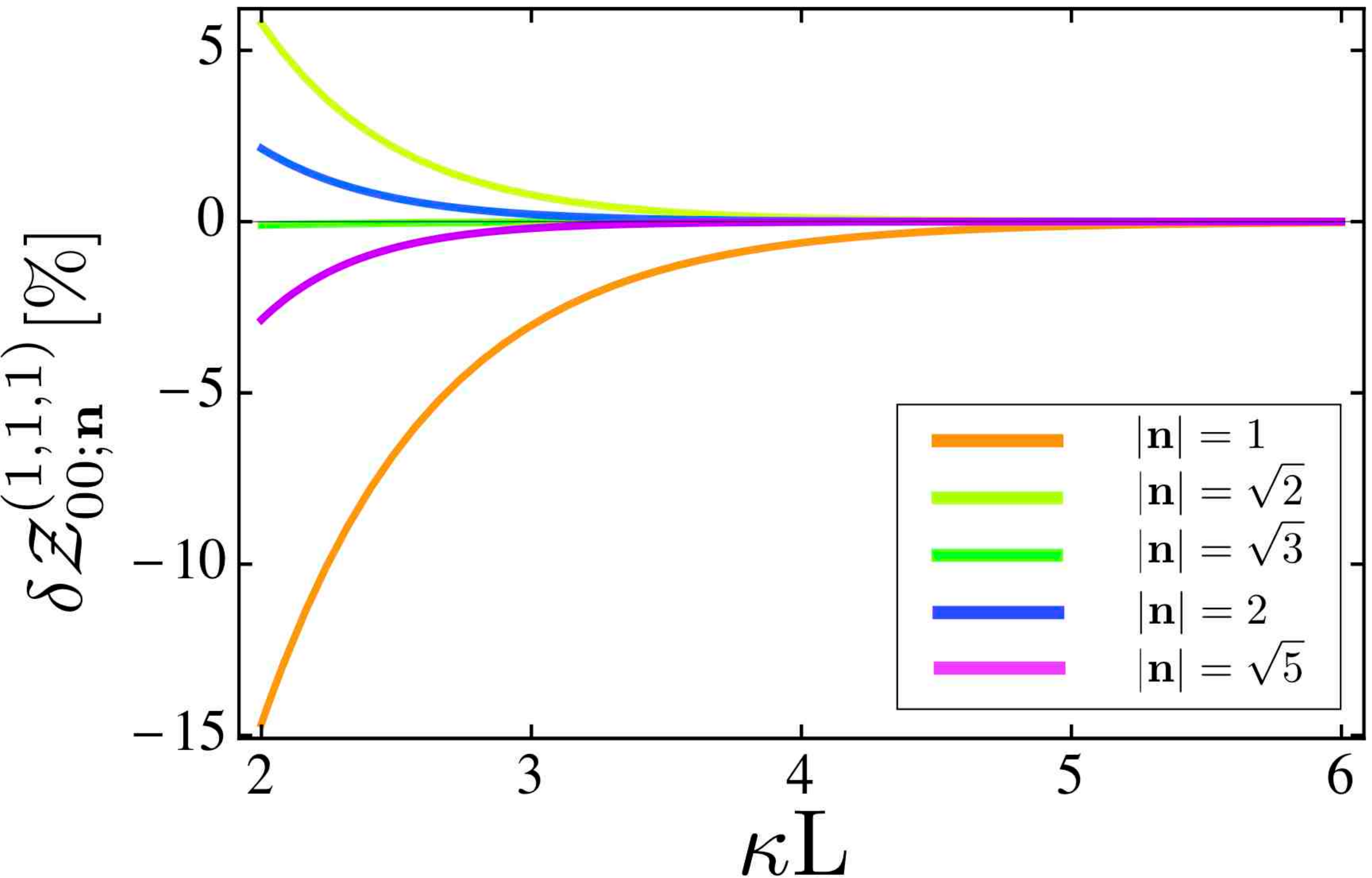}}
\caption{The quantities
  $\delta\mathcal{Z}^{\mathbf{d}}_{00;\mathbf{n}}
\equiv\frac{1}{\mathcal{Z}_{00}^{\mathbf{d}}}(\mathcal{Z}_{00}^{\mathbf{d}}-\mathcal{Z}_{00;\mathbf{n}}^{\mathbf{d}})$
(in percent) as a function of $\kappa {\rm L}$ for different
boosts. $\mathcal{Z}_{00;\mathbf{n}}^{\mathbf{d}}$ 
denotes the value of the $\mathcal{Z}_{00}$-function when the sum in
Eq.~(\ref{c00-exp}) is truncated 
to a maximum shell $\mathbf{n}$.}
\label{fig:dZ}
\end{center}
\end{figure}

Previous works~\cite{Beane:2006mx, Beane:2012vq,  Beane:2013br,
  Yamazaki:2012hi, Borasoy:2006qn, Lee:2008fa, Bour:2012hn, Davoudi:2011md},
have proposed extracting the infinite-volume deuteron binding energy from the
FV spectra using the S-wave QC expanded around
the infinite-volume deuteron pole, $\kappa_d^{\infty}$,
retaining only a 
finite number of terms in the expansion of the $c_{LM}^{\mathbf{d}}$. 
Fig. \ref{fig:dZ} shows the quantity
$\delta\mathcal{Z}^{\mathbf{d}}_{00;\mathbf{n}}
\equiv\frac{1}{\mathcal{Z}_{00}^{\mathbf{d}}}(\mathcal{Z}_{00}^{\mathbf{d}}-\mathcal{Z}_{00;\mathbf{n}}^{\mathbf{d}})$
as a function of $\kappa {\rm L}$ for different boosts. $\mathcal{Z}_{00;\mathbf{n}}^{\mathbf{d}}$ denotes the value of the
$\mathcal{Z}_{00}$-function when the sum in Eq.~(\ref{c00-exp}) is truncated to
a maximum shell $\mathbf{n}$. 
For modest volumes, truncating the $\mathcal{Z}$-functions can lead to large deviations
from the exact values.

\newpage
\ 
\newpage

\section{Finite-Volume Deuteron Wavefunctions
\label{sec:Wavefunc}
}
\noindent
It is useful to visualize how the deuteron is distorted within a FV,
and in this appendix, 
based on the asymptotic FV wavefunction of the deuteron given in Eq.~(\ref{psi-V}), 
we show the mass density  in the $xz$-plane from selected wavefunctions.  
As the interior region is not 
described by the asymptotic form of the wavefunction, 
it is ``masked'' by a shaded disk in the following figures. 
In each figure,
the black straight lines separate adjacent lattice volumes that contain the
periodic images of the wavefunction.

\begin{figure}[!ht]
\begin{center}  
\subfigure[]{
\label{WF-A2-L10}
\includegraphics[scale=0.3]{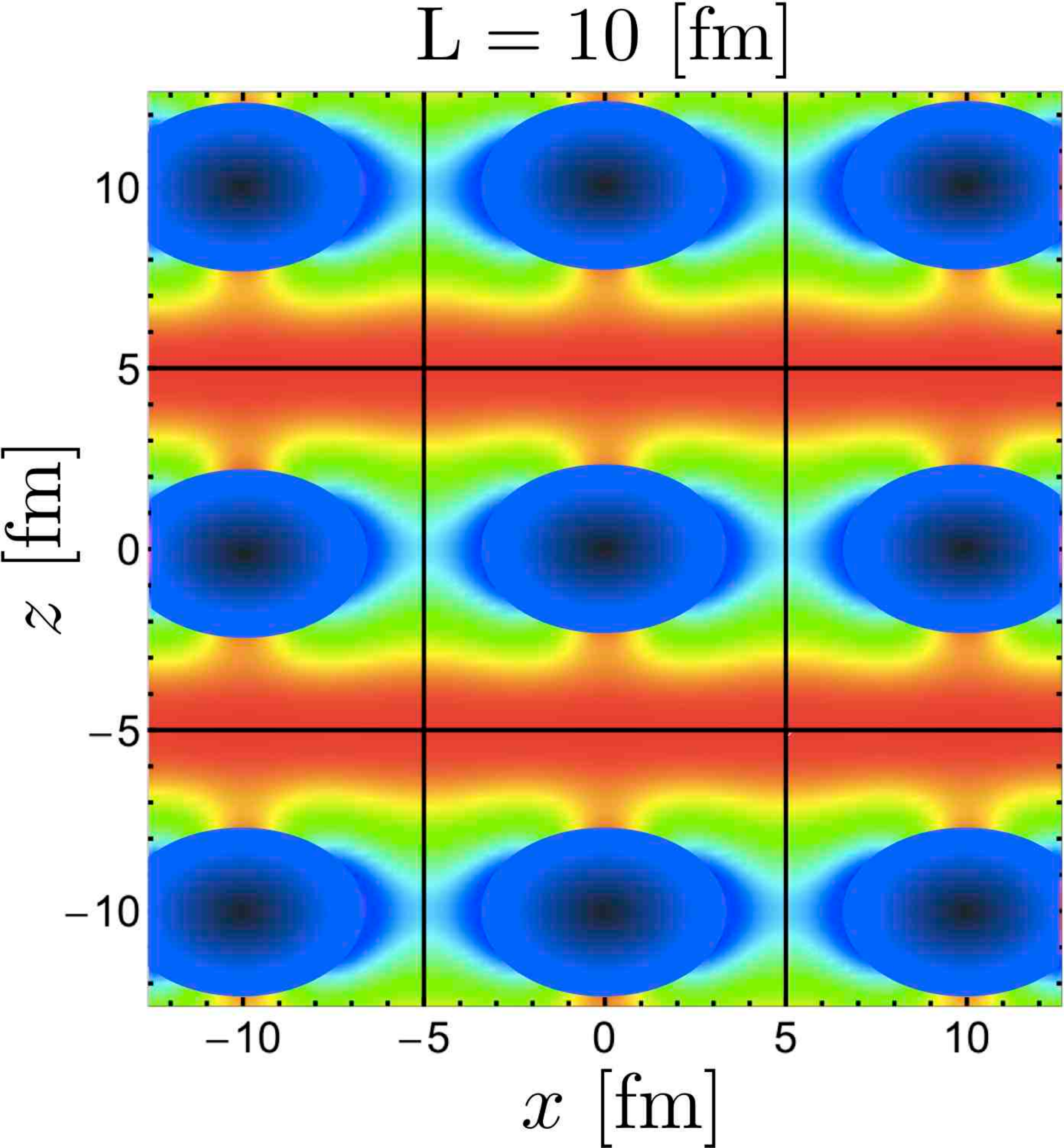}}
\subfigure[]{
\label{WF-A2-L15}
\includegraphics[scale=0.3]{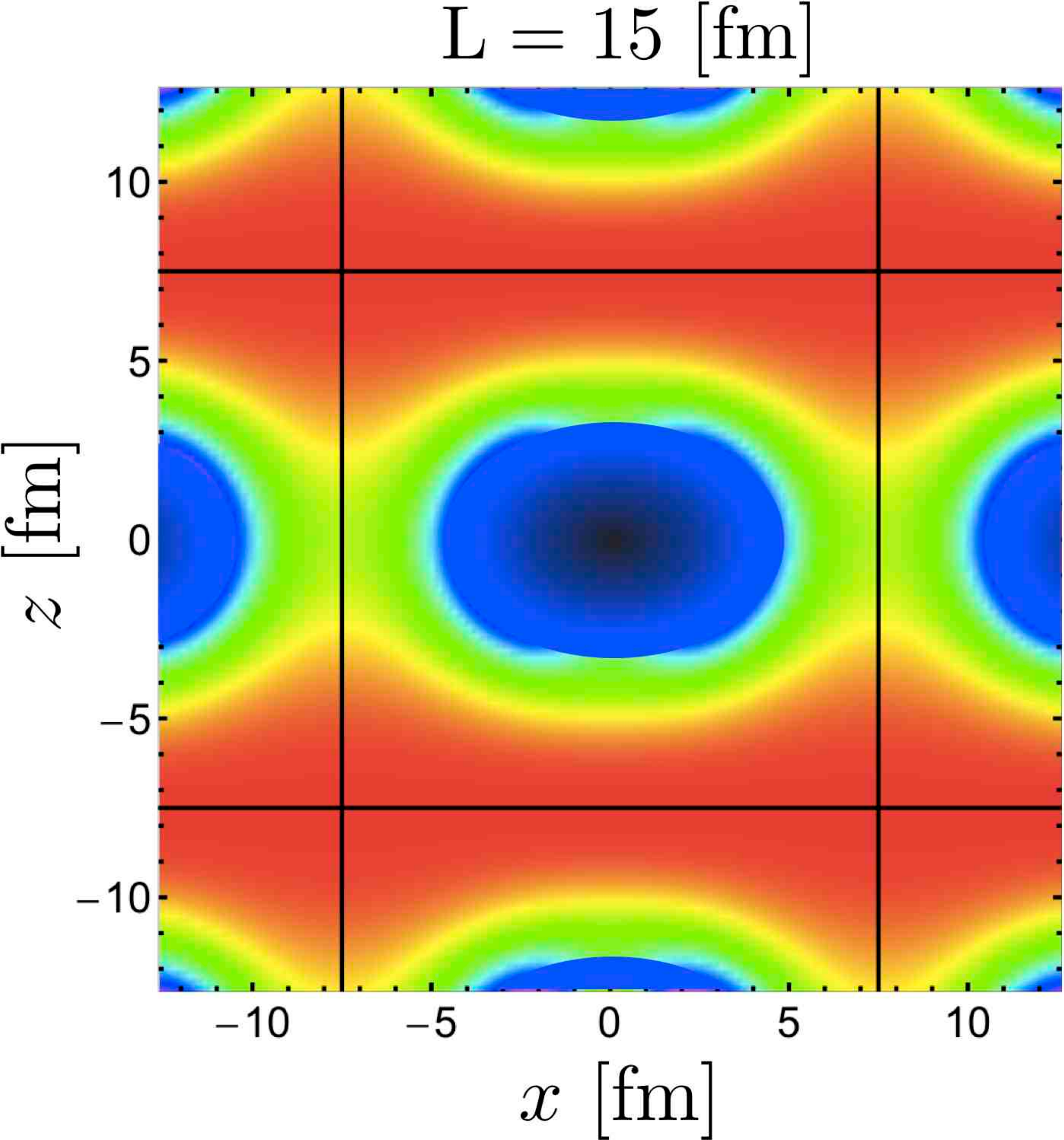}}
\subfigure[]{
\label{WF-A2-20}
\includegraphics[scale=0.3]{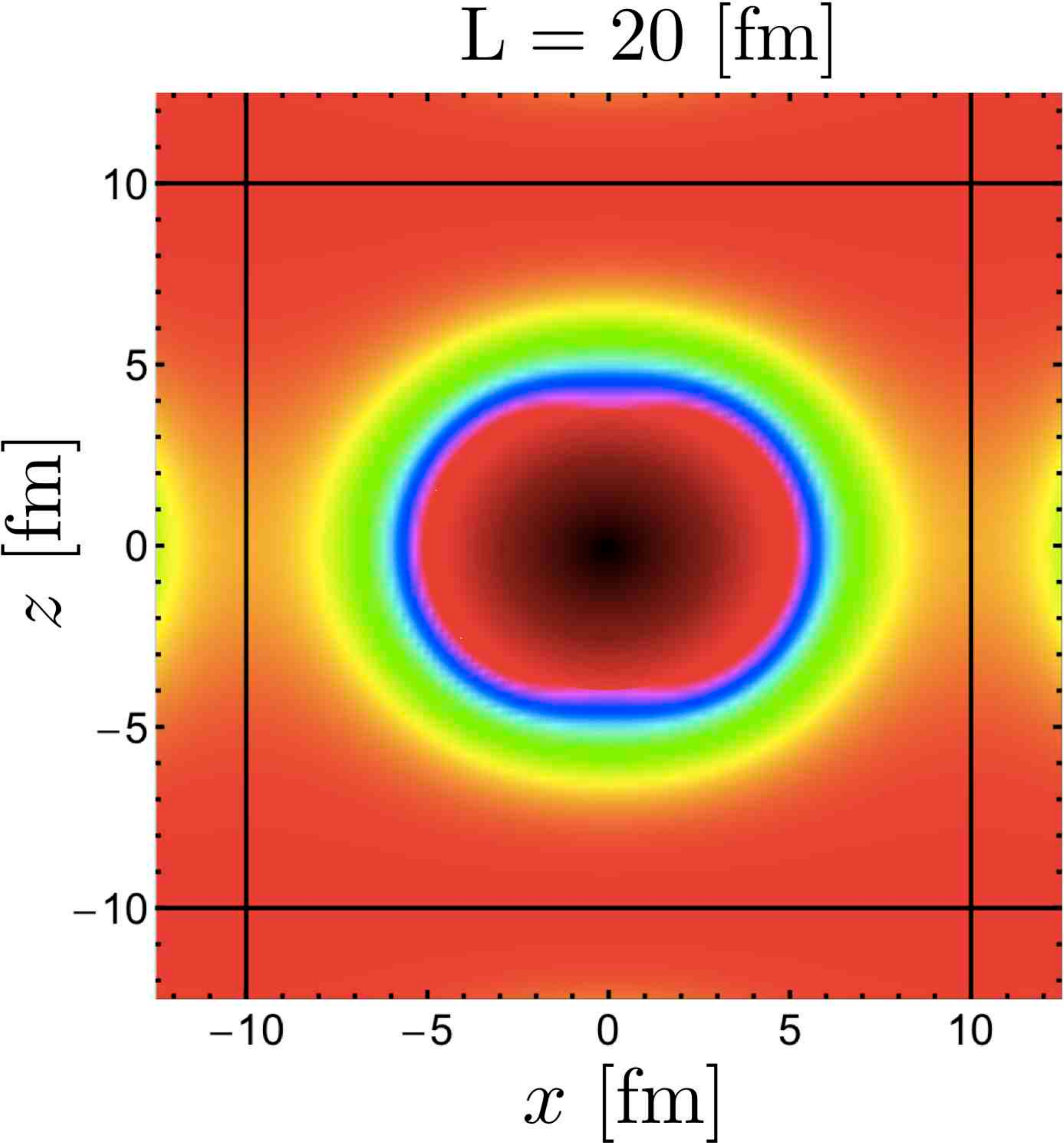}}
\subfigure[]{
\label{WF-A2-30}
\includegraphics[scale=0.3]{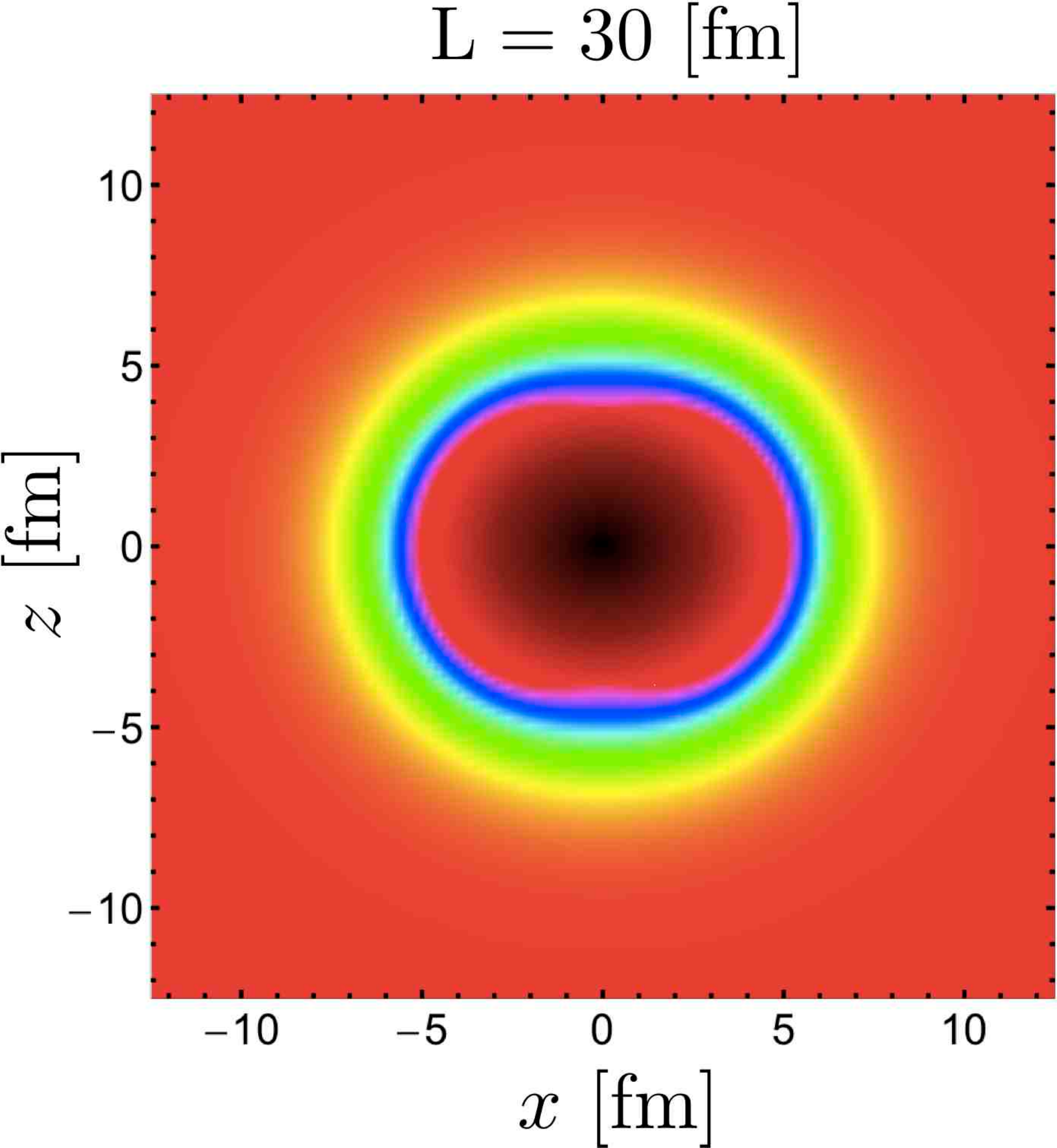}}
\caption{
The mass density  in the $xz$-plane from the $\mathbb{A}_2$ FV deuteron wavefunction with
  $\mathbf{d}=(0,0,1)$.
}
\label{WF-A2}
\end{center}
\end{figure}

\begin{figure}[!ht]
\begin{center}  
\subfigure[]{
\label{WF-E-L10}
\includegraphics[scale=0.3]{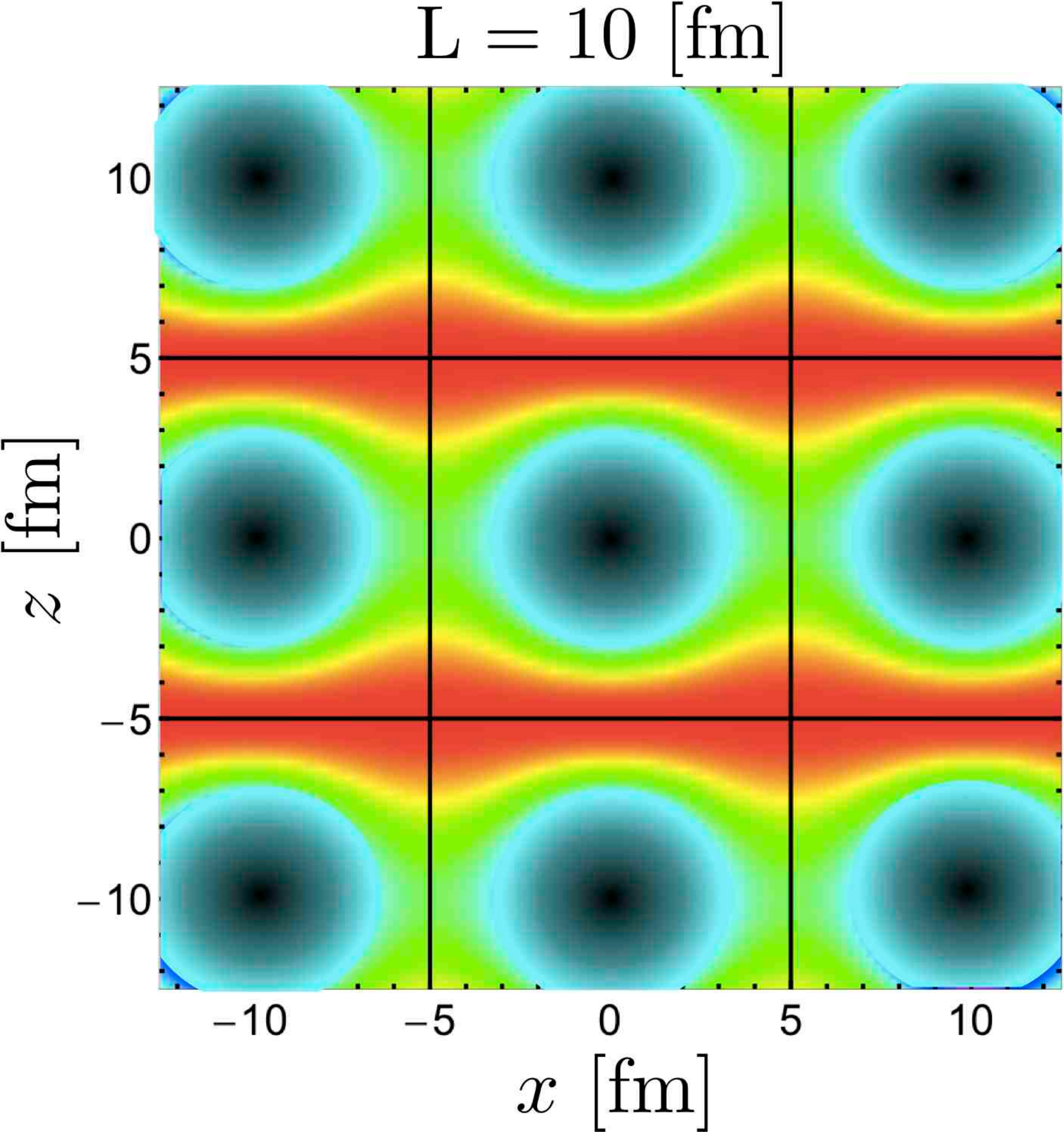}}
\subfigure[]{
\label{WF-E-L15}
\includegraphics[scale=0.3]{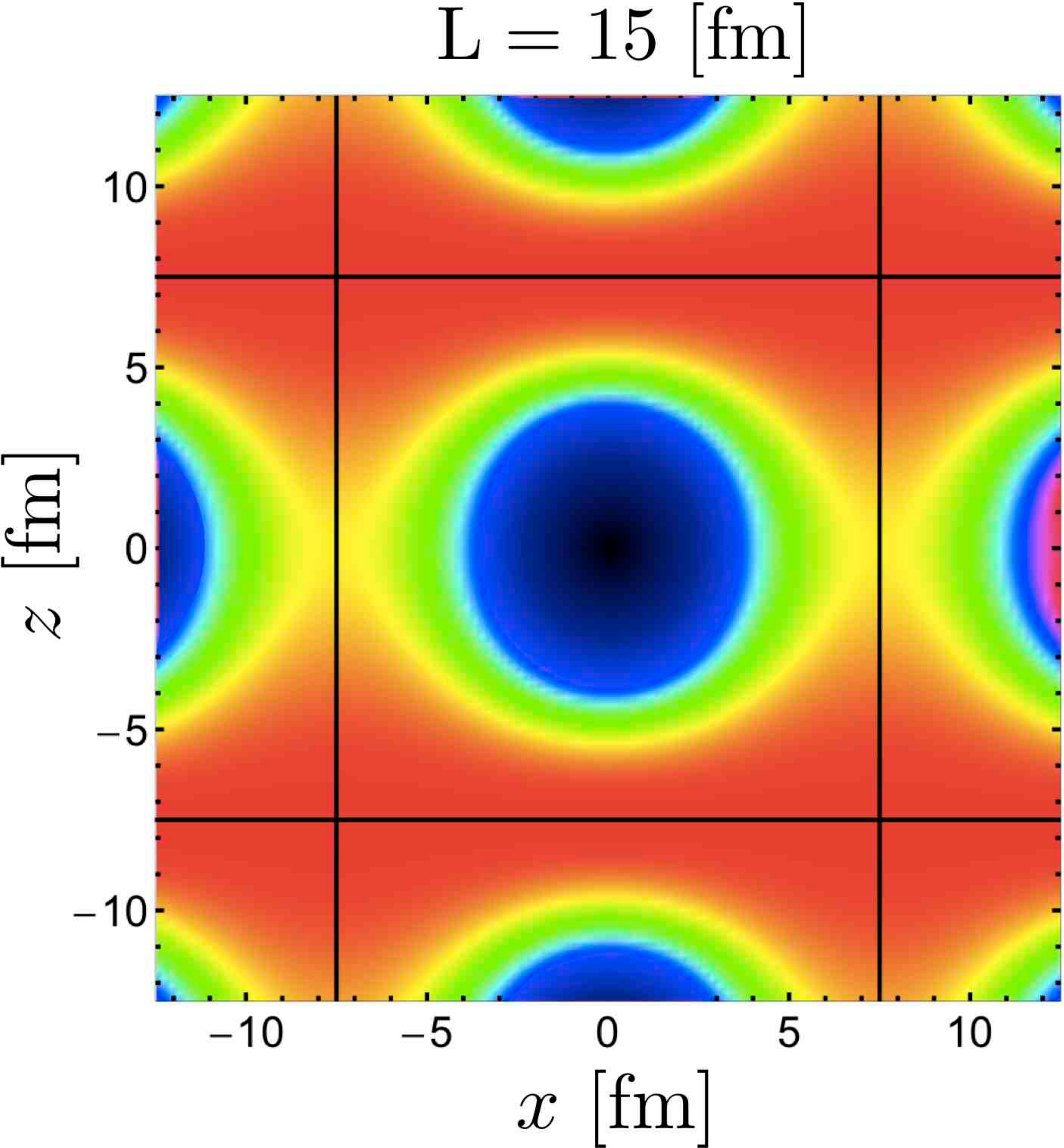}}
\subfigure[]{
\label{WF-E-20}
\includegraphics[scale=0.3]{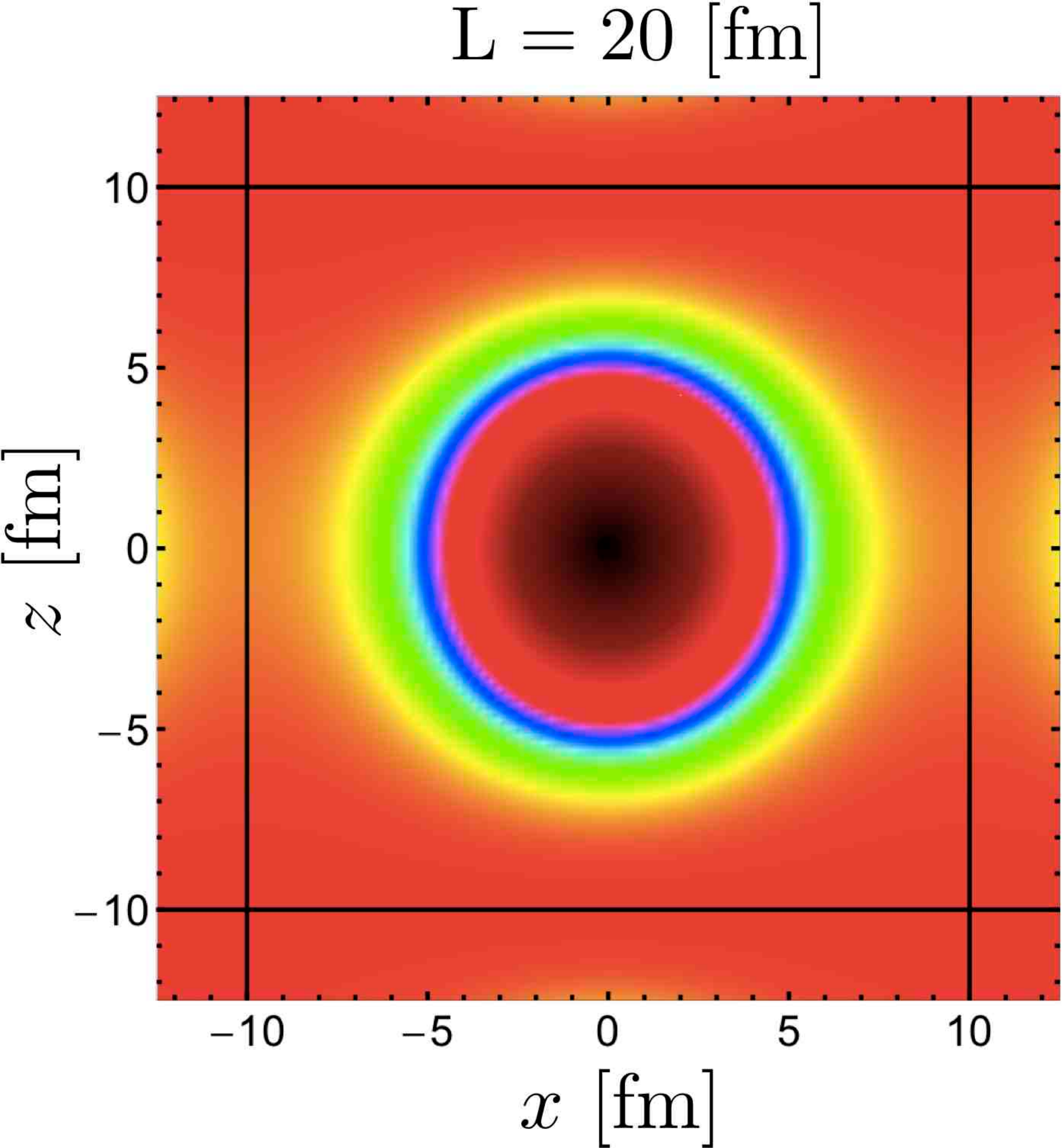}}
\subfigure[]{
\label{WF-E-30}
\includegraphics[scale=0.3]{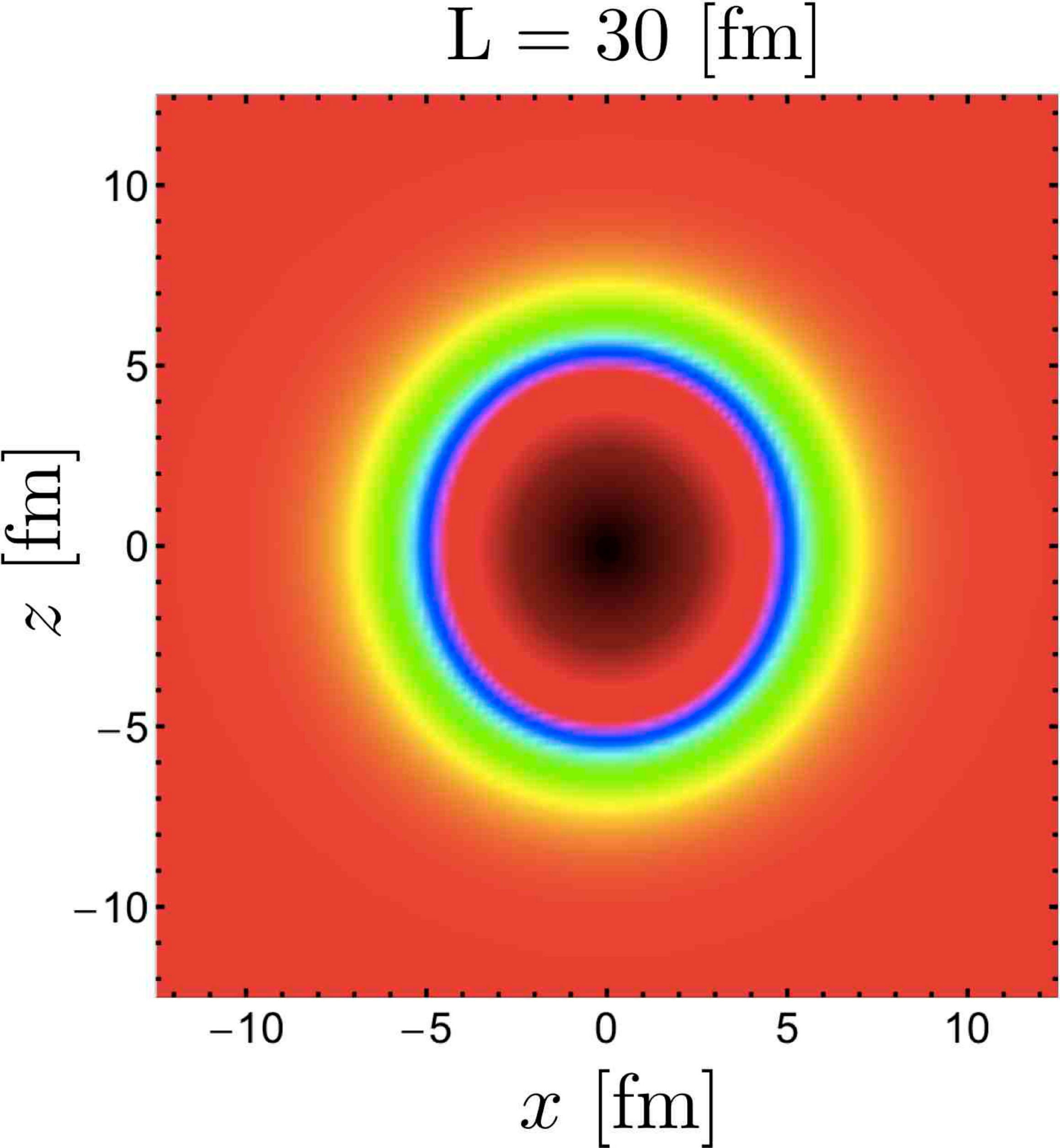}}
\caption{
The mass density  in the $xz$-plane from the $\mathbb{E}$ FV deuteron wavefunction with
  $\mathbf{d}=(0,0,1)$.
}
\label{WF-E}
\end{center}
\end{figure}

\begin{figure}[!ht]
\begin{center}  
\subfigure[]{
\label{WF-B1-L10}
\includegraphics[scale=0.3]{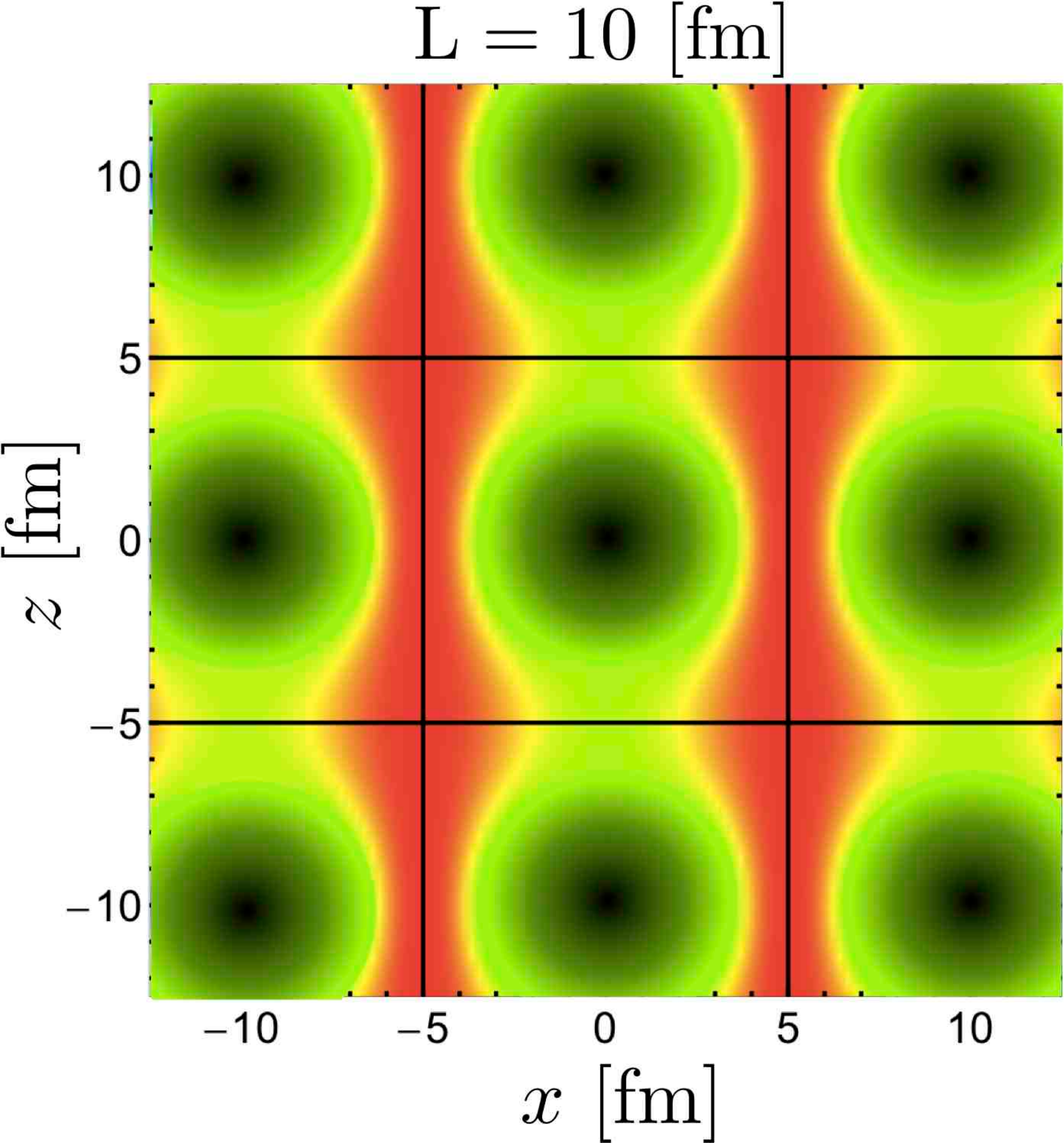}}
\subfigure[]{
\label{WF-B1-L15}
\includegraphics[scale=0.3]{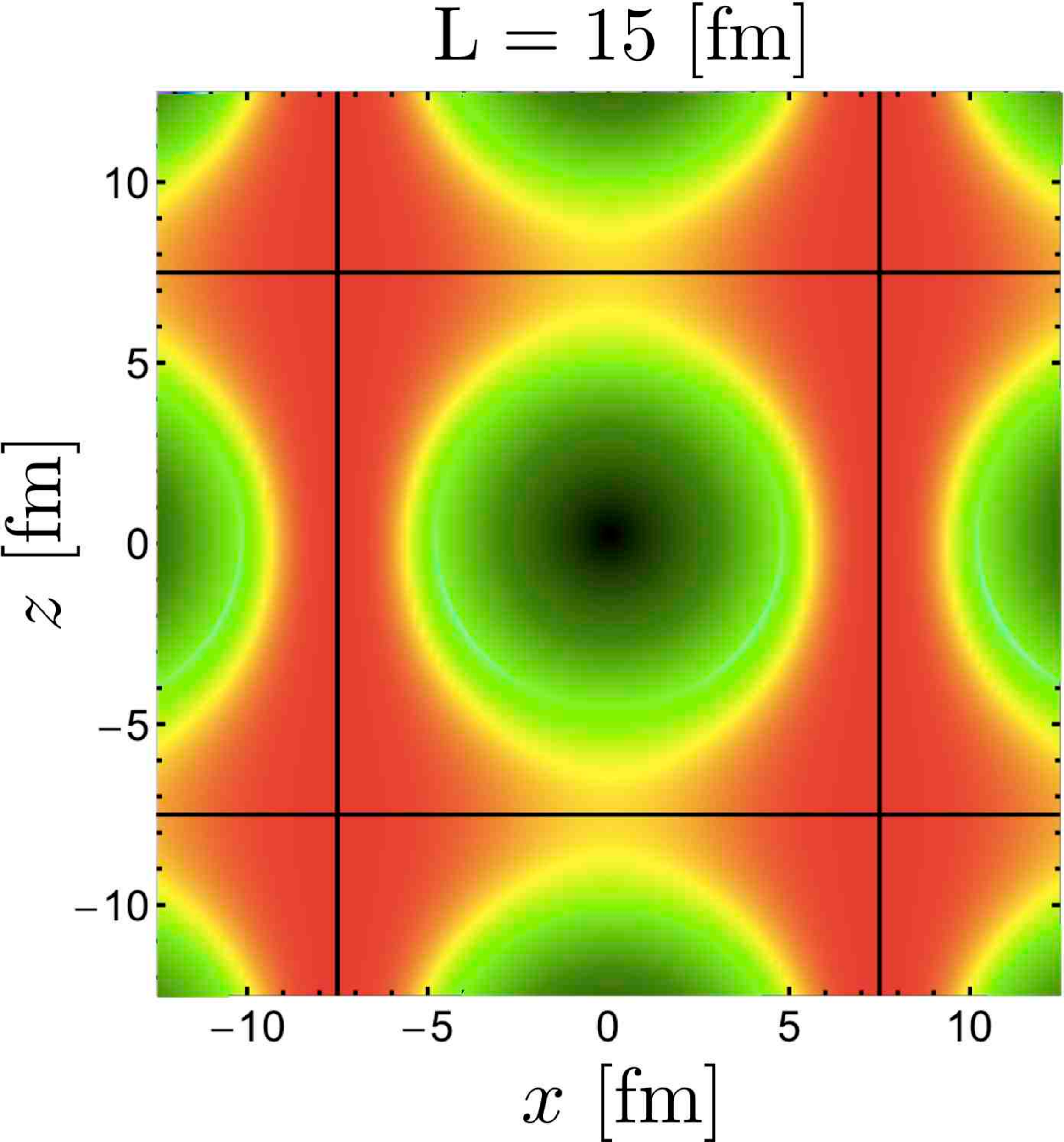}}
\subfigure[]{
\label{WF-B1-20}
\includegraphics[scale=0.3]{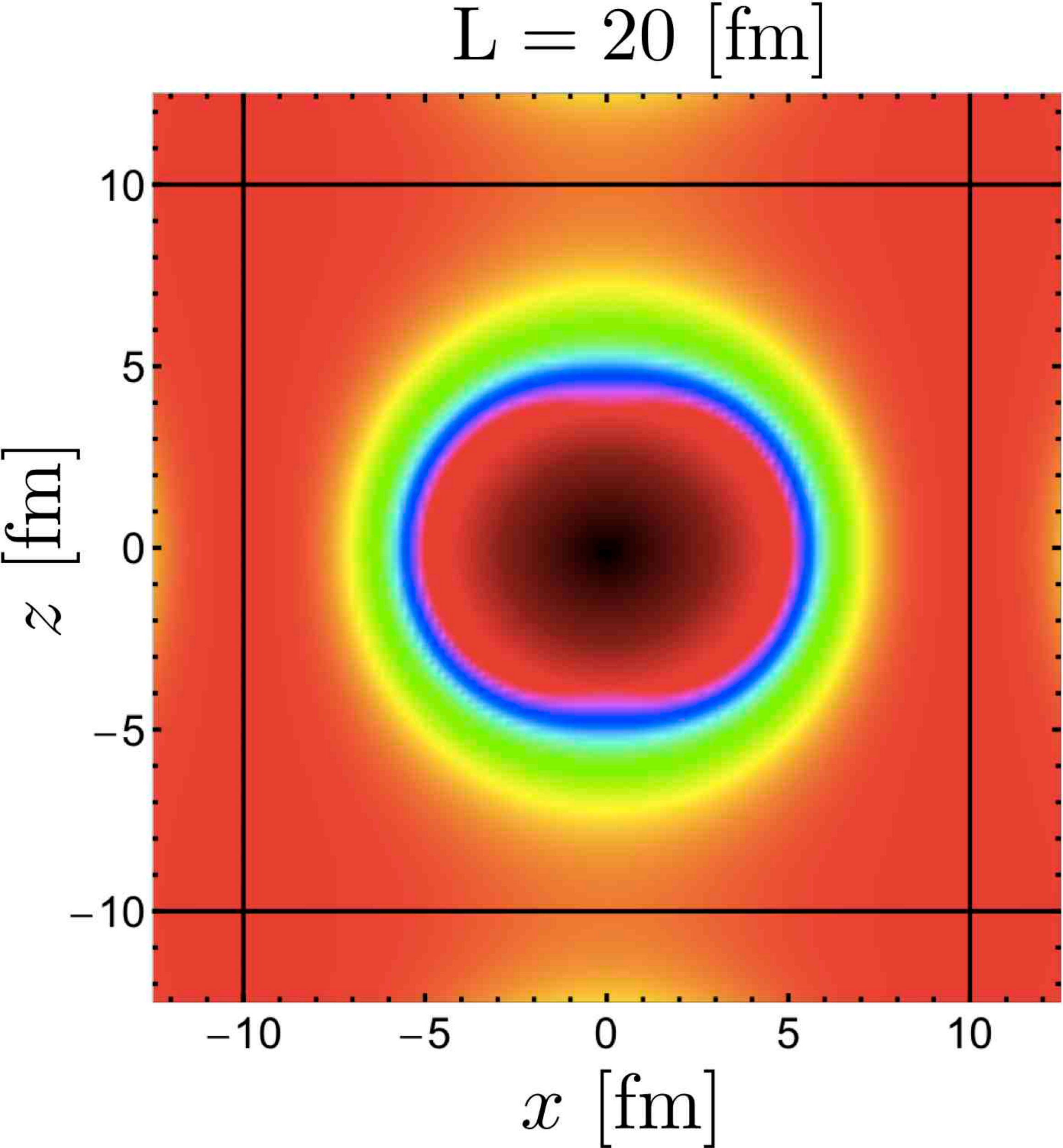}}
\subfigure[]{
\label{WF-B1-30}
\includegraphics[scale=0.3]{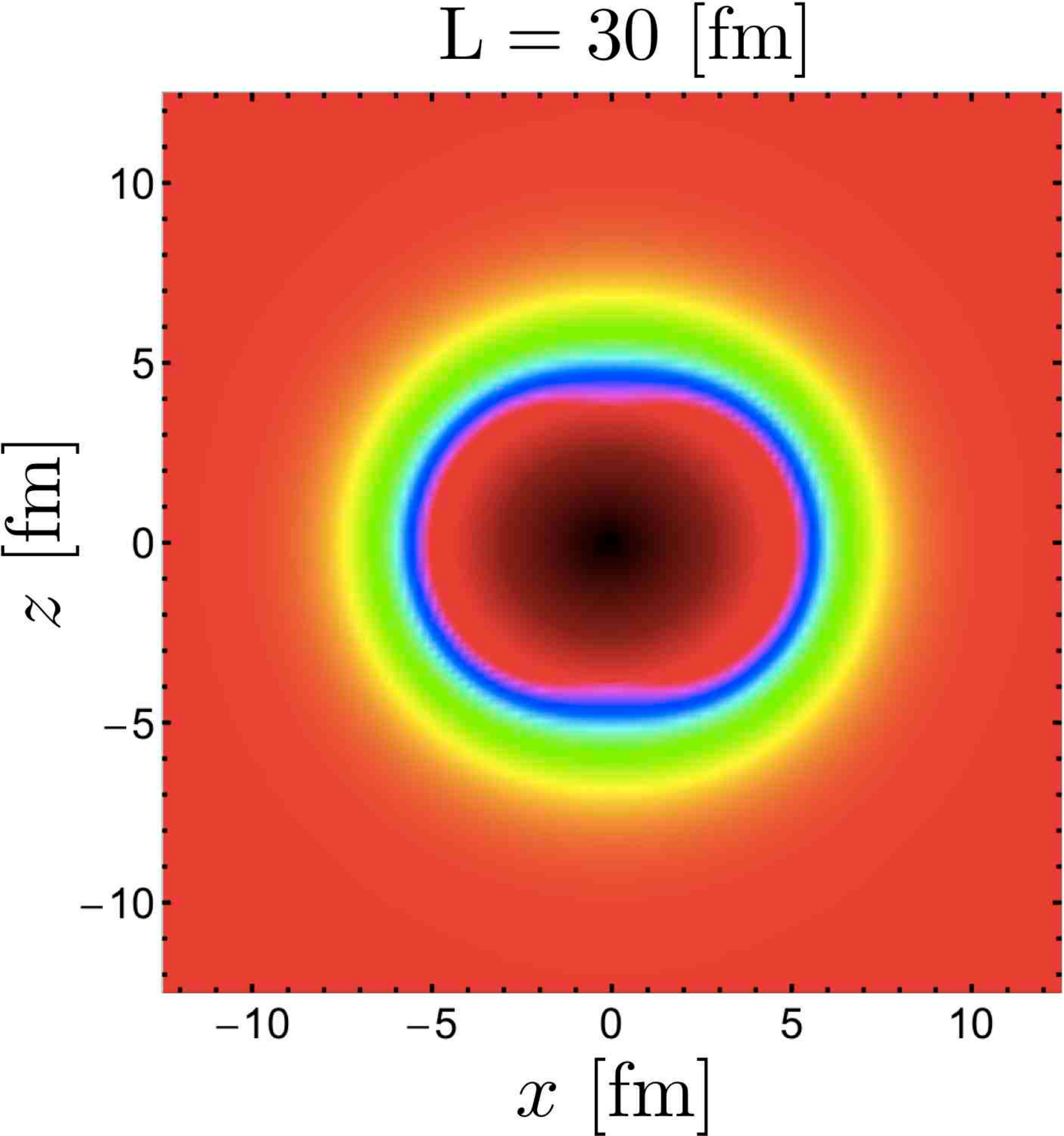}}
\caption{
The mass density  in the $xz$-plane from the $\mathbb{B}_1$ FV deuteron wavefunction with
  $\mathbf{d}=(1,1,0)$.
}
\label{WF-B1}
\end{center}
\end{figure}

\begin{figure}[!ht]
\begin{center}  
\subfigure[]{
\label{WF-B2B3-L10}
\includegraphics[scale=0.3]{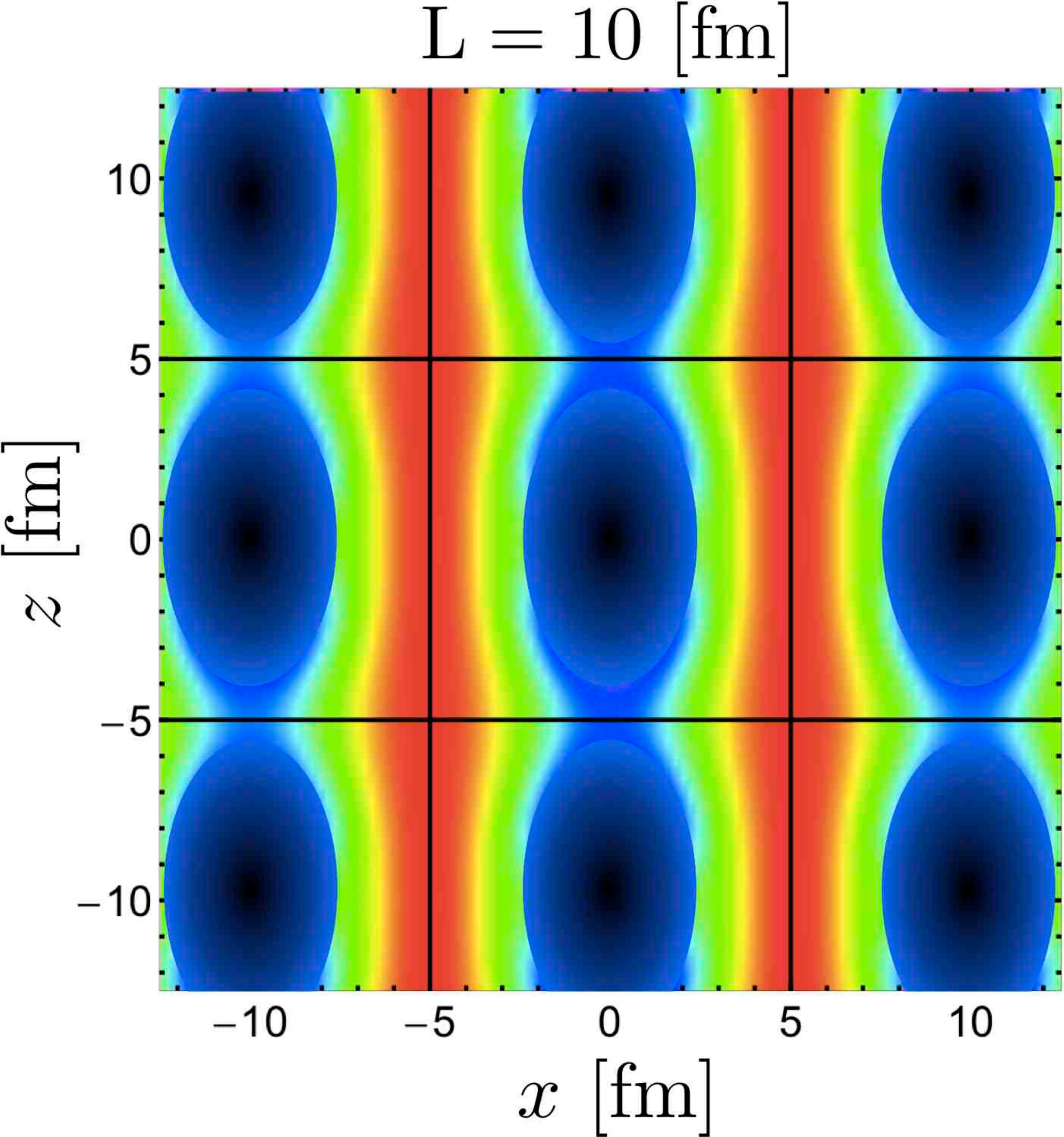}}
\subfigure[]{
\label{WF-B2B3-L15}
\includegraphics[scale=0.3]{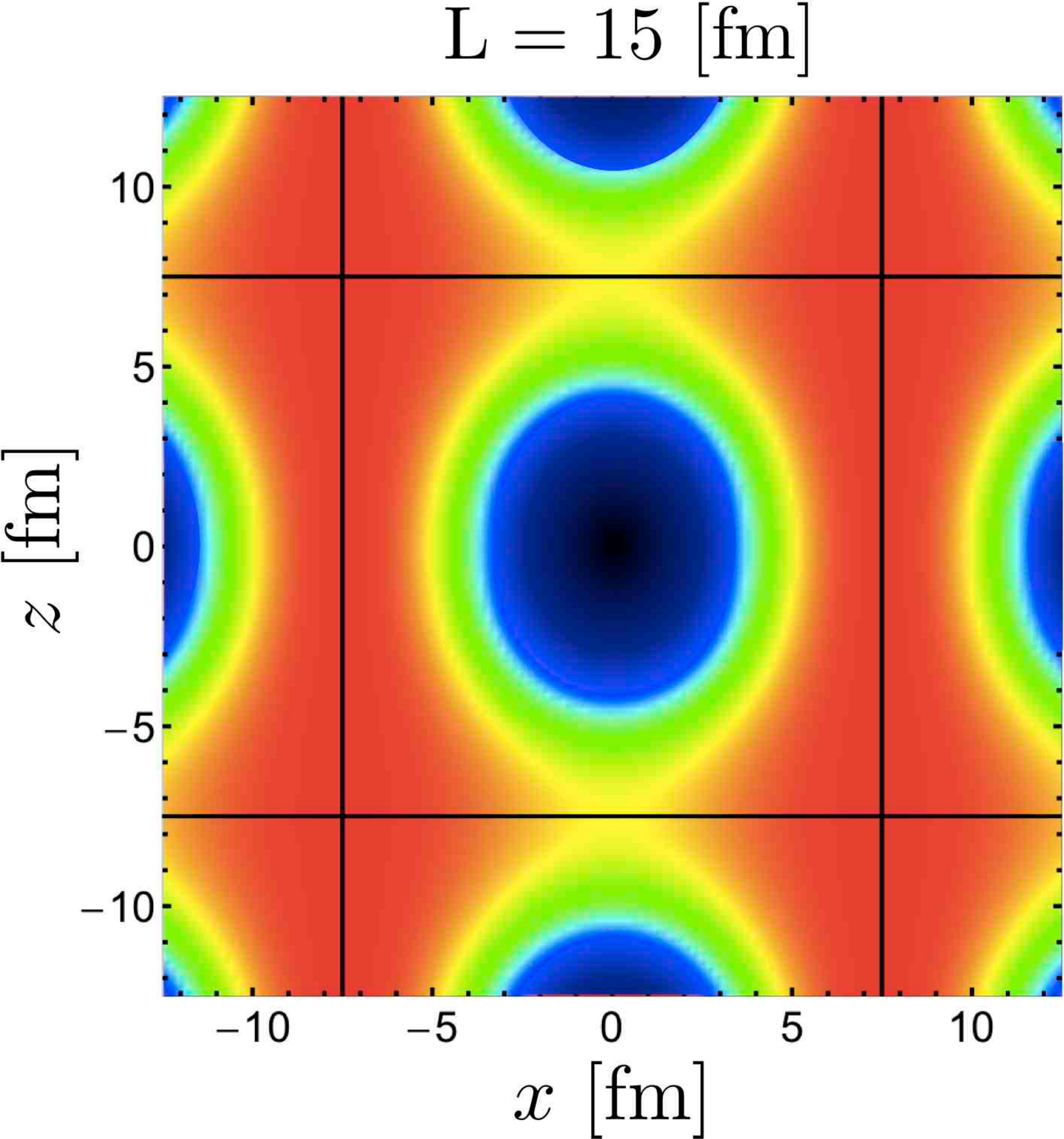}}
\subfigure[]{
\label{WF-B2B3-20}
\includegraphics[scale=0.3]{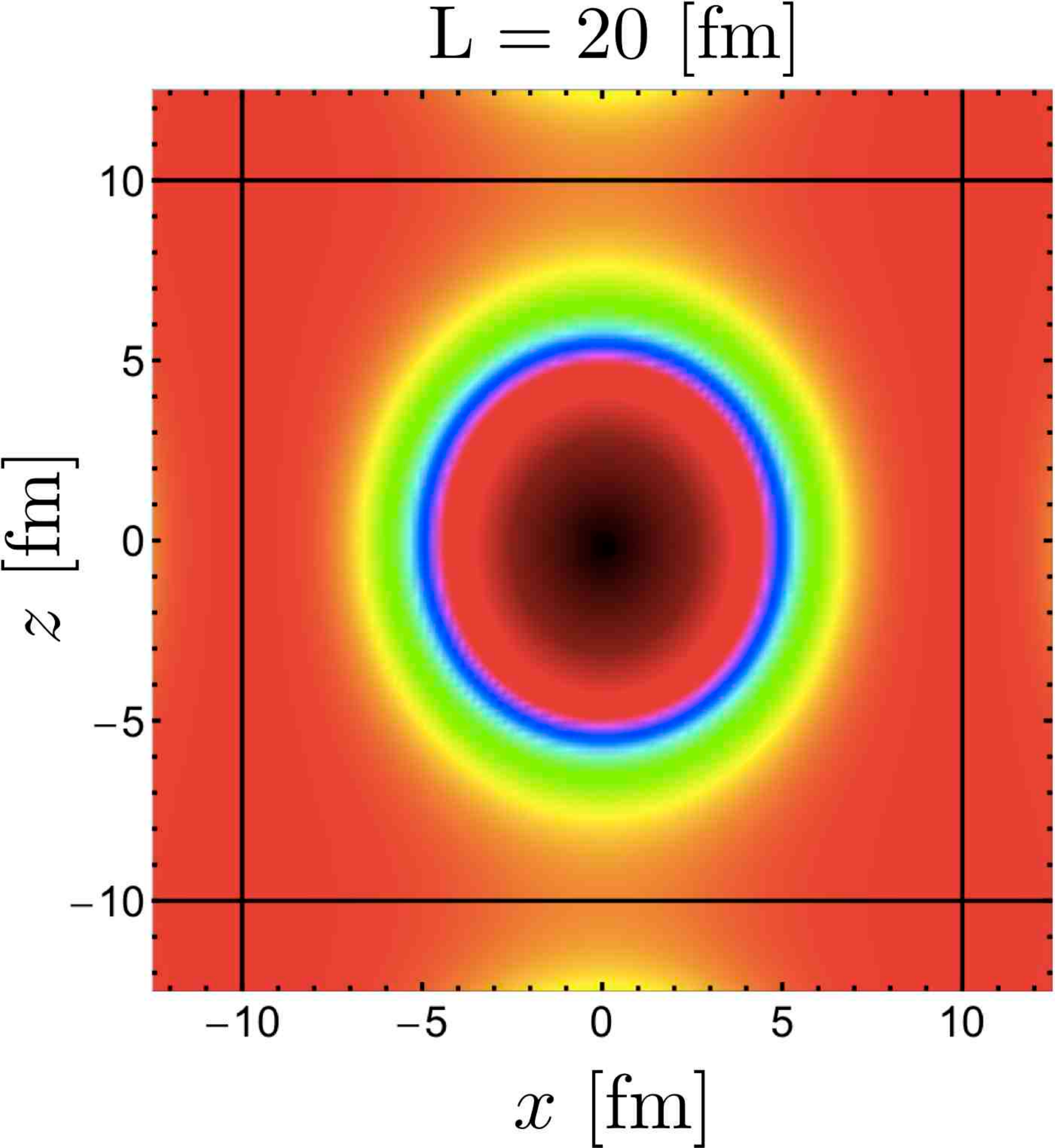}}
\subfigure[]{
\label{WF-B2B3-30}
\includegraphics[scale=0.3]{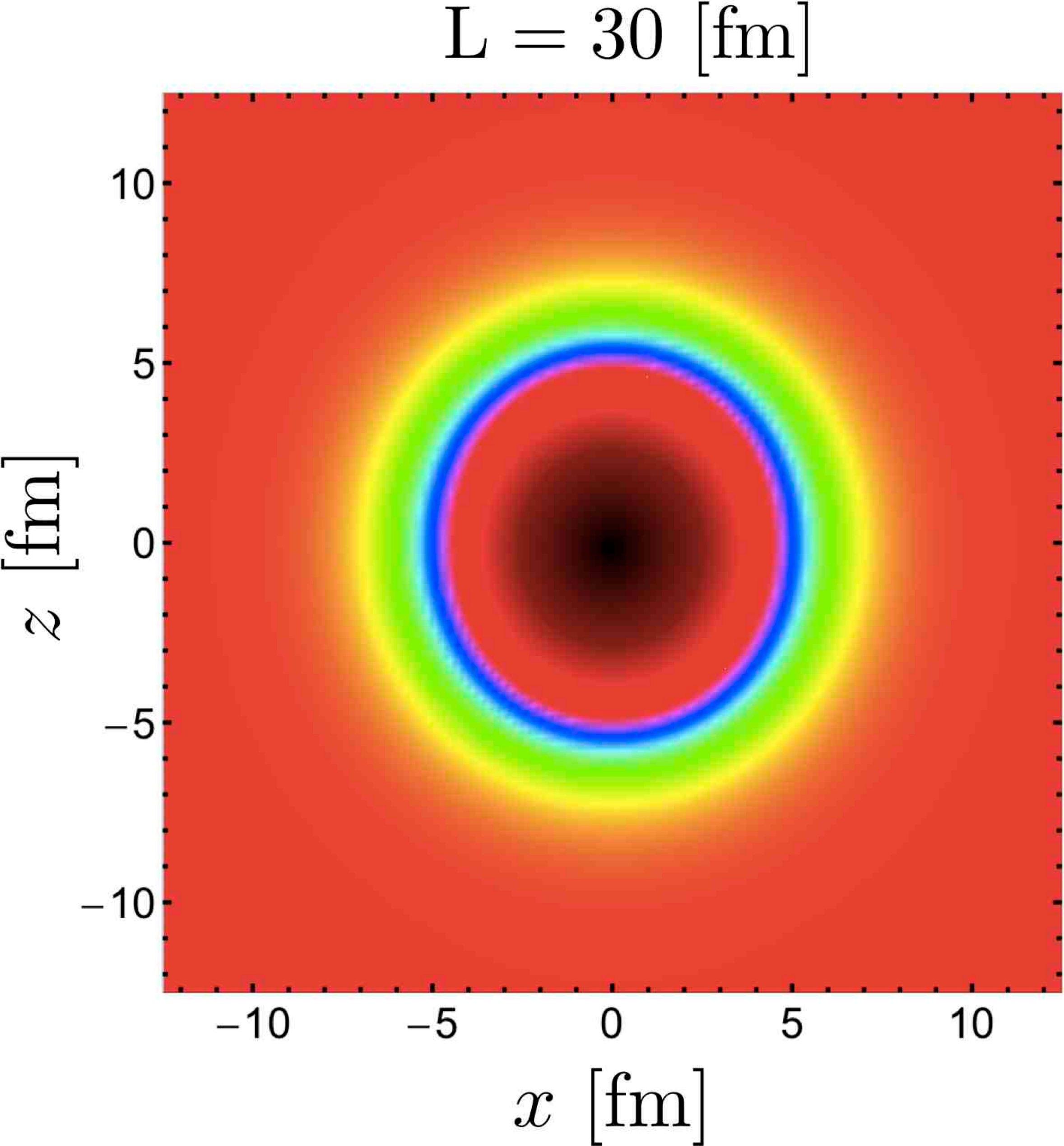}}
\caption{
The mass density  in the $xz$-plane from the $\mathbb{B}_2/\mathbb{B}_3$ FV deuteron wavefunction with
  $\mathbf{d}=(1,1,0)$.
}
\label{WF-B2B3}
\end{center}
\end{figure}

\begin{figure}[!ht]
\begin{center}  
\subfigure[]{
\label{WF-A2E-L10}
\includegraphics[scale=0.3]{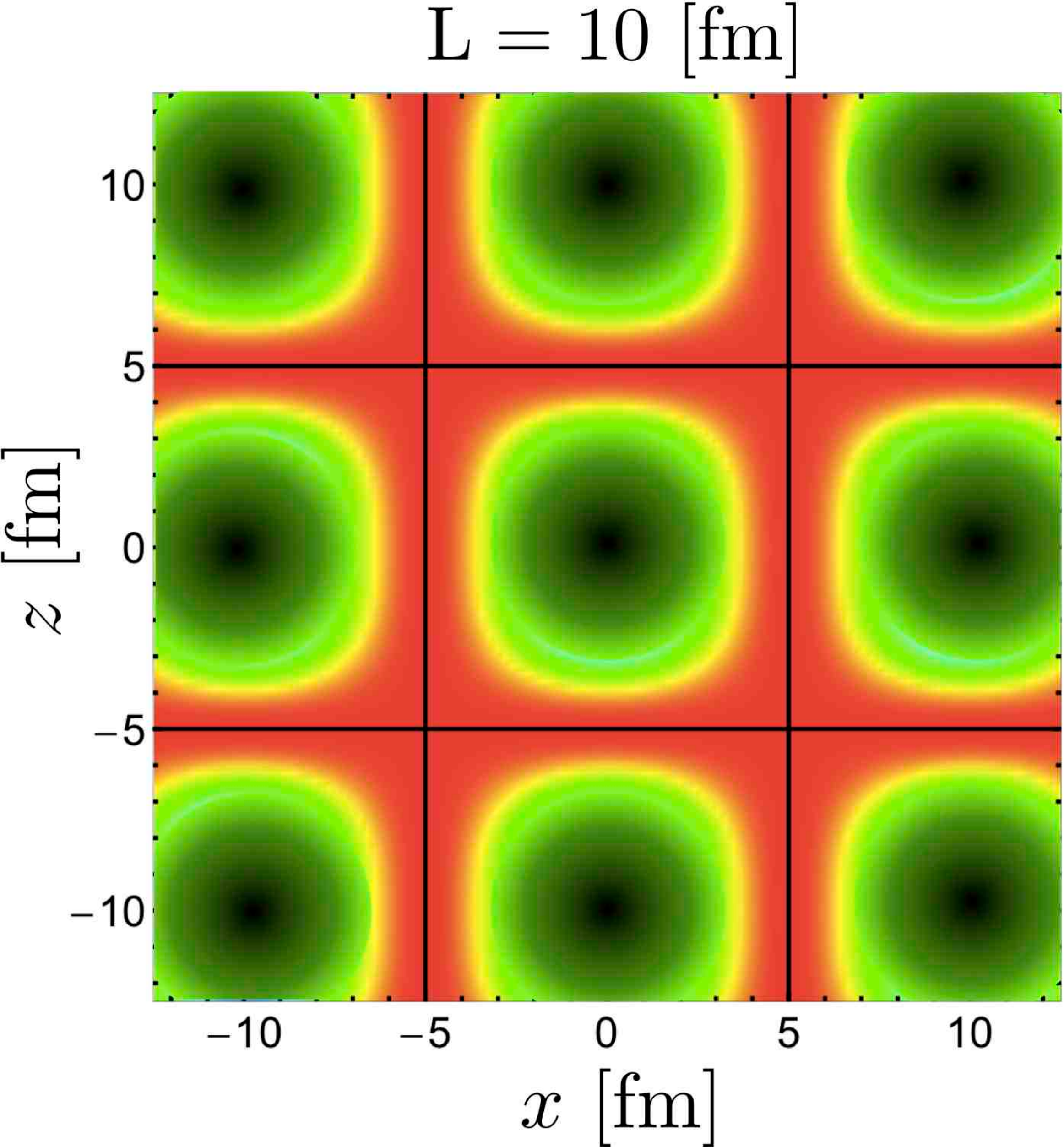}}
\subfigure[]{
\label{WF-A2E-L15}
\includegraphics[scale=0.3]{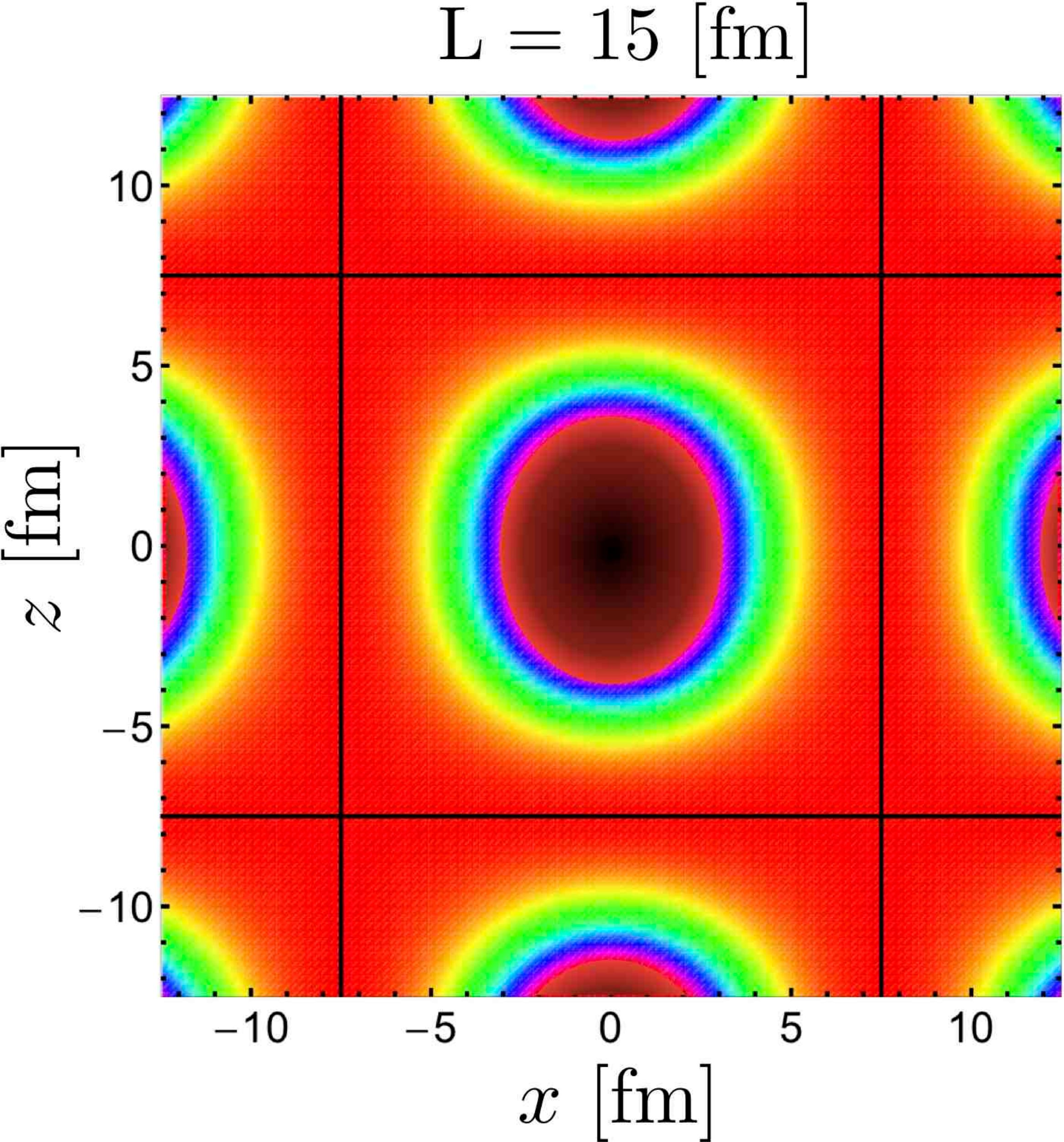}}
\subfigure[]{
\label{WF-A2E-20}
\includegraphics[scale=0.3]{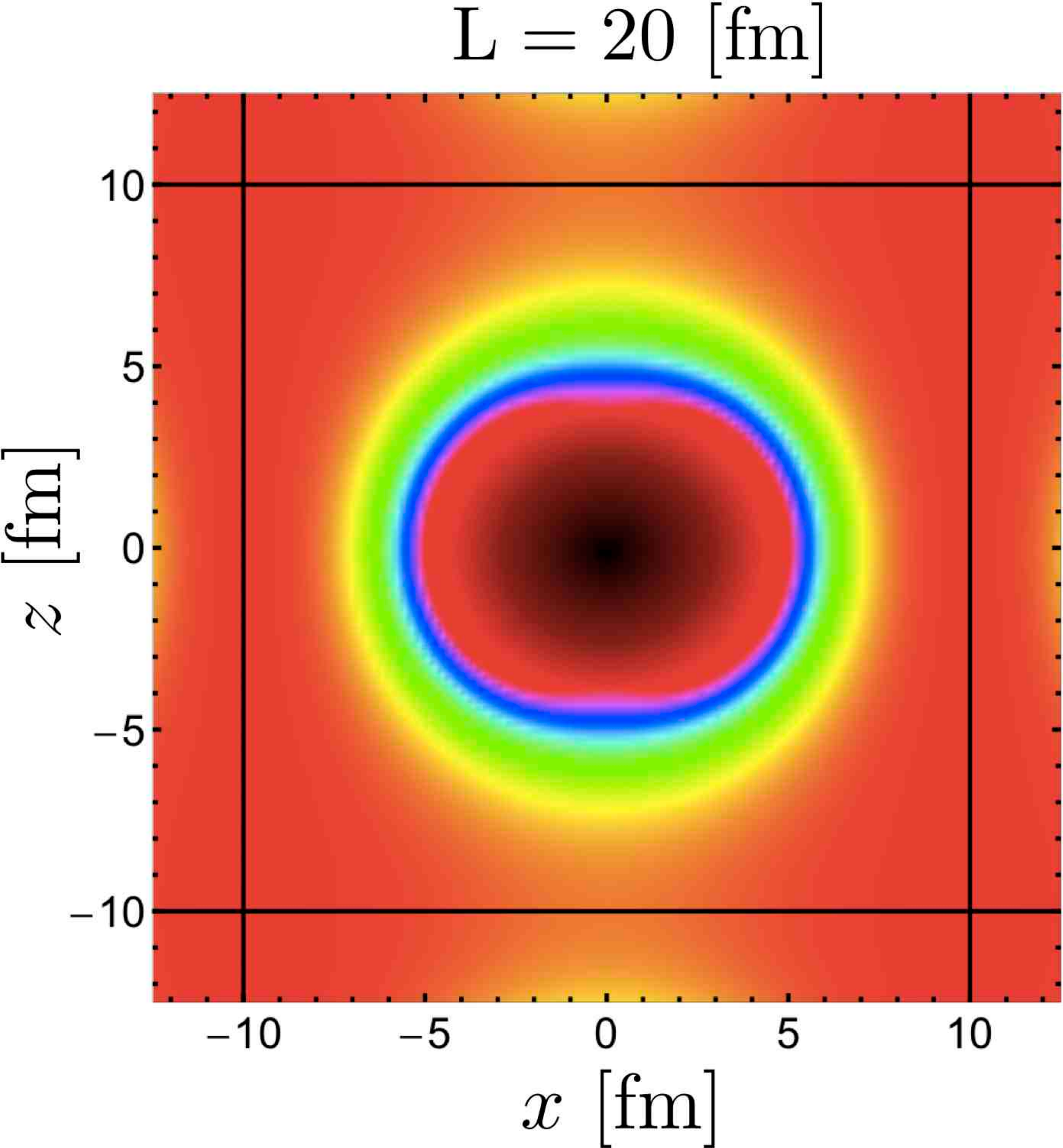}}
\subfigure[]{
\label{WF-A2E-30}
\includegraphics[scale=0.3]{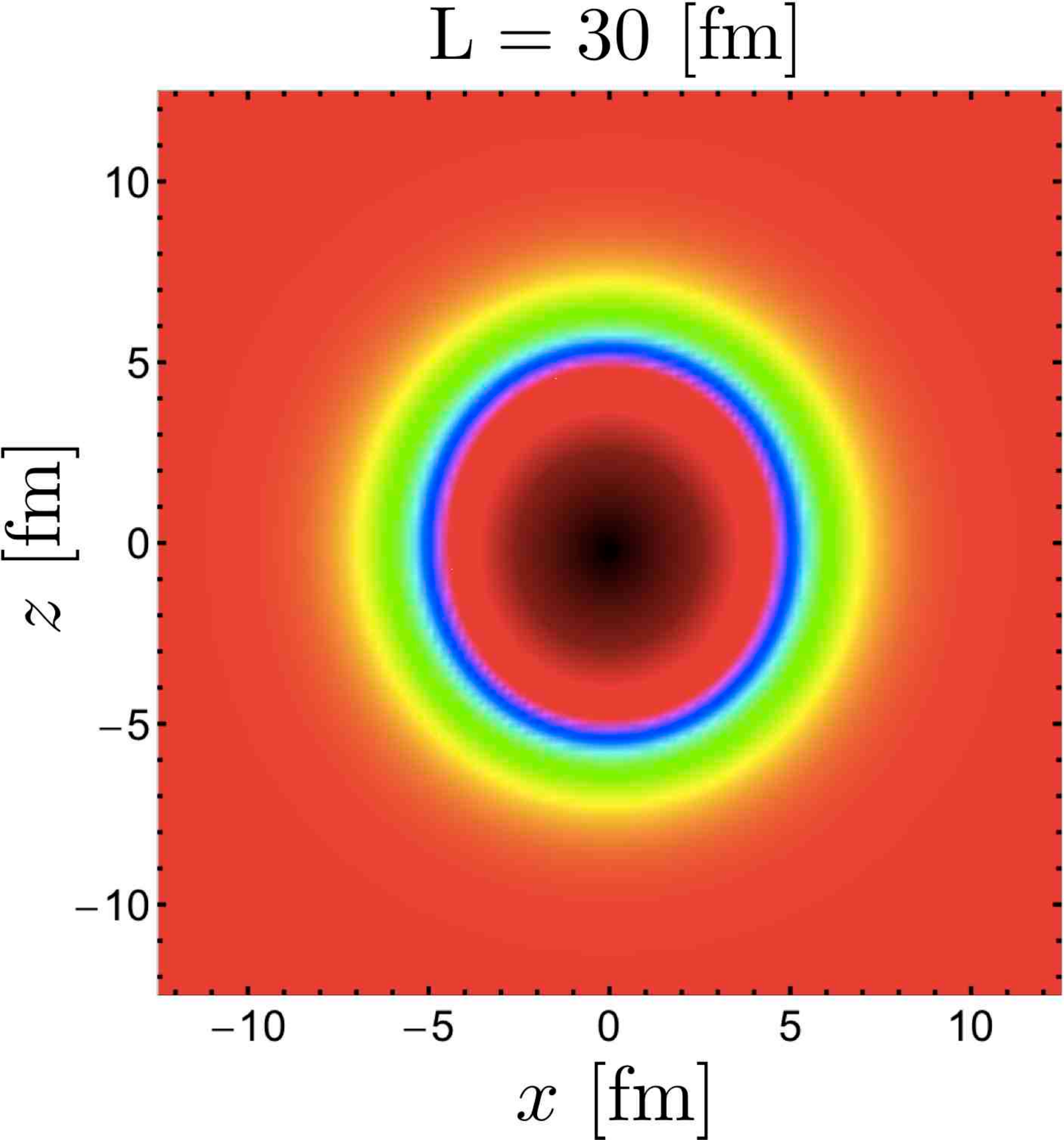}}
\caption{
The mass density  in the $xz$-plane from the $\mathbb{A}_2/\mathbb{E}$ FV deuteron wavefunction with
  $\mathbf{d}=(1,1,1)$.
}
\label{WF-A2E}
\end{center}
\end{figure}

\end{document}